\documentclass[12pt,preprint]{aastex}
\usepackage{epsf}
\usepackage{graphicx}
\usepackage{natbib}

\def\kms{\ifmmode {\rm km~s}^{-1}\else km~s$^{-1}$\fi}
\def\rsun{\ifmmode R_\odot\else $R_\odot$\fi}
\def\msun{\ifmmode M_\odot\else $M_\odot$\fi}
\def\msunyr{\ifmmode M_\odot~{\rm yr}^{-1}\else $M_\odot~{\rm yr}^{-1}$\fi}
\def\rsunv{\ifmmode \vec{r}_\odot\else $\vec{r}_\odot$\fi}
\def\vsunv{\ifmmode \vec{v}_\odot\else $\vec{v}_\odot$\fi}
\def\vsun{\ifmmode v_\odot\else $v_\odot$\fi}
\def\xsun{\ifmmode x_\odot\else $x_\odot$\fi}
\def\vej{\ifmmode v_{ej}\else $v_{ej}$\fi}
\def\veja{\ifmmode v_{ej,H}\else $v_{ej,H}$\fi}
\def\vejb{\ifmmode v_{ej,S}\else $v_{ej,S}$\fi}
\def\vejc{\ifmmode v_{ej,D}\else $v_{ej,D}$\fi}
\def\r0{\ifmmode \vec{r}_0\else $\vec{r}_0$\fi}
\def\rf{\ifmmode \vec{r}_f\else $\vec{r}_f$\fi}
\def\rstarv{\ifmmode \vec{r}_\star\else $\vec{r}_\star$\fi}
\def\v0{\ifmmode \vec{v}_0\else $\vec{v}_0$\fi}
\def\vf{\ifmmode \vec{v}_f\else $\vec{v}_f$\fi}
\def\vstarv{\ifmmode \vec{v}_\star\else $\vec{v}_\star$\fi}
\def\xdot{\ifmmode \dot{x}\else $\dot{x}$\fi}
\def\ydot{\ifmmode \dot{y}\else $\dot{y}$\fi}
\def\zdot{\ifmmode \dot{z}\else $\dot{z}$\fi}
\def\xdots{\ifmmode \dot{x}\else $\dot{x}$\fi}
\def\ydots{\ifmmode \dot{y}\else $\dot{y}$\fi}
\def\zdots{\ifmmode \dot{z}\else $\dot{z}$\fi}
\def\deg{\ifmmode ^{\rm o}\else $^{\rm o}$\fi}
\def\facR{\ifmmode f_{R}\else $f_R$\fi}
\def\abin{\ifmmode a_{bin}\else $a_{bin}$\fi}
\def\amin{\ifmmode a_{min}\else $a_{min}$\fi}
\def\amax{\ifmmode a_{max}\else $a_{max}$\fi}
\def\ma{\ifmmode M_1\else $M_1$\fi}
\def\mb{\ifmmode M_2\else $M_2$\fi}
\def\mbh{\ifmmode M_{bh}\else $M_{bh}$\fi}
\def\rclose{\ifmmode r_{close}\else $r_{close}$\fi}
\def\masyr{\ifmmode {\rm milliarcsec~yr^{-1}}\else milliarcsec~yr$^{-1}$\fi}

\begin{document}

\title{Predicted Space Motions for Hypervelocity and Runaway Stars: 
Proper Motions and Radial Velocities for the GAIA Era}

\author{Scott J. Kenyon}
\affil{Smithsonian Astrophysical Observatory,
 60 Garden St., Cambridge, MA 02138}
\email{skenyon@cfa.harvard.edu}

\author{Benjamin C. Bromley}
\affil{Department of Physics, University of Utah,
 115 S 1400 E, Rm 201, Salt Lake City, UT 84112}
\email {bromley@physics.utah.edu}

\author{Warren R. Brown}
\affil{Smithsonian Astrophysical Observatory,
 60 Garden St., Cambridge, MA 02138}
\email{wbrown@cfa.harvard.edu}

\author{Margaret J. Geller}
\affil{Smithsonian Astrophysical Observatory,
 60 Garden St., Cambridge, MA 02138}
\email{mgeller@cfa.harvard.edu}

\clearpage

\begin{abstract}
We predict the distinctive three dimensional space motions of 
hypervelocity stars (HVSs) and runaway stars moving in a realistic 
Galactic potential.  For nearby stars with distances less than 
10~kpc, unbound stars are rare; proper motions alone rarely isolate
bound HVSs and runaways from indigenous halo stars.  At large 
distances of 20--100~kpc, unbound HVSs are much more common 
than runaways; radial velocities easily distinguish both from 
indigenous halo stars.  Comparisons of the predictions with 
existing observations are encouraging.  Although the models fail 
to match observations of solar-type HVS candidates from SEGUE, 
they agree well with data for B-type HVS and runaways from other 
surveys.  Complete samples of $g \lesssim$ 20 stars with GAIA 
should provide clear tests of formation models for HVSs and 
runaways and will enable accurate probes of the shape of the 
Galactic potential.

\end{abstract}


\section{INTRODUCTION}
\label{sec: intro}

From the Galactic Center (GC) to the Magellanic Clouds, three dimensional 
(3D) space motions yield interesting information on the mass distribution 
and stellar populations in the Local Group. At the GC, proper motion and 
radial velocity data for several dozen bright O-type and B-type stars orbiting 
Sgr A$^*$ reveal the existence of a black hole with a mass of roughly 
$4 \times 10^6$ \msun\ \citep[e.g.,][]{genzel2010,morris2012}. 
For the LMC, 3D motions of several thousand stars allow measures of the 
orientation of the stellar disk and the mass contained within $\sim$ 9~kpc 
\citep{vdmarel2014}. On distance scales intermediate between these two 
extremes, accurate space motions of large groups of stars bound to the
Milky Way measure 
(i) the rotation of the Galactic bulge \citep[e.g.,][]{soto2014},
(ii) the kinematics of nearby OB associations in the Galactic disk 
\citep[e.g.,][]{dezeeuw1999,reid2014}, and
(iii) the frequency of streams of stars in the Milky Way halo \citep{kopos2013}.

Unbound stars ejected from the Milky Way can also probe Galactic structure.
HVSs are ejected from the GC when a close binary system passes within the 
tidal boundary of the central supermassive black hole\footnote{HVS ejections 
also occur when a single or binary star passes too close to a binary black hole 
\citep[e.g.,][]{yu2003}.  Here, we focus on the original Hills (1988) mechanism
for a single black hole at the GC.} 
\citep[SMBH;][]{hills1988}.  During this passage, one component of the binary 
becomes bound to the SMBH; to conserve energy, the other is ejected at velocities 
ranging from a few hundred to a few thousand \kms. Robust identification of 
unbound HVSs in the halo enables more accurate measurements of the total mass 
of the Milky Way \citep[e.g.,][]{brown2010a,gnedin2010}.  Because HVSs leave 
the GC on nearly radial orbits, measuring the 3D trajectories of unbound HVSs 
in the halo constrain the anisotropy of the Galactic potential 
\citep[e.g.,][]{gnedin2005,yu2007}. 

Space motions of runaway stars may provide additional constraints on the 
Galactic potential \citep[e.g.,][]{martin2006}.  Produced when one component 
of a binary system explodes as a supernova \citep{zwicky1957,blaauw1961} or 
when a star receives kinetic energy through dynamical interactions with several 
more massive stars \citep[e.g.,][]{pov1967,leon1991}, high velocity runaways
have space motions and spatial distributions distinct from HVSs 
\citep{martin2006,bromley2009}. Separating unbound HVSs from unbound runaways 
should enable more rigorous constraints on the mass of the Galaxy and any 
anisotropy in the Galactic potential.

Realizing these possibilities requires robust predictions for the space motions 
of HVSs and runaways moving through a realistic Galactic potential. Here, we 
focus on calculations in an axisymmetric potential.  Our results demonstrate
that proper motions (radial velocities) isolate nearby (distant) HVSs and 
runaways from indigenous stars. Unique variations of proper motion and radial 
velocity with Galactic longitude and latitude enable new ways to identify high 
velocity stars. For observed high velocity stars, comparisons with the models 
indicate a mix of HVSs and runaways, with a strong preference for an HVS origin 
among the most distant stars.

\section{OVERVIEW}
\label{sec: overview}

To predict proper motions and radial velocities for HVSs and runaways, we 
consider both analytic models and numerical simulations.  For stars with
specific trajectories, analytic models allow us to derive the variations in 
proper motion and radial velocity as a function of position in the Galaxy. 
Numerical simulations yield predictions for the distributions of positions
and space motions for specific models of HVSs and runaways.

We begin in \S\ref{sec: an-model} with a formal discussion of the analytic 
model.  After defining cartesian, cylindrical, and spherical coordinate 
systems (\S\ref{sec: an-defs}), we derive radial and tangential velocities 
for stars (i) orbiting the Galaxy (\S\ref{sec: an-orb}) and 
(ii) moving radially away from the GC (\S\ref{sec: an-rad}). Features in the 
behavior of the radial and tangential velocities with distance and Galactic 
coordinates provide a basis for differentiating the two types of motion.

Readers more interested in results than techniques can use the figures in
\S\ref{sec: an-model} as a guide and concentrate on \S3.3, where we summarize 
the relative value of radial velocities and proper motions for identifying 
HVSs and runaways among indigenous stars.  For stars at distances $d \gtrsim$ 
20~kpc from the Sun, radial velocities separate orbital motion from radial 
motion.  For nearby stars ($d \lesssim$ 20~kpc), tangential velocities may
discriminate ejected stars from bulge and disk stars, but probably cannot
isolate ejected stars from halo stars.

In \S\ref{sec: sims}, we describe numerical techniques for simulating 
HVSs and runaways moving through the Galaxy. Our procedures follow those 
discussed in \citet{bromley2006}, \citet{kenyon2008}, and \citet{bromley2009}. 
Here, we focus on the input gravitational potential for the Galaxy 
(\S\ref{sec: sims-pot}), the initial conditions (\S\ref{sec: sims-init}), 
and the integration technique (\S\ref{sec: sims-tech}).

We discuss results in four sections. We start by considering the fraction of
ejected stars which reach the outer Galaxy with Galactocentric distances 
$r \gtrsim$ 60~kpc and high Galactic latitude, $|b| \gtrsim$ 
30\deg\ (\S\ref{sec: halo-frac}). With their large ejection velocities, 25\% of 
HVSs reach the outer halo. Much smaller ejection velocities prevent a large fraction 
of runaways from leaving the inner disk.  For supernova-induced (dynamically 
ejected) runaways, only 1\% (0.25\%) reach Galactocentric distances of 60~kpc. 
Roughly 0.1\% of either type of runaway achieves Galactocentric distances of 
60~kpc and $|b| \gtrsim$ 30\deg.

In \S\ref{sec: res1}, we examine distributions of the proper motion $\mu$ and 
radial velocity $v_r$ for complete samples of stars produced in simulations 
of HVSs and runaways.  After exploring the density of stars in the $d - \mu$ 
(\S\ref{sec: res1-mud}) and the $d - v_r$ (\S\ref{sec: res1-vrd}) planes, we 
examine the distributions of radial velocity and proper motion in specific 
distance bins and distributions of proper motion for all stars in each 
simulation (\S\ref{sec: res1-vrmuhist}) and the density of stars as a function 
of Galactic coordinates (\S\ref{sec: res1-mul}). \S\ref{sec: res1-summ} briefly 
summarizes the highlights of these simulations.

To establish predictions for surveys with GAIA and other facilities, we 
continue by constructing magnitude-limited samples of HVSs and runaways
for 1~\msun\ and 3~\msun\ stars (\S\ref{sec: res2}). In the $v_r - \mu$ plane,
magnitude-limited samples of nearby, mostly bound HVSs and runaways with 
d $\lesssim$ 10~kpc have nearly identical distributions, complicating attempts 
to isolate these stars from the indigenous halo population.  Among 3~\msun\ stars, 
high velocity HVSs easily distinguish themselves from high velocity runaways.

Comparisons between the numerical results and observations of  several sets 
of high velocity stars complete our analysis (\S\ref{sec: obs}). HVS and
runaway models yield a poor match to observations of solar-type HVS candidates 
from SEGUE \citep[\S\ref{sec: obs1};][]{palla2014}. However, the models provide 
an excellent match to observations of B-type HVS candidates 
\citep[\S\ref{sec: obs2};][]{brown2014}, 
nearby B-type runaways \citep[\S\ref{sec: obs3};][]{silva2011}, 
and miscellaneous HVS and runaway star candidates from other surveys 
\citep[\S\ref{sec: obsm};][]{edel2005,heber2008,till2009,irrg2010,zheng2014}. 
Although radial velocities easily separate unbound HVSs and runaways from
indigenous halo stars, kinematic data alone are not sufficient to isolate 
bound HVSs or runaways from halo stars (\S\ref{sec: obs-halo}).  Combined 
with estimates of production rates (\S\ref{sec: obs-rates}), these results 
suggest that ejections from the GC are the source of the highest velocity 
stars in the Galactic halo.

Our exploration of the space motions of HVSs and runaways concludes with
a brief discussion and summary (\S\ref{sec: disc-summ}).

\section{ANALYTIC MODEL}
\label{sec: an-model}

\subsection{Definitions}
\label{sec: an-defs}

To establish a framework for analyzing numerical simulations, we consider an
analytic model for the proper motions of stars with simple trajectories in 
the Galaxy.  In a cartesian coordinate system with an origin at the GC, 
stars have positions $(x, y, z)$ and velocities $(v_x, v_y, v_z)$.  The 
distance from the GC to the star is $r$; the space velocity of the star 
relative to the GC is $v$. The angle of the position vector of the star 
relative to the $x$ axis is $\theta$; the angle relative to the $x$--$y$
plane is $\phi$.  To distinguish these angles from standard galactic longitude 
and latitude, we call $\theta$ ($\phi$) the GC longitude (GC latitude). 

In this convention, we specify coordinates in both the cartesian 
$(x, y, z)$ and spherical $(r, \theta, \phi)$ systems \citep[see also][]{bt2008}. 
These systems are appropriate for stars in the Galactic bulge or halo, where the 
potential is roughly spherically symmetric.  To make a clear link 
with the cylindrical coordinate system more appropriate for the 
Galactic disk, we define the cylindrical radius $\rho^2 = x^2 + y^2$.  
In this system, we specify coordinates with $(\rho, \theta, z)$.

\begin{figure}
\includegraphics[width=6.5in]{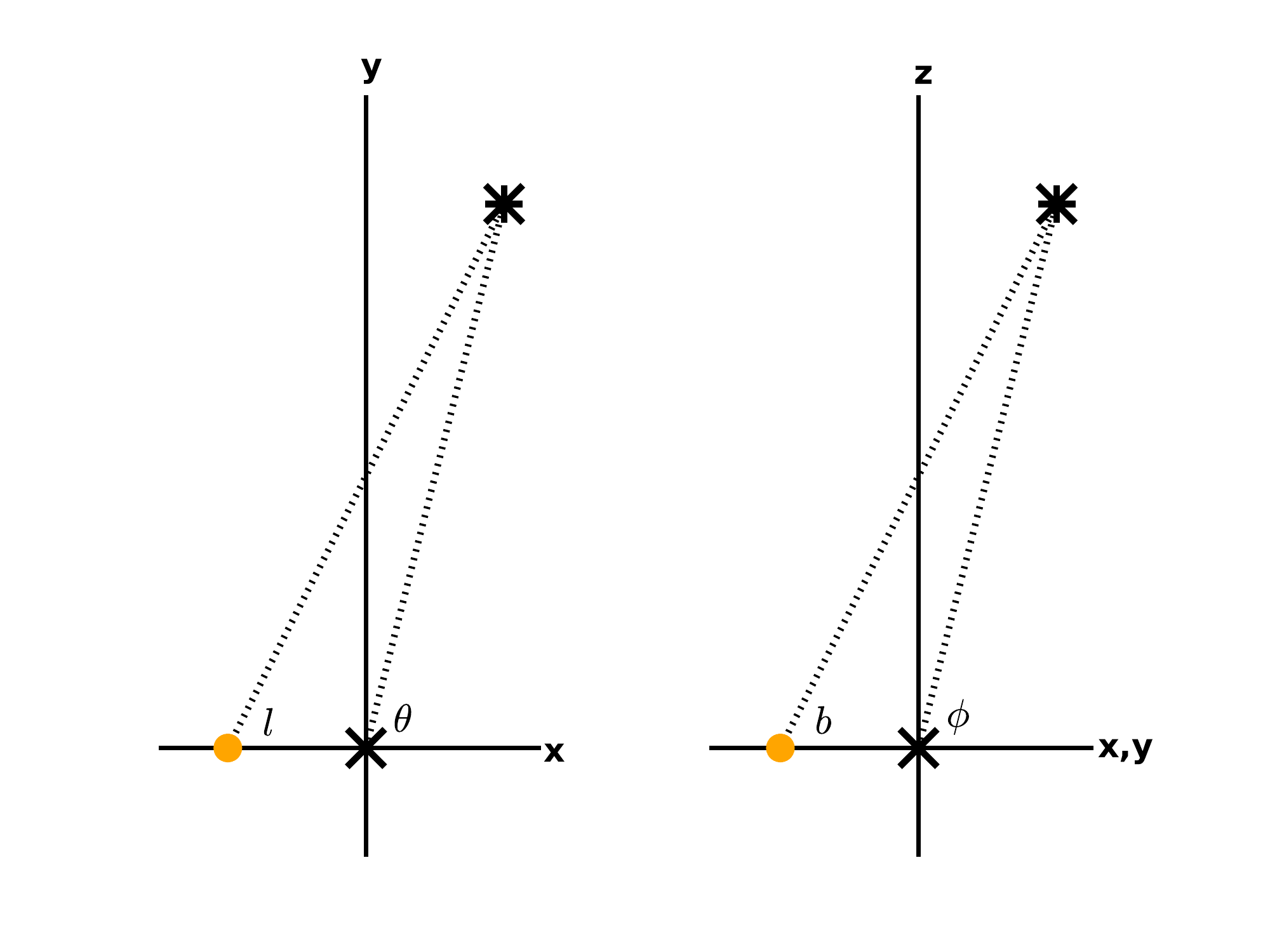}
\vskip 2ex
\caption{\label{fig: coord}
Schematic of the coordinate system. The `X' at the origin indicates the 
position of the GC. The Sun is represented as the orange dot along the 
$x$-axis. A star is represented as an asterisk. 
}
\end{figure}

To connect these coordinates to the heliocentric galactic system, we assign 
the Sun a position $(-\rsun, 0, 0)$ and a velocity $(0, \vsun, 0)$, where
\rsun\ = 8~kpc is the distance of the Sun from the GC 
\citep[e.g.,][]{bovy2012}
and \vsun\ is the space velocity of the Sun relative to the GC 
(Fig.~\ref{fig: coord}).  
Each star then has a distance $d = ( (x+\rsun)^2 + y^2 + z^2)^{1/2}$ from 
the Sun and a relative velocity 
$v_{rel} = (v_x^2 + (v_y - \vsun)^2 + v_z)^{1/2}$.
In this system, the galactic longitude $l$ of the star is the angle -- 
measured counter-clockwise in the $x-y$ plane -- from a line connecting 
the Sun to the GC, $l = {\rm tan^{-1}} (x~{\rm tan}~\theta / (x + \rsun))$.  
The galactic latitude measures the height of the star above the galactic 
plane, $b = {\rm sin^{-1}}(z/d) = {\rm sin^{-1}} (r~{\rm sin}~\phi / d)$. 
For $r \gg \rsun$, $\theta \approx l$ and $\phi \approx b$.  

\begin{figure}
\includegraphics[width=6.5in]{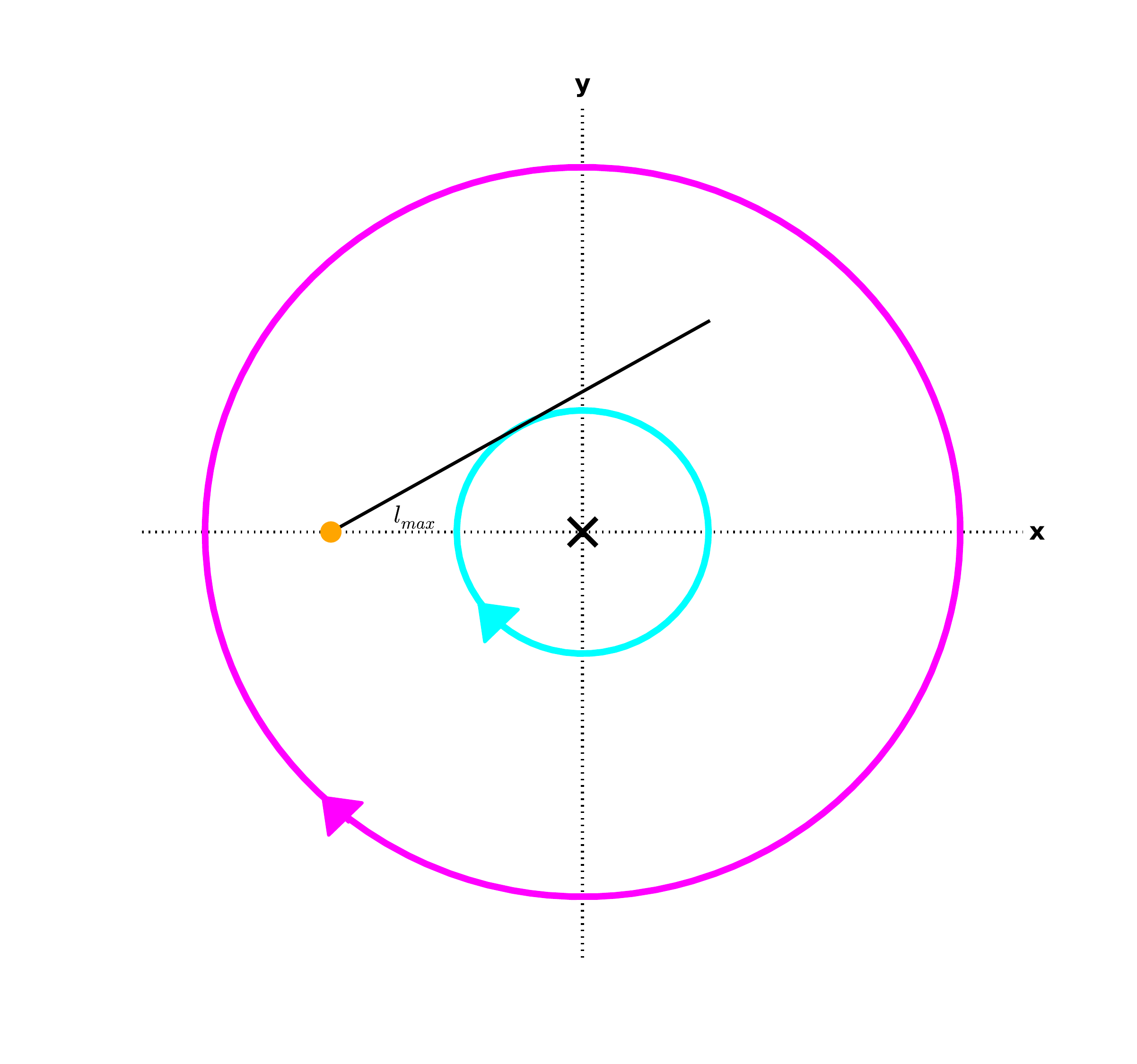}
\vskip 2ex
\caption{\label{fig: circ}
Schematic of stars on circular orbits of the GC. The `X' at the origin 
indicates the position of the GC. The Sun is represented as the orange 
dot along the $x$-axis. Stars follow circular orbits around the GC in
the direction of the arrows.
}
\end{figure}

Although these coordinate systems are clearly defined, angles in the 
heliocentric galactic system span a smaller range than in the pure GC
system (Fig.~\ref{fig: circ}). For stars with positions $r < \rsun$, the 
range of $\theta$ ($-\pi$ to $\pi$) is larger than the range of $l$
($-l_{max}$ to $l_{max}$), where 
\begin{equation}
l_{max} = {\rm sin^{-1}} (\rho / \rsun) ~ .
\label{eq: lmax}
\end{equation}
When $l = l_{max}$, $\theta_l = {\rm cos}^{-1} (-\rho / \rsun)$.  For 
each $l < l_{max}$, there are two values\footnote{Among other examples, 
this classic degeneracy in $l$ plagues H~I maps of the Galaxy.} of $\theta$.

We derive the radial velocity $v_r$, the tangential velocity $v_t$, and 
the proper motion $\mu$ in the heliocentric frame. For all stars, 
$v_{rel}^2 = v_r^2 + v_t^2$.  The radial velocity is 
\begin{equation}
v_r = v_x~{\rm cos}~l ~ {\rm cos}~b + (v_y - \vsun)~{\rm sin}~l~{\rm cos}~b + v_z~{\rm sin}~b ~ .
\end{equation}
We separate the tangential velocity into two components, 
$v_l$ and $v_b$, where $v_t^2 = v_l^2 + v_b^2$.
The component along the direction of galactic longitude is:
\begin{equation}
v_l = -v_x~{\rm sin}~l + (v_y - \vsun)~{\rm cos}~l ~ .
\end{equation}
The latitude component is
\begin{equation}
v_b = -(v_x~{\rm cos}~l + (v_y - \vsun)~{\rm sin}~l)~{\rm sin}~b + v_z~{\rm cos}~b ~ .
\end{equation}
For stars with $b \equiv$ 0 and no motion in the $z$-direction, $v_b = 0$. Although 
each component of the tangential velocity has a clearly-defined sign convention, we
plot the absolute magnitude when we combine the two components into the tangential
velocity, $v_t = | v_t | = (v_l^2 + v_b^2)^{1/2}$. 

The standard definition for the proper motion is
\begin{equation}
\mu = { v_t \over {4.74 ~ d} } ~ ,
\label{eq: pm}
\end{equation}
where $v_t$ is measured in \kms\ and $d$ is in pc. To set the proper motion in 
the heliocentric galactic frame, $\mu_l = v_l / (4.74 ~ d)$ and $\mu_b = v_b / (4.74 ~ d)$,
where the velocities are in \kms. Positive (negative) proper motions are in the 
direction of increasing (decreasing) $l$ or $b$.

\subsection{Simple Trajectories}
\label{sec: an-traj}

Within this framework, we consider several simple stellar motions to explore the
variation of $v_r$ and $v_t$ with position in the Galaxy. Most motions are composed
of both a circular component and a radial component.  Starting with stars following
circular orbits around the GC inside and outside the solar circle (\S\ref{sec: an-orb}), 
we derive the behavior of $v_r$ and $v_t$ with $\theta$, $l$, and $b$ for stars with 
total velocity $v$. In this simple 
example, we set $z$ = 0 and work in a coordinate system where $r = \rho$.  The maximum 
tangential velocity is then fixed at $v_{t,max} = v + \vsun$; the maximum radial velocity
falls with $r$ inside the solar circle and then grows with $r$ outside the solar
circle. At large $r$, $v_{r,max} \approx \vsun$. Continuing with stars on purely 
radial orbits (\S\ref{sec: an-rad}), we explore motions in the spherical coordinate 
system appropriate for the bulge and the halo.  For stars inside the solar circle, 
the maximum radial and tangential velocities are $v_{r,max} \approx v$ and 
$v_{t,max} \approx v + \vsun$.  Extrema in $v_r$ lie at $l \approx$ 0; stars have 
maximum $v_t$ at $l \approx l_{max}$. Well outside the solar circle, the maximum 
radial velocity ($v_{r,max} \approx v + \vsun$) is much larger than the maximum 
tangential velocity ($v_{t,max} \approx \vsun$). At intermediate $r \approx$ 8--20~kpc, 
there is a smooth transition from small $v_{r,max}$ and large $v_{t,max}$ to large 
$v_{r,max}$ and small $v_{t,max}$.

\subsubsection{Circular Orbital Motion}
\label{sec: an-orb}

For stars following simple circular orbits around the GC, $v_x = v~{\rm sin}~\theta$, 
$v_y = -v~{\rm cos}~\theta$, and $v_z$ = 0 (Fig.~\ref{fig: circ}). Although stars 
in the thin and thick disks have finite vertical distances from the Galactic plane
and non-zero motion out of the Galactic plane, we set $z \equiv 0$ and ignore any
out-of-plane motion here.  Thus, our radial coordinate $r$ is identical to the
standard cylindrical coordinate $\rho$. With $\phi = b = 0$, the heliocentric radial 
and tangential velocities are
\begin{equation}
v_r = v~{\rm sin}~\theta~{\rm cos}~l - (v~{\rm cos}~\theta + \vsun)~{\rm sin}~l ~
\label{eq: vr-orbx}
\end{equation}
and 
\begin{equation}
v_t = -v~{\rm sin}~\theta~{\rm sin}~l - (v~{\rm cos}~\theta + \vsun)~{\rm cos}~l ~ .
\label{eq: vt-orbx}
\end{equation}
In this system, $v_b$ = 0 and $v_t = v_l$.  Using trigonmetric identities, we can 
simplify these to:
\begin{equation}
v_r = v~{\rm sin}~(\theta - l) - \vsun~{\rm sin}~l ~
\label{eq: vr-orb}
\end{equation}
and 
\begin{equation}
v_t = -v~{\rm cos}~(\theta - l) - \vsun~{\rm cos}~l ~ .
\label{eq: vt-orb}
\end{equation}
For convenience, we can eliminate $\theta$ in the expression for the radial velocity,
\begin{equation}
v_r = \left ( {\rsun\ \over r} v - \vsun \right ) {\rm sin}~l ~ .
\label{eq: vr-orbl}
\end{equation}

\begin{figure}
\includegraphics[width=6.5in]{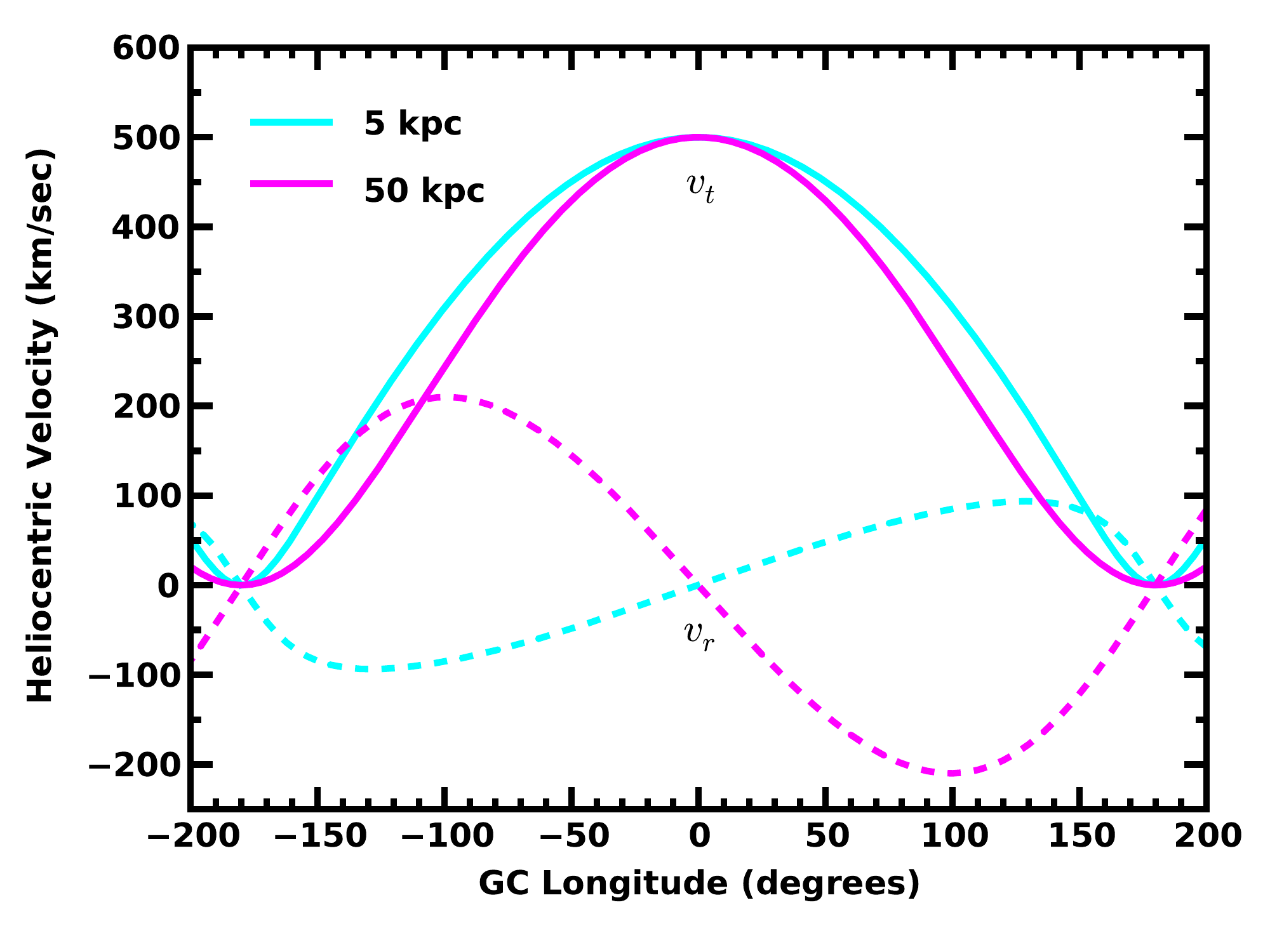}
\vskip 2ex
\caption{\label{fig: v-orbgc}
Radial (dashed curves) and tangential (solid curves) velocity for stars
with $r$ = 5~kpc (cyan curves) and $r$ = 50~kpc (magenta curves) on
circular orbits orbiting the Galaxy with $v$ = 250 \kms. In a frame 
centered on the GC, stars have velocities exactly anti-parallel (parallel)
to the motion of the Sun at GC longitude $\theta$ = 0 ($\pm \pi$). 
At $\theta = 0$ ($\theta = \pm \pi$), stars have maximum $|v_t|$ and 
$v_r$ = 0 ($v_r = v_t $ = 0).
}
\end{figure}

Fig.~\ref{fig: v-orbgc} illustrates the variation of $v_r$ (dashed curves) and 
$v_t$ (solid curves) with GC longitude for stars with $v$ = 250 \kms\ and $r$ = 5~kpc 
(cyan lines) and $r$ = 50~kpc (magenta lines). In this configuration, stars on the 
opposite side of the Galaxy from the Sun ($\theta = 0$) have no net 
radial velocity and a maximum tangential velocity of $v + \vsun$.  For 
\vsun\ = 250~\kms, $v_{t,max}$ = 500 \kms. Stars on the near side of the Galaxy
($\theta = \pm \pi$) have no net radial or tangential velocity. Thus, the 
minimum tangential velocity is $v_{t,min}$ = 0. 

The behavior of $v_r$ depends on $r$. For all $r$, $v_r = 0$ at $\theta = 0$ and 
$\pm \pi$.  When $r < \rsun$, maximum positive $v_r$ is at $\theta = -\theta_l$ 
($l = -l_{max}$). Maximum negative $v_r$ is at $\theta = +\theta_l$.  With a maximum 
radial velocity, $v_{r,max} = \pm ( v - \vsun r / \rsun )$, the amplitude of the 
radial velocity variation declines from roughly 250 \kms\ at $r \approx$ 0 to 
roughly zero at $r \approx \rsun$.  Outside the solar circle, the extrema in 
$v_r$ lie at $l = \pm \pi/2$ ($\theta = {\rm cos^{-1}} (-\rsun / r)$). With 
$v_{r,max} = v \rsun / r - \vsun$, the amplitude of the $v_r$ variation grows
from zero at $r \approx \rsun$ to \vsun\ at $r \gg \rsun$.  Thus,
\begin{equation}
v_{r,max} = \left\{
\begin{array}{lll}
\pm \left ( v - {r \over \rsun} \vsun \right ) & & r < \rsun \\
\\
\pm \left ( {\rsun \over r} v - \vsun \right ) & & r \ge \rsun \\
\end{array}
\right.
\label{eq: vr-maxo}
\end{equation}

\begin{figure}
\includegraphics[width=6.5in]{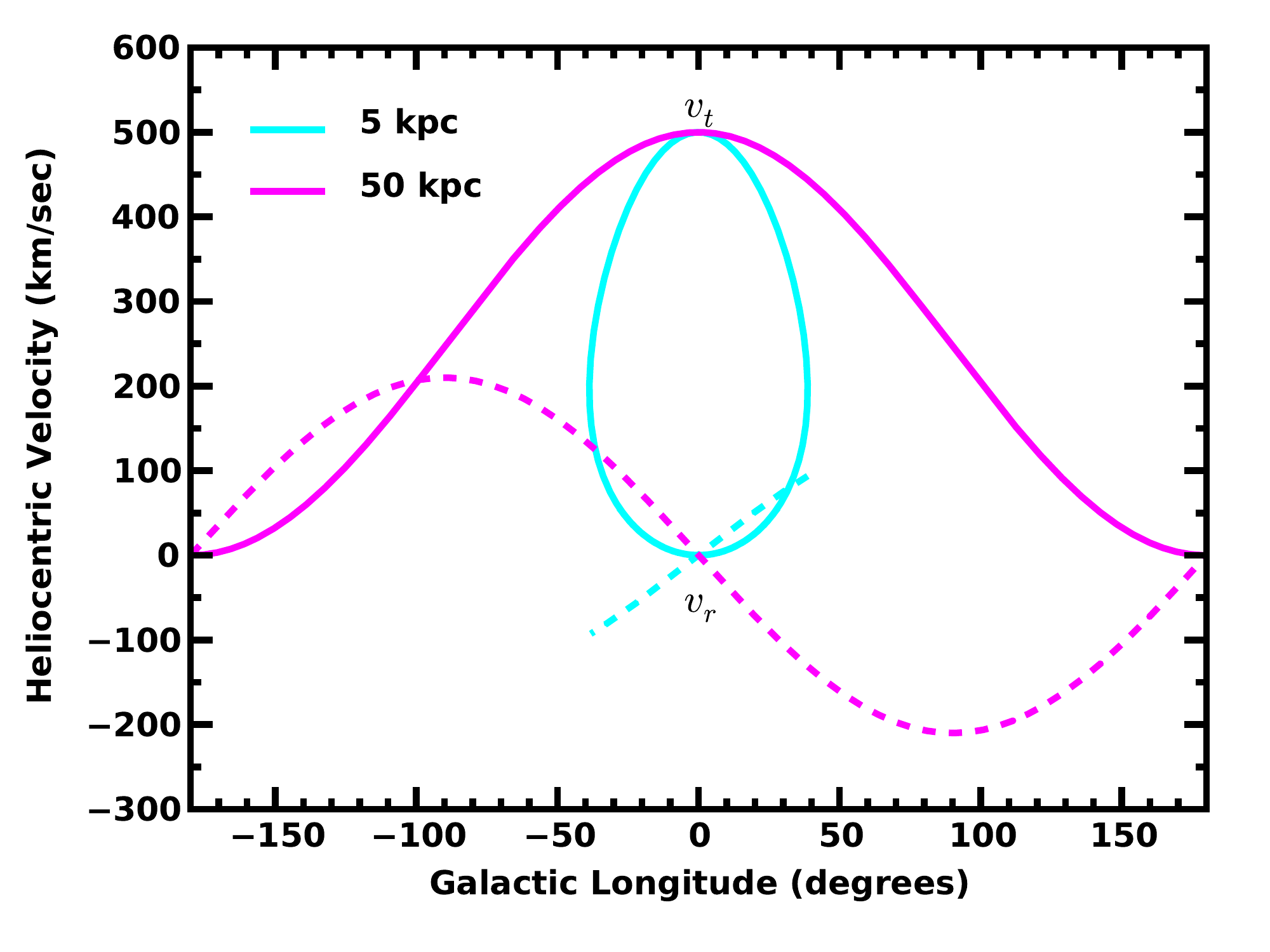}
\vskip 2ex
\caption{\label{fig: v-orbhel}
As in Fig. \ref{fig: v-orbgc} in the heliocentric galactic frame. In a frame
centered on the Sun, stars inside the solar circle (cyan lines) have a maximum 
galactic longitude, $l_{max}$ (eq. [\ref{eq: lmax}]). The tangential
velocity ($|v_t|$) reaches a maximum and a minimum at $l$ = 0. The radial 
velocity achieves extreme values at $\pm l_{max}$. Outside the solar circle 
(magenta lines), $v_r$ and $v_t$ follow simple sinusoids, with extreme
values at $\pm \pi/2$ ($v_r$) and at 0 and $\pm \pi$ ($v_t$).
}
\end{figure}

In the heliocentric galactic frame, the variation of $v_r$ and $v_t$ with $l$ is
somewhat different (Fig.~\ref{fig: v-orbhel}). For stars inside the solar circle,
$v_t$ follows an egg-shaped loop with minimum and maximum velocity at $l = 0$.
Here, $l_{max}$ sets the maximum extent of the loop in galactic longitude. Thus,
the `egg' widens at larger $r$, reaching $l_{max} \approx \pm \pi / 2$ at
$r \approx \rsun$. Outside the solar circle, $v_t$ varies sinusoidally with $l$,
with maxima of $v$ + \vsun\ = 500 \kms\ at $l = \pm \pi$ and a minimum of zero at
$l = 0$.

The radial velocity also follows simple trajectories. At small $r$, $v_r$ varies along
a curved line with extrema of $v_{r,max}$ (eq. [\ref{eq: vr-maxo}]) at $l_{max}$. 
As $r$ grows, the curves extend to larger $l$ but have smaller maxima. At large $r$,
$v_r$ follows a simple sinusoid, with extreme values set by \vsun\ at $l = \pm \pi/2$
and zero-crossings at $l = 0$ and $l = \pm \pi$.

\subsubsection{Radial Motion}
\label{sec: an-rad}

\begin{figure}
\includegraphics[width=6.5in]{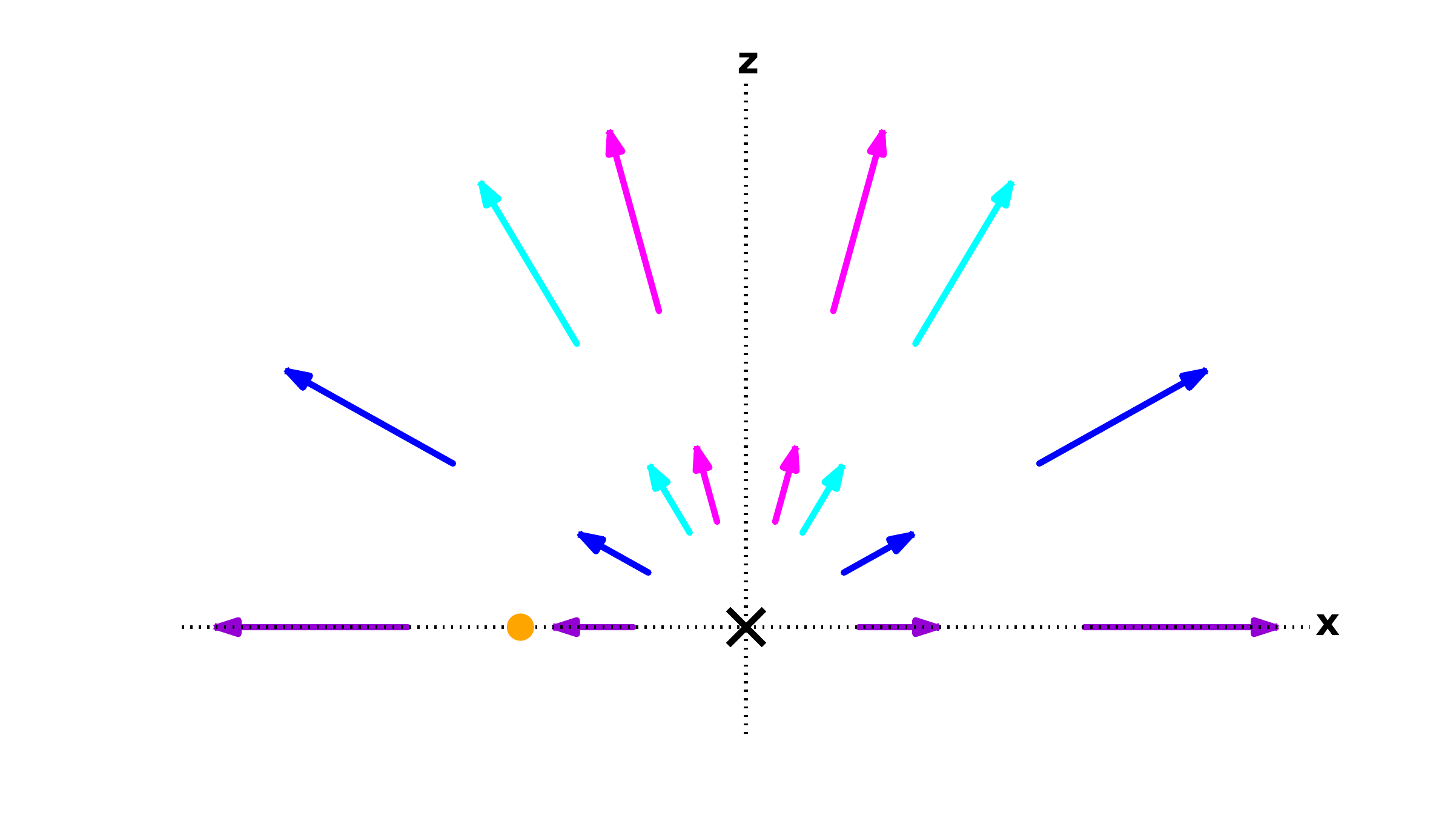}
\vskip 2ex
\caption{\label{fig: rad}
Schematic of stars on radially outflowing orbits from the GC. The `X' at the origin 
indicates the position of the GC. The Sun is represented as the orange dot along 
the $x$-axis. Colored arrows indicate velocity vectors for stars with 
$\phi$ = 0\deg\ (violet), $\phi$ = 30\deg\ (blue), $\phi$ = 60\deg\ (cyan), and
$\phi$ = 75\deg\ (magenta). Outside the solar circle (outer set of arrows), all
stars move away from the Sun. Inside the solar circle (inner set of arrows), 
stars with low galactic latitude on the near side of the GC have some component 
of their motion towards the Sun.
}
\end{figure}

Stars moving radially away from the GC have subtly different behavior. To infer
conclusions appropriate for stars in the bulge or the halo, we consider stars 
with a broad range of GC latitude. In the GC frame, outflowing stars have constant 
$\phi$ for all $r$ (see Fig.~\ref{fig: rad}). In the heliocentric frame, nearby stars 
have larger $b$ than more distant stars.  For stars with $\phi > 0$, $r > \rho$. Thus, 
distant stars with large $b$ may lie inside the solar circle.

With $v_x = v~{\rm cos}~\theta~{\rm cos}~\phi$, 
$v_y = v~{\rm sin}~\theta~{\rm cos}~\phi$, and $v_z = v~{\rm sin}~\phi$, 
the heliocentric radial and tangential velocities are
\begin{equation}
v_r = v~{\rm cos}~(\theta - l) ~ {\rm cos}~\phi ~ {\rm cos}~b - \vsun~{\rm sin}~l~{\rm cos}~b + v~{\rm sin}~\phi ~ {\rm sin}~b ~ ,
\label{eq: vr-rad}
\end{equation}
\begin{equation}
v_l = v~{\rm sin}~(\theta - l)~ {\rm cos}~\phi - \vsun~{\rm cos}~l ~ ,
\label{eq: vl-rad}
\end{equation}
and 
\begin{equation}
v_b = -v~{\rm cos}~(\theta - l)~ {\rm cos}~\phi ~ {\rm sin}~b - \vsun~{\rm sin}~l~{\rm sin}~b + v~{\rm sin}~\phi ~ {\rm cos}~b ~ .
\label{eq: vb-rad}
\end{equation}

When stars move radially outward through the Galactic plane, $\phi = b = 0$. 
With $v_b = 0$, the equations for radial and tangential velocity are then very 
simple: $v_r = v~{\rm cos}~(\theta - l) - \vsun~{\rm sin}~l$ and
$v_t = v_l = v~{\rm sin}~(\theta - l) - \vsun~{\rm cos}~l$.  For stars inside
the solar circle, the maximum galactic longitude is $l = l_{max}$. 

When $\phi > 0$, the variations of $v_r$ and $v_t$ with $\theta$ and $l$ are more complex. 
Aside from having a non-zero $v_b$, the amplitude of both velocity components declines 
with cos~$\phi$. For stars inside the solar circle, the maximum $l$ scales with $\phi$:
\rsun\ sin $l_{max}$ = $r~{\rm cos}~\phi$. Thus, stars inside the solar circle at large 
$\phi$ have a smaller range in $l$ than stars with small $\phi$.

\begin{figure}
\includegraphics[width=6.5in]{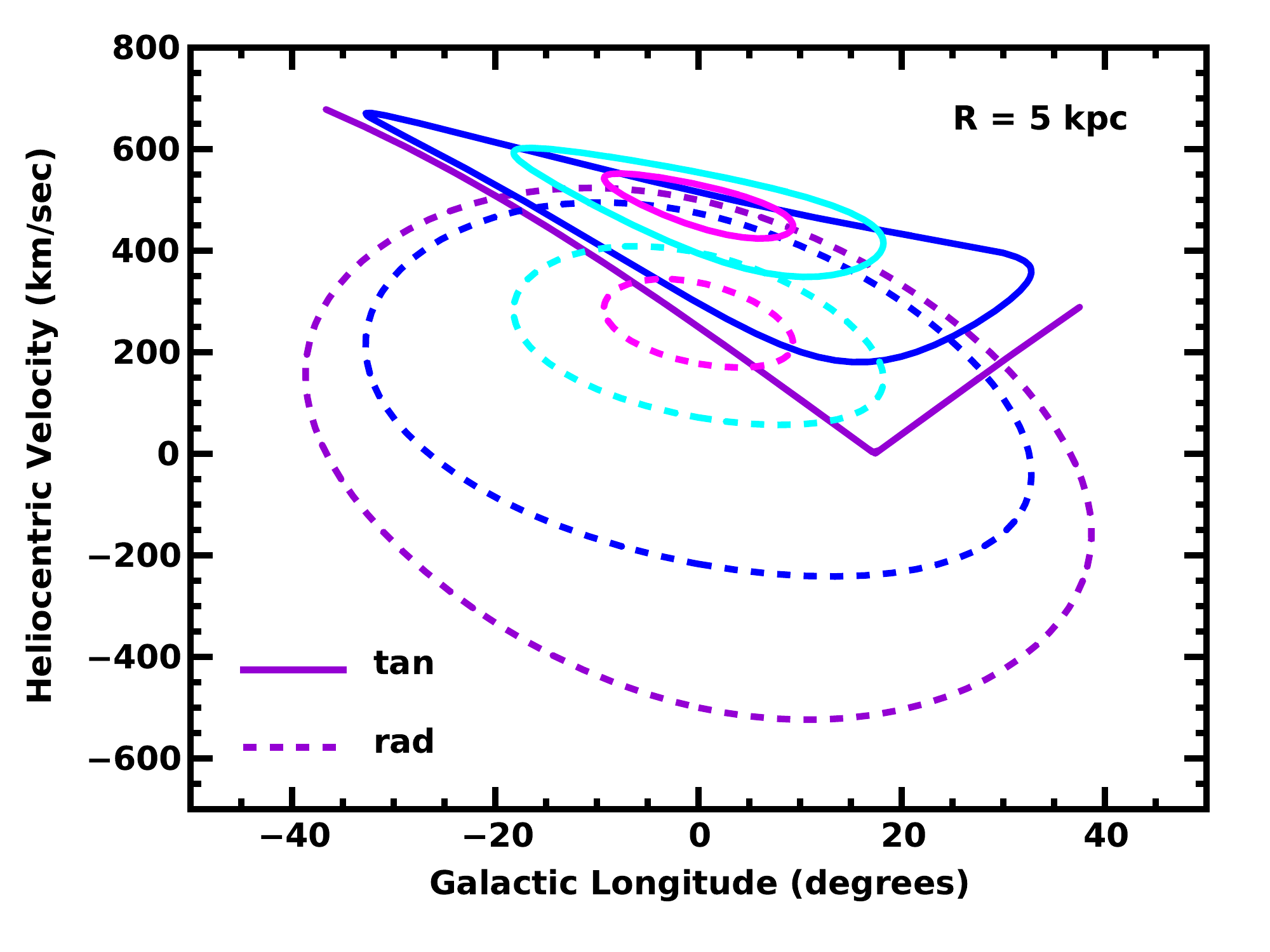}
\vskip 2ex
\caption{\label{fig: v-rad5hel}
Radial ($v_r$, dashed curves) and tangential ($|v_t|$, solid curves) velocity 
as a function of galactic longitude ($l$) for stars at $r$ = 5~kpc and 
$\phi$ = 0\deg\ (violet), 30\deg\ (blue), 60\deg\ (cyan), and 
75\deg\ (magenta) moving radially away from the GC with $v$ = 500~\kms.  
}
\end{figure}

Fig. \ref{fig: v-rad5hel} shows the variation of $v_r$ (dashed curves) and $v_t$ (solid
curves) as a function of $l$ for stars with $r$ = 5~kpc, $v$ = 500~\kms, and $\phi$ = 
0\deg\ (violet curves), 30\deg\ (blue curves), 60\deg\ (cyan curves), and 75\deg\ (magenta 
curves).  With $l_{max} \propto \phi$, curves at larger $\phi$ have a smaller extent in 
$l$.  For stars at $r$ = 5~kpc, the maximum galactic latitude is $b \approx$ 30\deg--40\deg. 
Both sets of curves follow loops in the $l,v$ plane. For $\phi$ = 0, the $v_t$ curve 
folds back on itself.

When stars lie inside the solar circle, the radial velocity varies symmetrically about 
an average velocity $v~{\rm sin}~\phi ~ {\rm sin}~b$. This average increases with $\phi$,
reaching $v_{r,avg} = \vsun~{\rm cos}~b + v~{\rm sin}~b$ at $\phi$ = 90\deg. With 
$b \approx$ 30\deg, $v_{r,avg} \approx \vsun$.  The amplitude of the $v_r$ variation 
scales with ${\rm cos}~\phi$ and thus declines markedly from $\Delta v_r \approx v$
to $\Delta v_r \approx$ 0 among the sequence of four curves. 

The variation of $v_t$ with $l$ is not symmetric. At $\phi$ = 0\deg, the tangential
velocity ranges from a minimum of 0 to a maximum close to $v + \vsun$, roughly 700 \kms.  
At large $\phi$, $v_t$ approaches a constant value of roughly 
$v~{\rm cos}~b + \vsun~{\rm sin}~b$ $\approx v$ for $v \gg \vsun$ and 
$b \approx$ 30\deg--40\deg.

\begin{figure}
\includegraphics[width=6.5in]{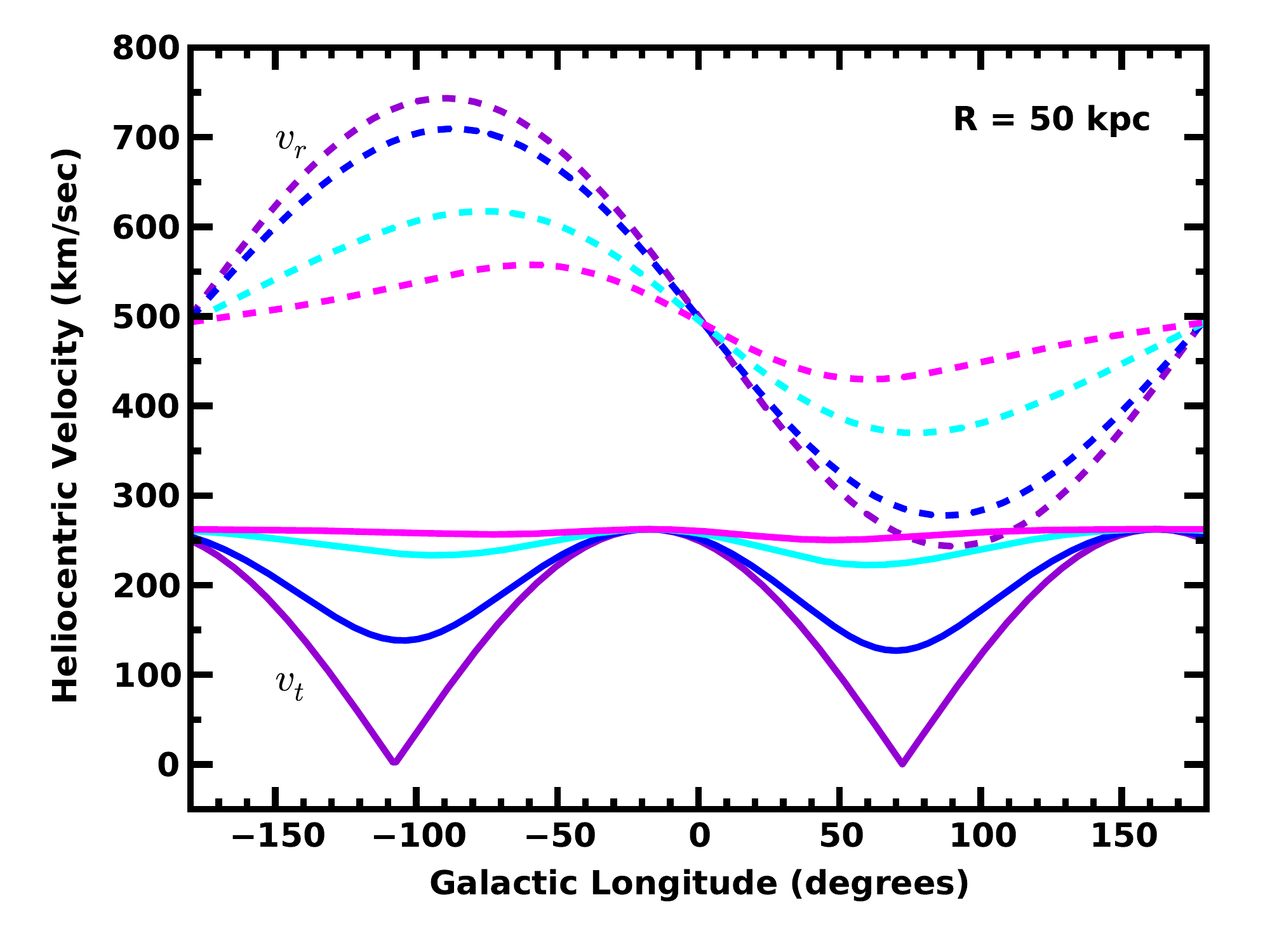}
\vskip 2ex
\caption{\label{fig: v-rad50hel}
As in Fig. \ref{fig: v-rad5hel} for stars with $r$ = 50~kpc.
}
\end{figure}

Well outside the solar circle ($r$ = 50~kpc), the motions are much simpler 
(Fig.~\ref{fig: v-rad50hel}). At large $r$, $v_r$ varies roughly sinusoidally 
with $l$ about an average velocity of $v_{r,avg} \approx v$.  The amplitude
of this variation decreases with $b$, reaching a constant $v_r \approx v$ when 
$r \gg \rsun$ and $b \approx$ 90\deg. Stars reach a minimum (maximum) $v_r$ 
at $l = \pm \pi/2$.

The tangential velocity has a smaller amplitude and different phasing with $l$.
At large $b$, the tangential velocity is roughly \vsun, a result of reflex solar 
motion. For stars close to the plane $\phi \approx 0$ and $b \approx$ 0, the
tangential velocity consists of two sinusoids with amplitudes of $v$ and $\vsun$
($v_t \approx v_l = v~{\rm sin}~(\theta - l) - \vsun~{\rm cos}~l$; 
eq. [\ref{eq: vl-rad}]). Thus, the minimum $v_t$ is small and approaches $v_t = 0$ 
at $b$ = 0.  Because the Sun is offset from the GC, the phase of minimum $v_t$ 
is offset from $\pm \pi/2$. The solar motion and position in the galaxy produce 
an offset of roughly $-$20\deg\ in longitude.

For stars with 5~kpc $\lesssim r \lesssim$ 50~kpc, there is a smooth transition 
between the behavior shown in Figs.~\ref{fig: v-rad5hel}--\ref{fig: v-rad50hel}.
Stars with in-plane distances less than \rsun\ ($\rho < \rsun$) follow the 
trajectories in Fig.~\ref{fig: v-rad5hel}.  Stars outside this limit follow the 
trajectories in Fig.~\ref{fig: v-rad50hel}.  

For an ensemble of stars with $r \approx$ 16~kpc, as an example, stars with 
$\phi \lesssim \pi / 3$ have $\rho \gtrsim \rsun$ and follow the trajectories
in Fig.~\ref{fig: v-rad50hel}. Stars at larger $\phi$ have the closed loop 
trajectories in Fig.~\ref{fig: v-rad5hel}.

\begin{figure}
\includegraphics[width=6.5in]{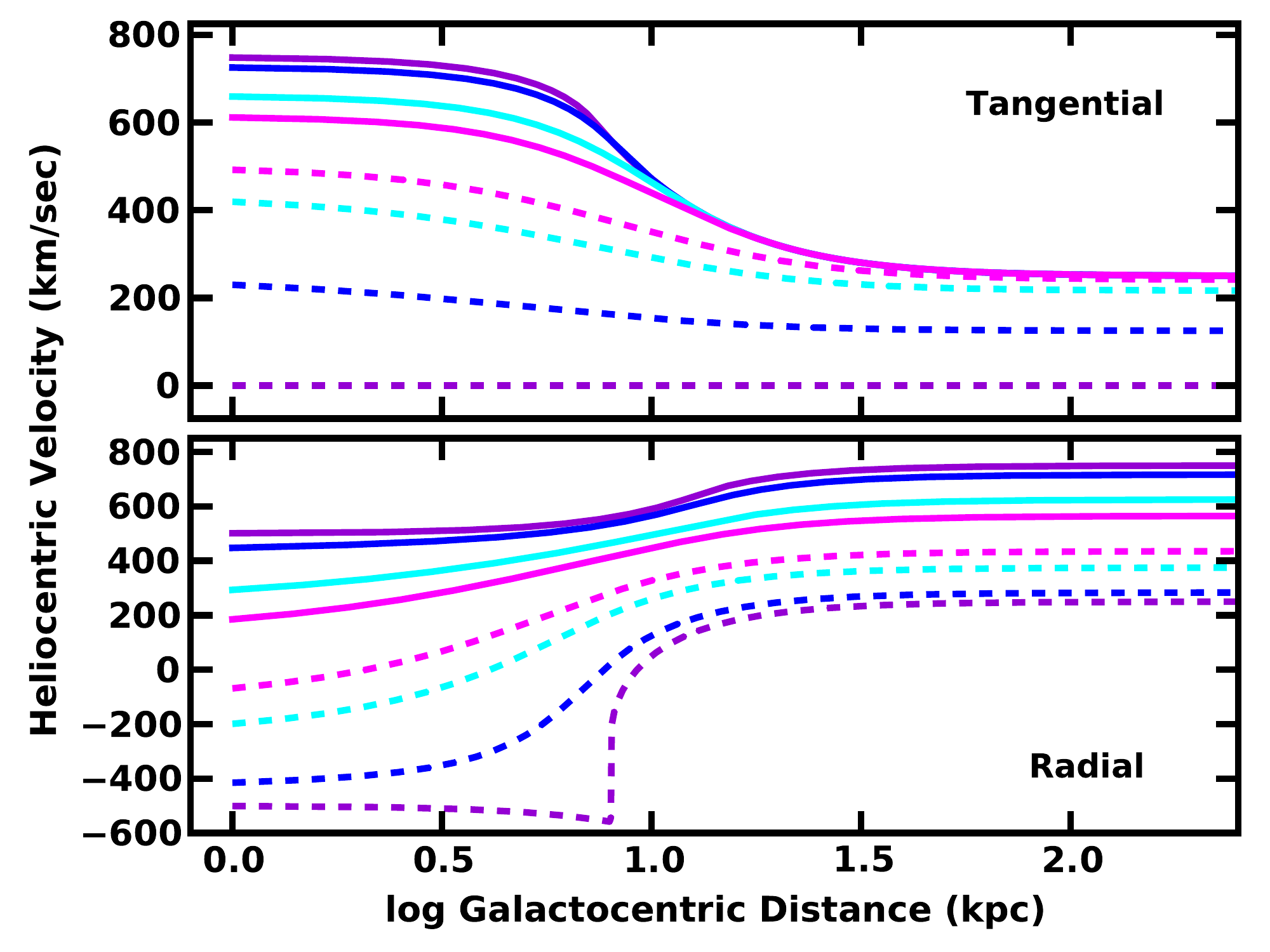}
\vskip 2ex
\caption{\label{fig: v-rad-gcd}
Maximum (solid curves) and minimum (dashed curves) radial velocity
(lower panel) and tangential velocity (upper panel) as a function 
of Galactocentric distance ($r$) for stars at
$\phi$ = 0\deg\ (violet), 30\deg\ (blue), 60\deg\ (cyan), and 
75\deg\ (magenta) moving radially away from the GC with $v$ = 500~\kms.  
}
\end{figure}

To illustrate this transition in more detail, Fig.~\ref{fig: v-rad-gcd} shows
the variation of the maximum and minimum radial velocity (lower panel) and 
tangential velocity (upper panel) as a function of $r$ and $\phi$ for stars 
on radial orbits with $v$ = +500 \kms\ relative to the GC.  At small $\phi$, 
the minimum $v_t$ is close to zero for all $r$. This minimum increases with 
$\phi$ until $v_t = \vsun$.  The maximum $v_t$ is roughly constant at 
500--750 \kms\ at small $r$ and then decreases smoothly to \vsun\ at large $r$. 

The extrema in $v_r$ have similar trends. Inside the solar circle, stars on
the near side of the GC all move towards the Sun and have large negative $v_r$.
On the far side of the GC, all stars move away from the Sun. Thus, the range 
in $v_r$ is largest for stars inside the solar circle. Because $v_r$ scales 
with cos~$\phi$, stars at small (large) $\phi$ have the largest (smallest) 
range in $v_r$.

Outside the solar circle, all stars in this example move away from the Sun.  
The minimum radial velocity is then always larger than zero, producing the large 
increase in minimum $v_r$ at $r = \rsun$. Somewhat counterintuitively, the 
maximum $v_r$ also grows with $r$. Stars with $r \gtrsim \rsun$ have the
largest velocity with respect to the Sun when they move radially outward 
roughly along the $y$-axis. As $r$ increases, the angle between the $y$-axis
and the line-of-sight from the Sun to the star decreases. Thus, the radial 
component of the relative velocity grows with $r$, reaching $v + \vsun$ 
when $r \gg \rsun$.

\begin{figure}
\includegraphics[width=6.5in]{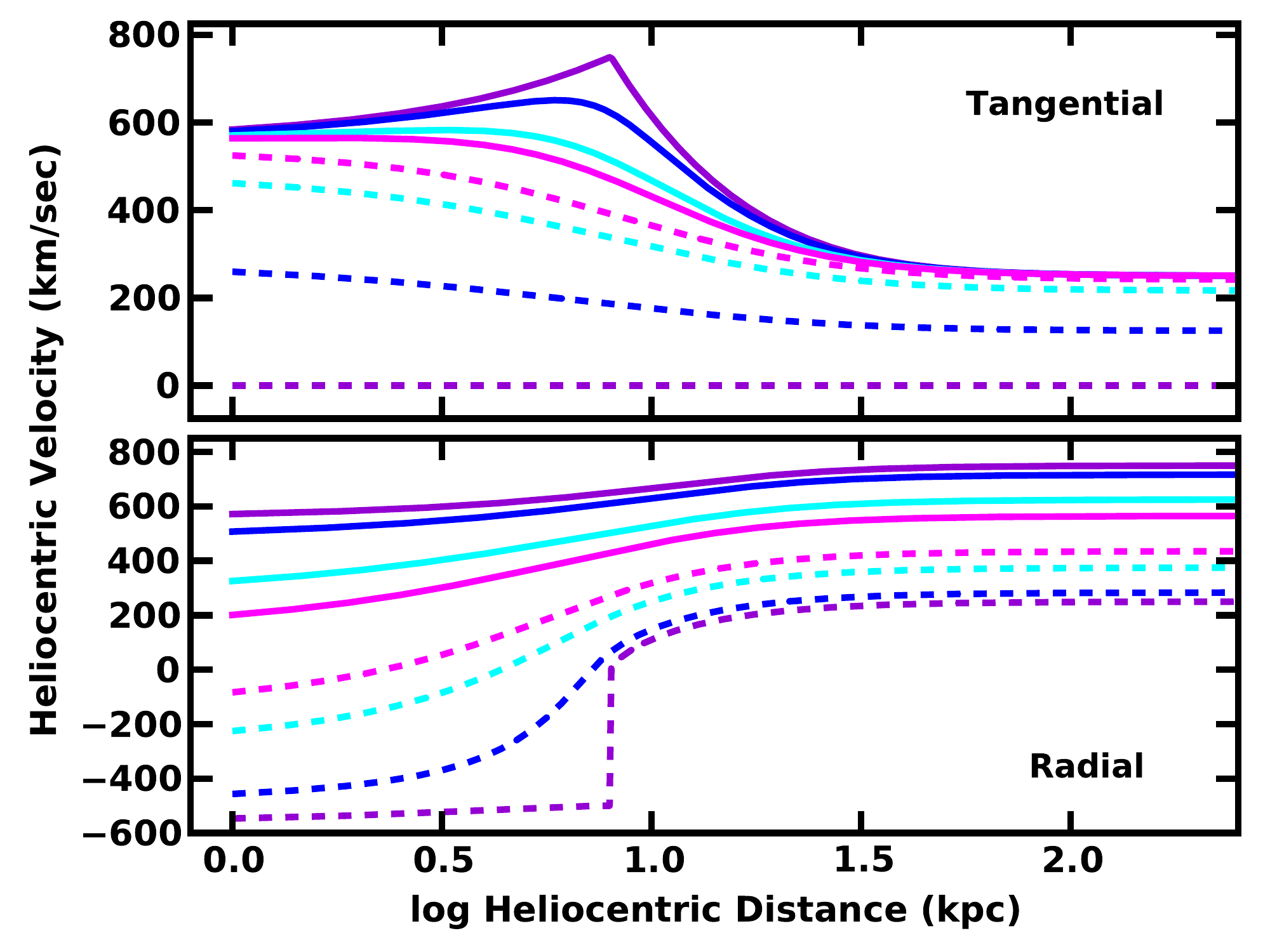}
\vskip 2ex
\caption{\label{fig: v-rad-hd}
As in Fig. \ref{fig: v-rad-gcd} for the heliocentric distance. 
}
\end{figure}

Fig.~\ref{fig: v-rad-hd} repeats Fig.~\ref{fig: v-rad-gcd} for heliocentric
distance $d$. Aside from a clear discontinuity in the minimum $v_r$ for
$\phi$ = 0 at $d = \rsun$, the behavior in $v_r$ is almost identical to
Fig.~\ref{fig: v-rad-gcd}. The
minimum $v_r$ crosses from negative to positive $v_r$ at $d = \rsun$. The
maximum $v$ slowly increases to $v + \vsun$ at large $r$. The variation 
in the minimum $v_t$ is also similar, a slowly decreasing function of 
increasing $r$.  

The maximum in the tangential velocity, however, exhibits a clear maximum 
for stars with $\phi \lesssim$ 30\deg\ at $d$ = 8 kpc. Stars close to the
GC produce this maximum. When GC stars move radially outward in the 
$y$--$z$ plane, their tangential velocity is at a maximum. For $\phi = 0$,
this peak in $v_t$ is a sharp feature. Although still visible for 
$\phi \approx$ 5\deg\ to 30\deg, the feature vanishes for larger $\phi$.

\subsection{Summary}
\label{sec: an-summ}

Despite the simple stellar motions in these examples, the behavior of
$v_r$ and $v_t$ with $d$, $l$, and $b$ is amazingly rich.  For stars 
inside the solar circle, circular orbital motion and radial outflow 
produce large maximum tangential velocities, $v_{t,max} \approx v + \vsun$.
These maxima occur at distinct galactic longitudes: $l \approx$ 0 for
stars orbiting the GC and $l \approx -l_{max}$ for stars moving radially
away from the GC. Thus, proper motion measurements offer some promise 
for distinguishing high velocity stars ejected from the GC from stars 
on circular orbits around the GC.

Outside the solar circle, circular orbital motion is also distinct from 
purely radial motion. For stars orbiting the GC, the maximum $v_t$ is 
independent of $r$. However, the maximum tangential velocity of radially 
outflowing stars gradually declines with $r$ until 
$v_{t,max} \approx \vsun$. This maximum $v_t$ changes little with $l$.  
For distant stars, orbital motion yields a larger $v_{t,max}$ than
radial motion.

Trends of $v_r$ with $r$ and $d$ are opposite those of $v_t$. Inside the
solar circle, stars on circular orbits have smaller and smaller $v_r$ at 
larger and larger $r$. For stars moving radially away from the GC, $v_r$ 
has clear minima and maxima of $\pm v$ at $l \approx \pm l_{max}$.  Outside 
the solar circle, stars on circular orbits have maximum radial velocity 
$v_{r,max} \approx \vsun$ at large $r$.  Stars moving radially away from 
the GC have much larger maximum $v_r$, with $v_{r,max} \approx v + \vsun$.  
Thus, radial velocity measurements excel at separating distant stars on 
roughly circular orbits from high velocity stars moving radially outwards 
from the GC.

To conclude this section, we derive predicted proper motions for stars
moving radially away from the GC (Fig.~\ref{fig: mu-rad}). Close to the
Sun, proper motions are large, roughly 100 milliarcsec yr$^{-1}$. The 
range in proper motions is small (large) for stars at high (small)
galactic latitude. At large distances ($d \approx 50-100$~kpc), the
maximum proper motion of roughly 1 milliarcsec yr$^{-1}$ results from 
solar reflex motion.  

At intermediate distances, there is a small `peak' at $d \approx$ 8~kpc 
in the trend of $\mu \propto d^{-1}$. Stars inside the solar circle with
$l \approx l_{max}$ produce this peak. For $b \approx$ 0\deg--10\deg, the 
peak has the largest contrast with the general trend in proper motion
(see Fig.~\ref{fig: v-rad-hd}). At the largest galactic latitudes
($b \gtrsim$ 60\deg), the peak fades considerably.

These results demonstrate that radial velocities can isolate high velocity 
stars from the space motions of typical stars in the Galaxy.  Radial 
velocity measurements succeed at large $d$, where observations can easily 
separate HVSs or runaways with $v_r \gtrsim$ +300~\kms\ from normal halo stars 
with $| v_r | \lesssim$ 100~\kms\ \citep[e.g.,][and references therein]{brown2014}.

Inside the solar circle, proper motion measurements provide a clear path 
for isolating high velocity stars from bulge and disk stars orbiting the 
GC. Among B-type stars with $d \lesssim$ 1~kpc, typical proper motions are 
10--40~\masyr\ \citep[e.g.,][]{dezeeuw1999}. This motion is a factor of 
3--10 times smaller than the predicted motion for nearby HVSs and runaways 
with space velocities of 500~\kms\ (Fig.~\ref{fig: mu-rad}).  The observed 
velocity dispersion ($\sigma_r \approx$ 100~\kms) of stars in the Galactic 
bulge implies typical proper motions of 1--5~\masyr\ \citep{tre2002,soto2014}, 
smaller than the 10--20~\masyr\ predicted for high velocity HVSs and runaways 
escaping the inner Galaxy. 

For all distances, proper motions alone cannot easily separate ejected stars 
from indigenous halo stars. The maximum proper motions of typical halo stars 
with $d \approx$ 1--10~kpc, $\sim$ 30--50~\masyr\ \citep{kinman2007,kinman2012,
bond2010}, are comparable to the likely proper motions of typical ejected
stars. We return to this issue in \S\ref{sec: obs-halo} with a direct 
comparison between observations of halo stars and predictions from our 
numerical simulations.

\section{NUMERICAL SIMULATIONS}
\label{sec: sims}

To explore the space motions of high velocity stars in more detail, we now
consider a set of numerical simulations\footnote{For an analytical approach
to some aspects of our discussion, see \citet{rossi2013}.}. As in previous papers
\citep{bromley2006,kenyon2008,bromley2009}, we follow the dynamical evolution
of HVSs and runaways throughout their main sequence lifetimes in a realistic 
Galactic potential. Snapshots of the ensemble yield predictions for the
radial distributions of space density, proper motion, and radial velocity.
In contrast with previous discussions, we concentrate on observables in the
heliocentric frame instead of the Galactocentric frame.

Building a realistic ensemble of HVSs or runaways requires two steps.
For each star with main sequence lifetime $t_{ms}$, we generate initial 
position \r0\ and velocity \v0\ vectors, an ejection time $t_{ej}$, and 
an observation time $t_{obs}$, with $t_{ej} \le t_{obs} \le t_{ms}$. 
For a flight time $t_f = t_{obs} - t_{ej}$, we integrate the orbit of each 
star in the Galactic potential and record the final position \rf\ and 
velocity \vf\ vectors at $t_{obs}$.  Finally, we adopt a position and 
velocity for the Sun to derive a catalog of $d$, $v_r$, $v_t$, $\mu_l$, and 
$\mu_b$. Analyzing this catalog yields predictions for the observable parameters.

\subsection{Gravitational Potential of the Milky Way}
\label{sec: sims-pot}

As in \citet{kenyon2008}, we adopt a three component model for the Galactic 
potential $\Phi_G$ \citep[for other approaches, see][]{gnedin2005,dehnen2006,yu2007}:
\begin{equation}
\label{eq: phi}
\Phi_G = \Phi_b + \Phi_d + \Phi_h ~ ,
\end{equation}
where 
\begin{equation}
\label{eq: phib}
\Phi_b(r) = - G M_b / (r + a_b) ~ 
\end{equation}
is the potential of the bulge,
\begin{equation}
\label{eq: phid}
\Phi_d(\rho,z) = - G M_d /\sqrt{\rho^2 + [a_d + (z^2+b_d^2)^{1/2}]^2}
\end{equation}
is the potential of the disk, and
\begin{equation}
\label{eq: phih}
\Phi_h(r) = - G M_h \ln(1+r/r_h) / r
\end{equation}
is the potential of the halo \citep[e.g.,][]{hern1990,miya1975,nav1997}.

For the bulge and halo, we set $M_b = 3.76 \times 10^9 \msun$,
$M_h = 10^{12} \msun$, $r_b$ = 0.1~kpc, and $r_h$ = 20~kpc.
These parameters match measurements of the mass and velocity dispersion 
inside 1~kpc and outside 50~kpc \citep[see \S2.2 of][]{kenyon2008}. 

To match a circular velocity of 235~\kms\ at the position of the Sun
\citep[e.g.,][]{hogg2005,bovy2012,reid2014},
we adopt parameters for the disk potential $M_d = 6 \times 10^{10} \msun$,
$a_d$ = 2750~kpc, and $b_d$ = 0.3~kpc.  The complete set of parameters for 
the bulge, disk, and halo yields a flat rotation curve from 3--50~kpc.

\subsection{Initial Conditions}
\label{sec: sims-init}

To select \r0\ and \v0, we rely on published calculations for HVSs and runaways.
For HVSs, we consider a model where a single supermassive black hole at the 
GC disrupts close binary systems with semimajor axes \abin\ between \amin\ and 
\amax\ \citep{hills1988,kenyon2008,sari2010}.  Our choice of the minimum semimajor 
axis \amin\ minimizes the probability of a collision between the two binary 
components during the encounter with the black hole \citep{gins2007,kenyon2008}. 
Setting the maximum semimajor axis $\amax \approx$ 4~AU limits the number of low 
velocity ejections which cannot travel more than 10--100~pc from the GC and use
a substantial amount of computer time. 

\subsubsection{Hypervelocity Stars}
\label{sec: sims-hvs}

Numerical simulations of binary encounters with a single black hole demonstrate
that the probability of an ejection velocity \vej\ is a gaussian,
\begin{equation}
\label{eq: pvej}
p_H(\vej) \propto e^{(-(\vej - \veja)^2 / \sigma_v^2)}~ ,
\end{equation}
where the average ejection velocity is 
\begin{equation}\label{eq:vej} 
\veja = 1760 
\left(\frac{\abin}{\rm 0.1\ AU}\right)^{-1/2} 
       \left(\frac{\ma + \mb}{2~\msun}\right)^{1/3}
\left(\frac{\mbh}{3.5 \times 10^6~\msun}\right)^{1/6}
       \facR \ \   {\rm km~s^{-1}} ~ , ~
\end{equation}
and $\sigma_v \approx$ 0.2 \veja\ \citep{bromley2006}. Here
\ma\ (\mb) is the mass of the primary (secondary) star and \mbh\ is the mass 
of the central black hole.  The normalization factor $\facR$ depends on 
$r_{close}$, the distance of closest approach to the black hole:
\begin{eqnarray}\label{eq:facR}
\nonumber
    \facR & = & 0.774+(0.0204+(-6.23\times 10^{-4}
           +(7.62\times 10^{-6}+ \\
& &
(-4.24\times 10^{-8}
        +8.62\times 10^{-11}D)D)D)D)D,
\end{eqnarray}
where
\begin{equation}\label{eq:D}
D = D_0
\left(\frac{\rclose}{\abin}\right) ~ 
\end{equation}
and
\begin{equation} 
\label{eq: D0}
D_0 = \left[\frac{2 \mbh}{10^6 (\ma + \mb)}\right]^{-1/3}.
\end{equation}
This factor also sets the probability for an ejection, $P_{ej}$:
\begin{equation}
\label{eq:PE}
P_{ej} \approx 1 - D/175
\end{equation}
for $0 \le D \le 175$. For $D > 175$, $\rclose \gg \abin$; the binary
does not get close enough to the black hole for an ejection and
$P_{ej} \equiv 0$. 

To establish initial conditions, we select each HVS from a random
distribution of \abin, \rclose, and \vej.  The binaries have semimajor axes 
uniformly distributed in log \abin\ \citep[e.g.,][]{abt1983,duq1991,hea1998}. 
For binaries with $a$ = \amax, the maximum distance of closest approach is 
$r_{close,max} = 175 ~ \amax\ / D_0$. We adopt a minimum distance of closest 
approach $r_{close,min}$ = 1~AU.  Within this range, the probability of any 
\rclose\ grows linearly with $r$. Choosing two random deviates thus yields 
\abin\ and \rclose; \veja, $D$, and $P_{ej}$ follow from 
eqs.~(\ref{eq:vej}--\ref{eq:PE}).  Selecting a third random deviate 
from a gaussian distribution yields the ejection velocity. Two additional 
random deviates drawn from a uniform distribution spanning the main sequence
lifetime of the star fix $t_{ej}$ and $t_{obs}$.  To see whether this combination 
of parameters results in an ejection, we select a sixth random deviate, $P$, 
and adopt a minimum ejection velocity $v_{ej,min}$ = 600~\kms. Stars with 
smaller ejection velocities cannot escape the GC \citep{kenyon2008}.  When 
$P_{ej} \ge P$, $\vej \ge v_{ej,min}$, and $t_{ej} < t_{obs}$, we place the 
star at a random location on a sphere with a radius of 1.4 pc centered on the 
GC and assign velocity components appropriate for a radial trajectory from 
the GC.  Failure to satisfy the three inequalities results in a new selection 
of random numbers.

\subsubsection{Runaway Stars}
\label{sec: sims-run}

For runaway stars, we consider two analytic models for the ejection velocity. 
Following \citet{bromley2009}, we assume runaway companions of a supernova 
have an exponential velocity distribution:
\begin{equation}
\label{eq: pej-sn}
p_S(\vej) \propto e^{-\vej / \vejb} ~ ,
\end{equation}
where $\vejb \approx$ 150~\kms. For a minimum (maximum) velocity of ejected 
stars of 20~\kms\ (400~\kms), this distribution roughly matches simulations
of binary supernova ejections \citep{port2000}.

Predicted velocity distributions for stars ejected dynamically are much steeper 
\citep{perets2012}. To allow a reasonable number of high velocity ejections,
we adopt
\begin{equation}
\label{eq: pej-dy}
p_D(\vej) \propto \left\{
\begin{array}{lll}
   v_{ej}^{-3/2} & & v_{ej} \le \vejc \\
                                  \\
   v_{ej}^{-8/3} & & v_{ej} \ge \vejc \\
\end{array}
\right.
\end{equation}
where \vejc = 150 \kms\ \citep{perets2012}. In our standard calculations, we 
set a minimum ejection velocity of 20~\kms\ and a maximum ejection velocity of 
800~\kms.  To improve the accuracy of the statistics for the highest velocity 
runaways, we perform a second set of simulations with a minimum velocity of 
50~\kms. Together, these simulations yield a robust picture for the frequency 
and observable parameters for runaways produced by the dynamical ejection 
mechanism.

Both of these models yield small production rates for high velocity runaways.
To enable more robust comparisons with simulations of HVSs, we consider a `toy'
model where the ejection velocity is uniformly distributed between 400~\kms\ and
600~\kms. Thus, we use eqs. (\ref{eq: pej-sn}--\ref{eq: pej-dy}) to derive rates
for high velocity runaways and the toy model to understand the galactic 
distribution of the highest velocity runaways.

Establishing the initial conditions for runaways also requires a set of random
deviates. We assume the initial space density of runaways follows the space density 
of stars in the Galactic disk. Thus, the probability of ejecting a runaway from a
cylindrical radius $\rho_0$ is
\begin{equation}
\label{eq: pej-rho}
p(\rho_0) \propto \rho_0 e^{-\rho_0 / \rho_s} ~ ,
\end{equation}
where the scale length is $\rho_s$ = 2.4~kpc \citep{siegel2002,bovy2013}. 
We adopt a range for the initial radius, $\rho_0$ = 3--30~kpc \citep{brand2007}. 
Setting the position of the runaway requires two random deviates, one for 
$\rho_0$ and another for the initial longitude in the GC frame. In this 
approach, the initial height above the Galactic plane is $z = 0$.

Once $\rho_0$ is known, we choose a random deviate for \vej\ and two
random deviates for the ejection angles (spherical $\theta$ and $\phi$). 
Adding the velocity from Galactic rotation yields three velocity components.
We then choose a final random deviate for $t_{obs}$. In these simulations,
ejections occur on time scales much shorter than the lifetime of the ejected 
star. Thus, $t_{ej}$ = 0.

\subsection{Numerical Technique}
\label{sec: sims-tech}

To integrate the motion of each ejected star through the Galactic potential,
we use an adaptive fourth-order integrator with Richardson extrapolation
\citep[e.g.,][]{press1992,bk2006,bromley2009}. Starting from an initial 
position \r0\ with velocity \v0, the code integrates the full three-dimensional
orbit through the Galaxy, allowing us to track position and velocity as a 
function of time. We integrate the orbit for a time $t_f = t_{obs} - t_{ej}$,
which is  smaller than the main sequence lifetime of the ejected star. This 
procedure allows us to integrate millions of orbits fairly rapidly. Several 
tests demonstrate our approach yields typical errors of 0.01\% in position 
and velocity after 1--10~Gyr of evolution time.

\section{REACHING THE HALO}
\label{sec: halo-frac}

Before analyzing results from the simulations, it is useful to establish the
initial conditions which enable ejected stars to reach the Milky Way halo. 
Stars orbiting the galaxy have a circular velocity $v_c^2(r) = G M(r) / r$, 
where $M(r)$ is the mass inside radius $r$. For our adopted Milky Way potential,
$v_c \approx$ 235~\kms\ for disk stars with $r \approx$ 3--30~kpc. To reach the 
halo, ejected stars must have a total velocity comparable to the escape velocity, 
$v_{esc}(r)$. To set $v_{esc}(r)$, we calculate the velocity required for particles 
starting from radius $r$ to reach $r$ = 250~kpc with zero velocity. For HVSs 
ejected at $r$ = 1.4~pc, $v_{esc} \approx$ 913~\kms. At $r \approx$ 3--30~kpc, 
$v_{esc}(r) \approx$ 537 $(r / {\rm 10~kpc})^{-0.19}$~\kms. \citet{brown2014} 
quote a more accurate, polynomial approximation to $v_{esc}$ which is valid over 
a larger range of Galactocentric distances.

With the definitions in \S\ref{sec: sims-hvs}, many HVSs ejected from the GC 
reach the outer halo. In our simulations, roughly 18\% of HVSs have initial
velocities larger than $v_{esc}$. Another 6\% have ejection velocities,
$v_0 \approx$ 850--913~\kms, sufficient to reach $r \approx$ 60--100~kpc. 
For these speeds, typical travel times to reach the halo are 100--250~Myr 
\citep[see also][]{brown2014}. If most HVS ejections occur roughly in the middle
of the main sequence lifetime, stars with $t_{ms} \gtrsim$ 200--500~Myr escape 
the Galaxy as main sequence stars \citep[e.g.,][]{bromley2006,kenyon2008,rossi2013}.

Among the bound population of HVSs, most lie close to the GC. Roughly 60\% of
ejected stars have \vej\ = 600--750~\kms\ and maximum distances of 1~kpc from 
the GC. With their low Galactic latitudes, $b \lesssim$ 7\deg, detecting this
population requires infrared surveys.  Another 5\% have \vej\ = 755--780~kms; 
these stars have maximum distances of 5--20~kpc from the GC. Compared to the 18\% 
of unbound HVSs, the population of bound HVSs near the solar circle makes up a 
small fraction of all ejected stars.

Despite starting far from the GC, it is hard for runaways to reach the outer 
halo.  In the supernova ejection model, the maximum ejection velocity of 
400~\kms\ is smaller than the escape velocity of 660~\kms\ (430~\kms) at 
$r$ = 3~kpc (30~kpc).  Although the maximum velocity in the dynamical ejection 
model, 800~\kms, exceeds $v_{esc}$ at all locations in the disk, few runaways
achieve such large ejection velocities.  Typical velocities are smaller than
400~\kms.  Thus, {\it runaways need a boost from Galactic rotation to reach 
the halo} \citep[see also][]{bromley2009}. 

To quantify the fraction of runaways which can escape the disk and reach the
outer halo, we consider the initial velocity of an ejected star with rotational
velocity $\vec{v}_c$ and ejection velocity $\vec{v}_{ej}$. The angles between
the two velocity vectors are $\alpha$ (in the Galactic plane) and $\beta$ (out
of the plane). The initial velocity of the star is then
\begin{equation}
v_0^2 = v_c^2 + v_{ej}^2 + 2 v_c v_{ej} {\rm cos}~\alpha ~ {\rm cos}~\beta ~ .
\label{eq: vrun}
\end{equation}
For any $v_{ej}$, stars ejected along the direction of Galactic rotation
($\alpha = \beta = 0$) have the maximum initial velocity, $v_0 = v_c + v_{ej}$.
These stars have the best chance to reach the outer part of the Galaxy.  Stars 
ejected in the opposite direction ($\alpha = -\pi$, $\beta$ = 0) have the 
smallest initial velocity, $v_0 = v_c - v_{ej}$, and the worst chance to reach
the outer Galaxy. When stars are ejected perpendicular to the disk ($\beta$ = 
$\pi$/2), they have an intermediate velocity, $v_0 = (v_c^2 + v_{ej}^2)^{1/2}$, 
and a modest chance to reach the halo.  At other angles, $v_0$ has a constant 
value when $\alpha = -\beta$ (sin~$(\alpha + \beta)$ = 0), which defines a 
circle in the $\alpha - \beta$ plane.  

To calculate the fraction of runaways which can reach $r \gtrsim$ 60~kpc, we derive
the allowed range of $\alpha$ and $\beta$ for runaways with ejection velocity
$v_{ej}$ starting from distance $r$ from the GC. Integrating over the appropriate
probability distributions for $v_{ej}$ (eqs. [\ref{eq: pej-sn}--\ref{eq: pej-dy}]),
the initial position (eq. [\ref{eq: pej-rho}]), and the ejection angles yields 
the total fraction $f$ of runaways with initial distance $r$ that reach 
$r \gtrsim$ 60~kpc. To illustrate the difficulty of reaching the Galactic halo, 
we calculate $f$ for all runaways and those with maximum $b \gtrsim$ 30\deg.

\begin{figure}
\includegraphics[width=6.5in]{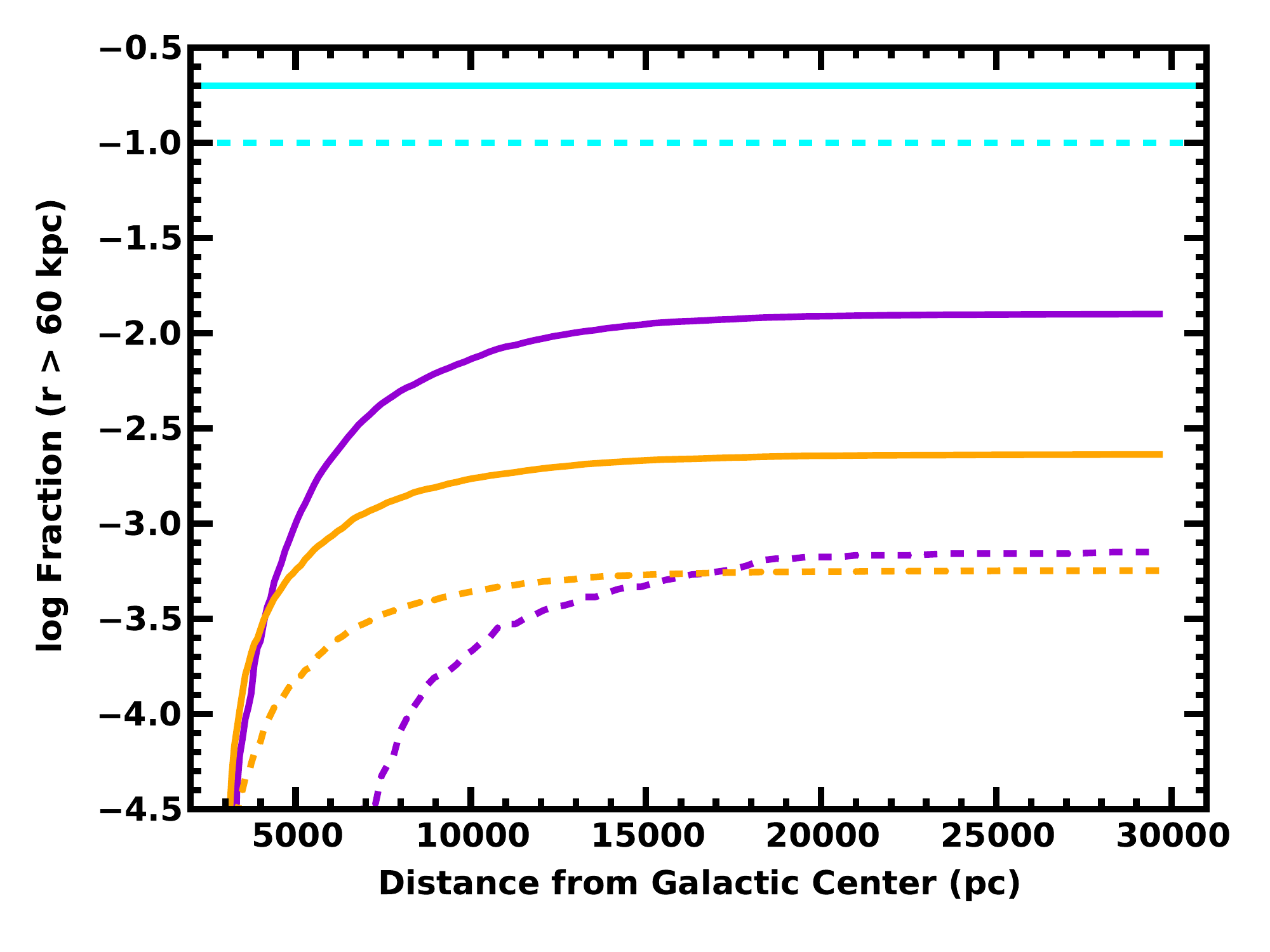}
\vskip 2ex
\caption{\label{fig: halo-frac}
Cumulative fraction of stars with final Galactocentric distances, $d \gtrsim$ 
60~kpc, as a function of their initial position in the disk for supernova-induced 
runaways (violet curves) and dynamically generated runaways (orange curves).
Solid lines plot results for all stars; dashed lines show results for halo
stars with $|b| \ge$ 30\deg. Cyan lines show the fraction of HVSs ejected 
from the GC which reach the halo (solid line: all $b$; dashed line: 
$|b| \ge$ 30\deg). 
}
\end{figure}

This exercise demonstrates that few runaways reach the outer Galaxy 
(Fig.~\ref{fig: halo-frac}). Roughly 1\% of all supernova-induced runaways
travel beyond 60~kpc (solid violet curve). Stars with initial positions 
$r \lesssim$ 10~kpc contribute nearly all of the ejected stars. Within this 
group, less than 10\% (0.1\% of all runaways) reach $d \gtrsim$ 60~kpc with
$|b| \gtrsim$ 30\deg. Although dynamical ejections into the outer Galaxy are 
more rare ($\sim$ 0.3\% of the total population), dynamical ejections into
the outer halo are as frequent as supernova-induced ejections, $\sim$ 0.06\% 
of all runaways.

The larger maximum ejection velocity in the dynamical model accounts for these
differences. Most runaways are ejected at 3--5~kpc, where the escape velocity 
is large. With a maximum $v_{ej}$ of 400~\kms, supernova-induced runaways 
require the maximum boost from Galactic
rotation to reach the outer Galaxy. Sacrificing some of this boost to eject
stars into the halo keeps stars from reaching the outer Galaxy. Few of these 
high $b$ runaways reach the outer halo. Dynamically ejected stars with ejection 
velocities of 600--800~\kms\ require little boost from Galactic rotation. 
These stars easily reach the outer halo. Compared to the supernova model, 
however, the dynamical model yields a smaller fraction of stars with high 
velocities. The lack of high velocity stars compensates for the relative
ease of reaching the halo, resulting in comparable fractions of high velocity
halo stars with $r \gtrsim$ 60~kpc in both models.

If the production rates for HVSs and both types of runaways are comparable, this 
analysis predicts that HVSs dominate the population of high velocity stars in the
outer halo. For every high speed runaway generated by a supernova or a dynamical 
interaction among massive stars, there should be roughly 100 HVSs. We will
re-consider this conclusion in \S\ref{sec: obs-rates} when we examine 
predicted production rates for each mechanism.

\section{COMPLETE SAMPLES OF STARS}
\label{sec: res1}

All ejection models yield populations of bound and unbound stars
\citep{bromley2006,kenyon2008,bromley2009}.  To explore the properties of
both populations, we consider simulations of 1~\msun\ and 3~\msun\ stars. 
Calculations with long-lived solar-type stars provide a sample of bound 
stars in the solar neighborhood and a sample of unbound stars with a
broad range of distances. While current facilities can probe the bound
population, most unbound stars are too distant and too faint for detailed
study. Although simulations with shorter-lived, more luminous 3~\msun\ stars 
yield a smaller sample of bound stars, the population of unbound stars is
well-matched to the sensitivity of GAIA and large ground-based optical 
telescopes.  These two sets of simulations allow us to derive general 
predictions for the bound and unbound populations.

The numerical simulations of the motions of HVSs and runaways through the Galaxy 
yield ensembles of $10^6 - 10^7$ stars with final positions \rf\ and velocities 
\vf\ relative to the GC.  These data represent a snapshot of all ejected stars 
still on the main sequence. The HVSs fill a spherical volume from the GC out to 
roughly 1~Mpc (34~Mpc) for 3~\msun\ (1~\msun) stars. Although runaways are more 
concentrated towards the Galactic disk \citep[e.g.,][]{bromley2009}, a few reach 
Galactocentric distances of $\sim$ 300~kpc (3~\msun) to 7~Mpc (1~\msun). To put
these results in perspective, the modern magnitude-limited surveys described in
\S7 can probe 1~\msun\ (3~\msun) stars to $d$ = 10~kpc (100~kpc).

To derive heliocentric observables for each star in a snapshot, we set the Sun 
at a position $(-\rsun, 0, 0)$ with velocity $(0, \vsun, 0)$ relative to the GC. 
We adopt \rsun\ = 8~kpc and \vsun\ = 235~\kms\ \citep[e.g.,][]{bovy2012} and 
divide each ensemble into 
five distance bins, $d \le$ 10~kpc, 10~kpc $< d \le$ 20~kpc, 20~kpc $< d \le$ 
40~kpc, 40~kpc $< d \le$ 80~kpc, and 80~kpc $< d \le$ 160~kpc.  Tables~1--2 list 
the median $v_{med}$, first and third quartile $v_{q1}$ and $v_{q3}$, average 
$v_{avg}$, and standard deviation for the radial ($\sigma_r$) and tangential 
($\sigma_t$) velocities in each simulation. Table~3 summarizes statistics for
the proper motion.  For runaways produced by dynamical interactions, we quote 
results for simulations with a minimum ejection velocity of 50~\kms.  Velocity 
distributions for calculations with a smaller ejection velocity of 20~\kms\ are 
fairly similar to those for supernova-induced runaways.

In the next subsections, we examine several broad trends in the variation 
of $\mu$ and $v_r$ with $d$, $l$, and $b$. After discussing predicted
distributions of stars in the $d - \mu$ (\S\ref{sec: res1-mud}) and $d - v_r$ 
(\S\ref{sec: res1-vrd}) planes, we describe predicted histograms for $v_r$ 
and $\mu$ in well-defined distances bins and for the complete ensemble of
stars in each simulation (\S\ref{sec: res1-vrmuhist}).  To isolate how observables 
depend on Galactic coordinates, we then discuss the distribution of stars in 
the $l - \mu$  plane for specific ranges of Galactic latitude 
(\S\ref{sec: res1-mul}). This section concludes with a brief summary of the
major results (\S\ref{sec: res1-summ}).

\subsection{Ejected Stars in the $d - \mu$ Plane}
\label{sec: res1-mud}

To investigate the distribution of stars as a function of $\mu$ and $d$, 
we construct a density diagram.  For stars with $|b| \ge$ 30\deg, we 
(i) divide the log~$d$--log~$\mu$ plane into bins spaced by 0.01 in log~$d$ 
and log~$\mu$, (ii) count the number of stars in each bin, and (iii) plot the 
relative number in a contour diagram. In each diagram, bright red represents 
the largest density; dark blue the smallest density.  The full range in 
relative density varies from a factor of 5--10 for 1~\msun\ runaways to a 
factor of 50--500 for 1--3~\msun\ HVSs.

\begin{figure}
\includegraphics[width=6.5in]{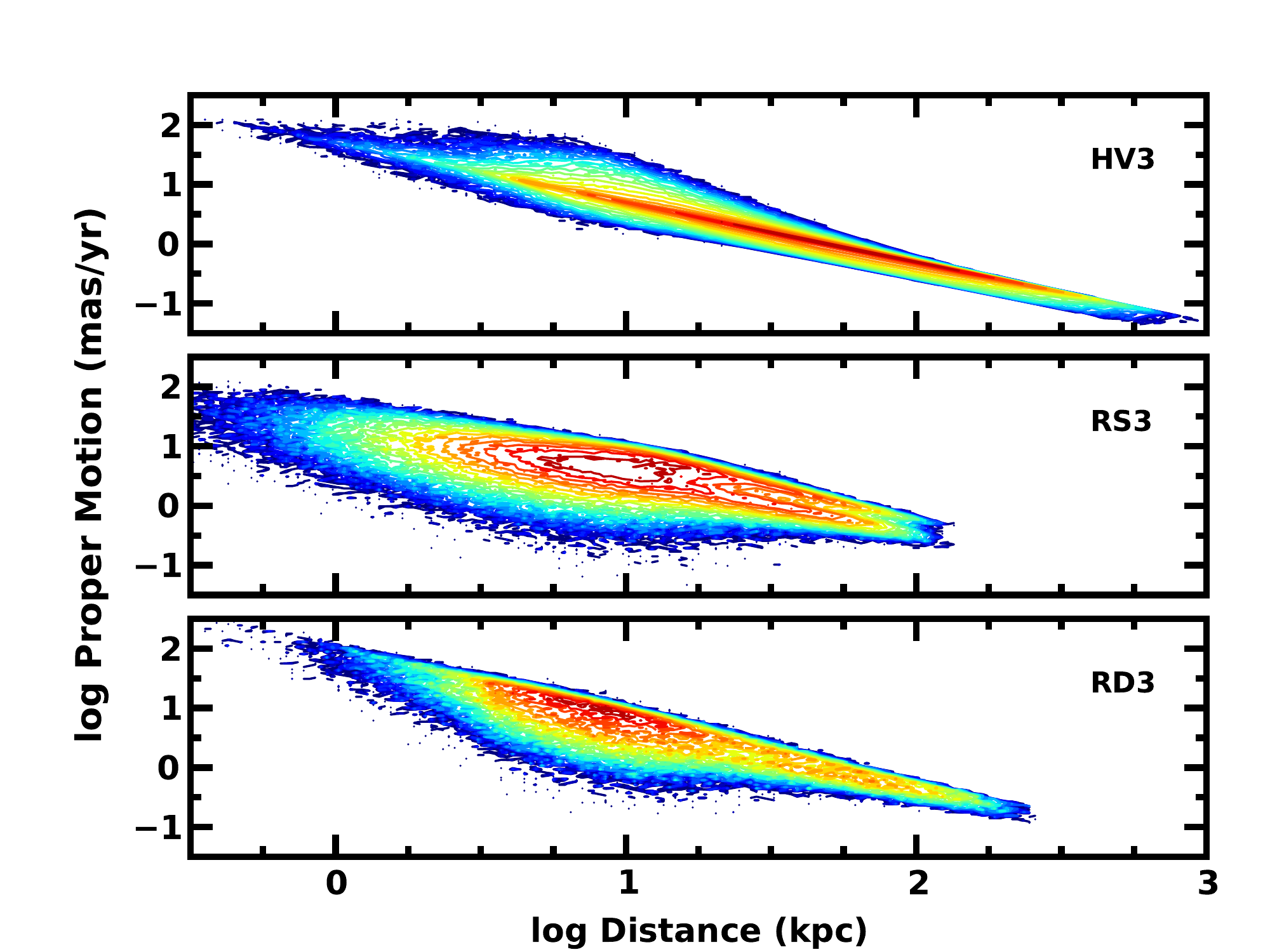}
\vskip 2ex
\caption{\label{fig: mud3}
Predicted density distributions of 3~\msun\ HVSs and runaways with $|b| \ge$ 
30\deg\ in the $\mu-d$ plane. In each panel, the density scale is logarithmic
with a minimum of 0.0 (displayed as dark blue). The maximum of the density 
(displayed as bright red) varies from 2.7 (HV3) to 1.7 (RS3) to 1.4 (RD3).
{\it Upper panel:} most 3~\msun\ HVSs lie along the linear $\mu(d)$ relation 
expected for reflex solar motion. HVSs ejected out the Galactic poles lead 
to a small concentration above this relation at $d \approx$ 10~kpc. 
{\it Middle panel:} compared to HVSs, 3~\msun\ runaways ejected during a supernova
show much more scatter about the linear $\mu(d)$ relation for reflex solar motion. 
At large distances, the bimodal proper motion distribution of 3~\msun\ runaways 
mirrors Galactic rotation. 
{\it Lower panel:} 3~\msun\ runaways ejected from dynamical interactions among
massive stars follow the linear $\mu(d)$ relation and display a bimodal proper
motion distribution at large $d$. Due to their larger ejection velocities, some
runaways reach larger distances. Runaways in the Galactic anti-center produce 
the ensemble of nearby stars with very small proper motions.
}
\end{figure}

\begin{figure}
\includegraphics[width=6.5in]{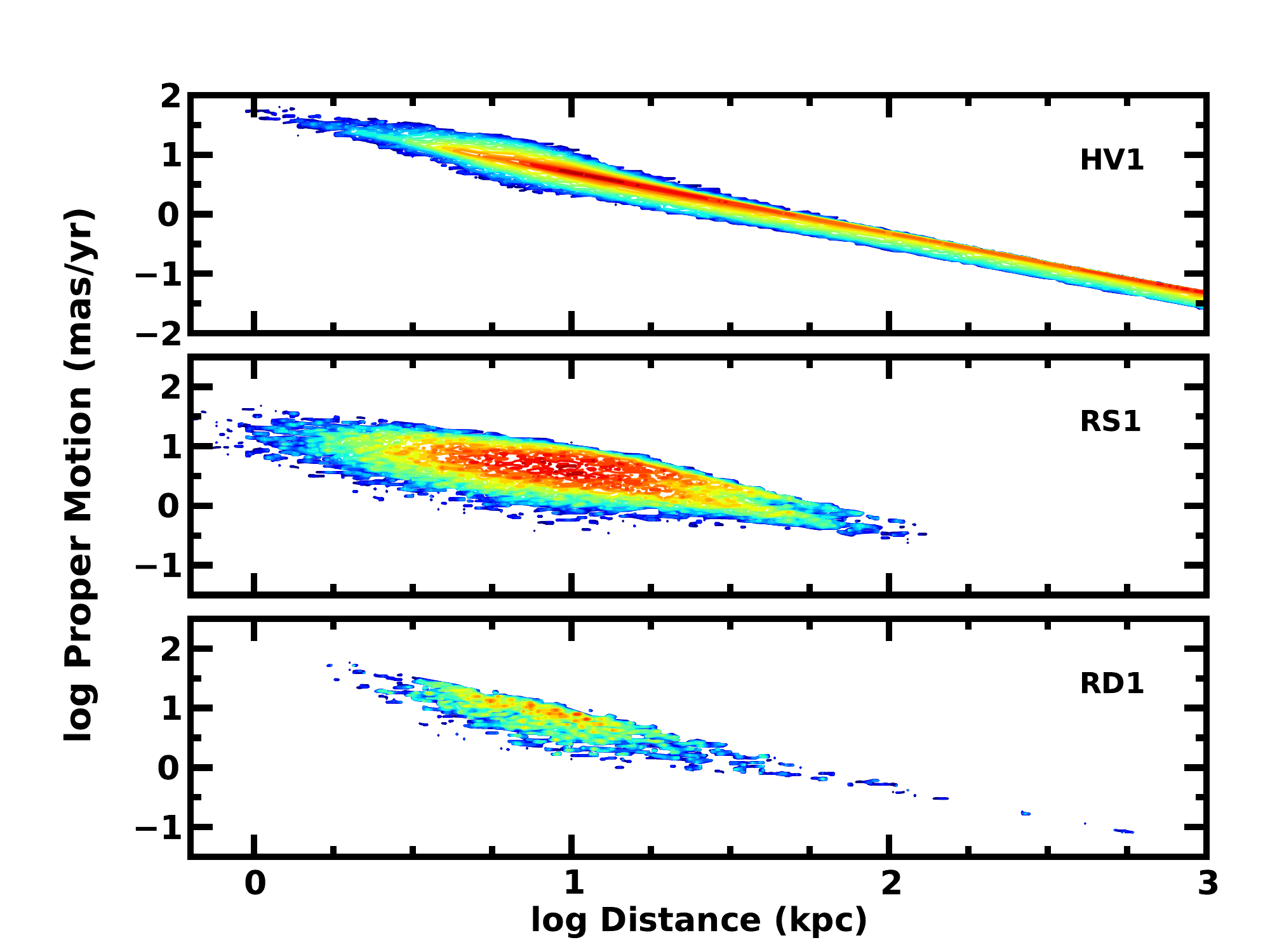}
\vskip 2ex
\caption{\label{fig: mud1}
As in Fig.~\ref{fig: mud3} for 1~\msun\ stars. The maximum of the density is
1.8 (HV1), 1.0 (RS1), or 0.7 (RD1).
{\it Upper panel:} the distribution of 1~\msun\ HVSs extends to large $d$,
with two density maxima at 10~kpc and at 2--3~Mpc.
{\it Middle panel:} Most supernova-induced runaways have $d \approx$ 10~kpc;
a few reach $d \gtrsim$ 80~kpc.
{\it Lower panel:} Few dynamically-generated runaways reach high Galactic latitude.
Most of these have $d \lesssim$ 10~kpc; a few reach $d \approx$ 100--200~kpc.} 
\end{figure}

Fig.~\ref{fig: mud3} shows predicted density distributions for 3~\msun\ HVSs 
and runaways.  Fig.~\ref{fig: mud1} plots predictions for 1~\msun\ stars.  
The HVS results assume stars ejected from equal mass binaries 
(1~\msun: $a_{bin}$ = 0.032--4~AU,
$t_{ms}$ = 10~Gyr;  3~\msun: $a_{bin}$ = 0.115--4~AU, $t_{ms}$ = 350~Myr).  For 
the runaway simulations, we adopt minimum ejection velocities of 20~\kms\ (supernova 
ejections) or 50~\kms\ (dynamical ejections). Eliminating the lower velocity 
dynamical ejections artificially enhances the density at large proper motions 
relative to small proper motions. This enhancement provides a clearer picture of 
the relative frequency of the highest velocity runaways.

\begin{figure}
\includegraphics[width=6.5in]{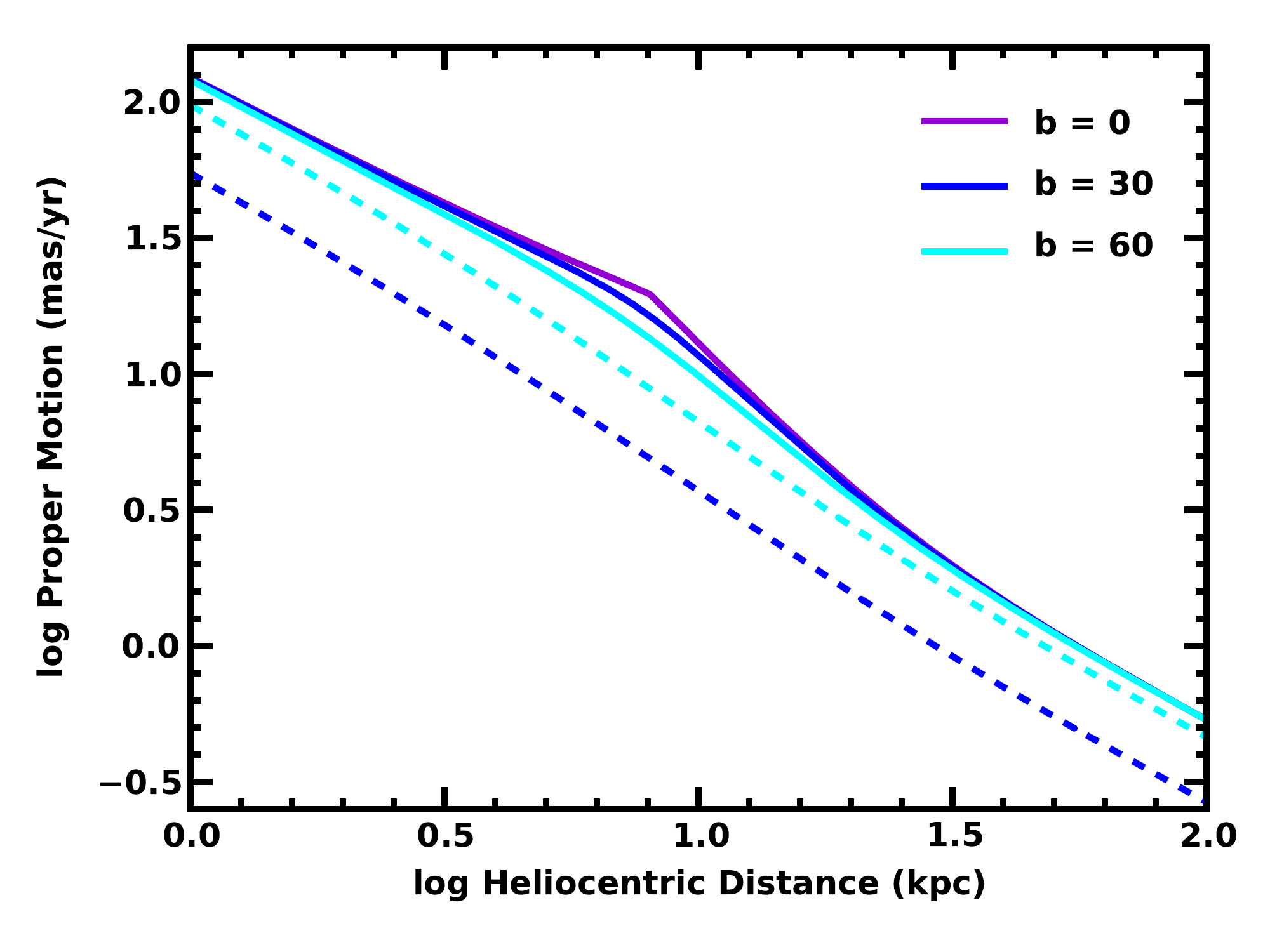}
\vskip 2ex
\caption{\label{fig: mu-rad}
Predicted proper motions for radially outflowing stars as a function
of heliocentric distance. The legend indicates the galactic latitude 
for each curve. Solid curves show the maximum proper motion at each $b$;
dashed curves show the minimum proper motion at each $b$.
}
\end{figure}

These results demonstrate that nearly all of the proper motions for 3~\msun\ HVSs 
result from reflex solar motion (Fig.~\ref{fig: mud3}, upper panel). Most HVSs fall 
close to the line 
\begin{equation}
\mu = 49.6 \left ( { d \over {\rm 1~kpc} } \right ) ~ \masyr ~ . 
\end{equation}
At fixed $d$, stars with smaller $\mu$ have smaller $b$ (see also 
Fig.~\ref{fig: mu-rad}).  Along the locus, the number of 3~\msun\ HVSs peaks 
at $d \approx$ 50~kpc. 

Above the $\mu(d)$ locus, there is a cloud of stars with $d \lesssim$ 10--20~kpc 
and $\mu \lesssim$ 100~\masyr. This group of mostly bound HVSs lies at all $b$ 
in the direction of the GC. Stars ejected along the $z$-axis produce this clump
of high proper motion stars (see Fig.~\ref{fig: mu-rad}).  

Although runaways generally follow the $\mu (d)$ relation expected for reflex solar
motion, the distribution about this relation is much more diffuse than for HVSs
(Fig.~\ref{fig: mud3}, middle and lower panels).  Galactic rotation produces this 
fuzziness. In HVS ejections, the distribution of ejection velocities is gaussian; 
the position of a star along the $\mu(d)$ relation is a simple function of this 
ejection velocity and the flight time. In runaway ejections, the ejection velocity 
consists of Galactic rotation plus a random velocity with a random angle relative 
to Galactic rotation.  This randomness creates a much larger dispersion of space 
velocities and much larger dispersion about the simple $\mu(d)$ relation.  

For $d \approx$ 20--100~kpc, galactic rotation also produces twin peaks in the
density at fixed distance. Separated by roughly 0.3 in log~$\mu$, these twin
density maxima are very prominent in the ensemble of supernova-induced runaways
(middle panel) and less prominent among the dynamically generated runaways
(lower panel). For runaways in the 
Galactic anti-center, the rotational component of their motion is parallel to the 
Sun's motion. These stars lie in the low proper motion peak.  Distant runaways in 
the direction of the GC are beyond the GC; the rotational component of their space 
motion is anti-parallel to the Sun's motion. These stars produce the high proper
motion peak. Nearby indigenous disk stars in the direction of the GC have rotational 
motions parallel to the Sun, eliminating the double-peaked aspect of the proper 
motion distribution.

The larger maximum velocities from dynamical ejections blur the double-peaked
distributions of proper motions identified in runaways from supernovae
(Fig.~\ref{fig: mud3}, lower panel). Despite the diffuse nature of the contour
diagram, Galactic rotation is clearly visible at 20--50~kpc.  As with HVSs and 
supernova-induced runaways, the width of the proper motion distribution narrows 
with increasing distance. 

Results for 1~\msun\ HVSs and runaways are similar (Fig.~\ref{fig: mud1}). The
HVSs closely follow the linear $\mu(d)$ relation out to $d$ = 30~Mpc 
(Fig.~\ref{fig: mud1}, upper panel).  The density of 1~\msun\ HVSs has two clear 
maxima at $d \approx$ 10~kpc and $d \approx$ 2--3~Mpc.  Stars at high $b$ closely 
follow the line (red contour); stars at $b \approx$ 30\deg\ occupy the blue contour 
below the line. At $d \approx$ 10~kpc, there is a group of stars with large $\mu$ 
above the red contour. High velocity ejections along the Galactic poles produce 
this collection of large proper motion stars (see also Fig.~\ref{fig: mu-rad}).

Despite their lower frequency, 1~\msun\ runaways also clearly follow the linear
$\mu(d)$ relation expected for solar reflex motion. Aside from having a shape similar 
to the contours for the 3~\msun\ runaways, the contours for 1~\msun\ runaways extend 
to slightly larger distances due to their longer main sequence lifetimes. 

The number and location of density peaks for HVSs in Fig.~\ref{fig: mud3}--\ref{fig: mud1} 
depend solely on stellar lifetime \citep[see also][]{bromley2006,kenyon2008}.  For 
ejected stars with infinite lifetimes, the space density $n$ is a simple power-law 
with distance from the GC, $n \propto r^{-2}$. Thus, the total number of HVSs grows 
monotonically with distance.  However, real HVSs have finite lifetimes. The total 
number falls at distances where the travel time exceeds the main sequence lfetime. 
For 1~\msun\ (3~\msun) HVSs with ejection velocities drawn from eq. (\ref{eq:vej}), 
lifetimes of 10~Gyr (350~Myr) result in peaks at 2--3~Mpc (50~kpc). The second peak 
in the density of 1~\msun\ stars results from bound stars with orbital periods smaller 
than the main sequence lifetime.  Continuous ejection of relatively low velocity HVSs 
over 10~Gyr produces a large concentration of bound 1~\msun\ HVSs with $d \lesssim$ 20~kpc. 
The short lifetimes of 3~\msun\ HVSs preclude a significant concentration of nearby 
HVSs.  

The space density of stars in the disk sets the density of runaways in these diagrams.
For stars with an exponential distribution of ejection velocities, unbound stars are 
very rare (\S\ref{sec: halo-frac}). Thus, most stars in the diagram are bound to the 
Galaxy. For bound stars at $|b| >$ 30\deg, the final in-plane distance from the GC, 
$\rho_f$, is similar to the initial distance, $\rho_0$. The space density of these 
stars follows the initial density, which is concentrated towards the GC. As a result, 
most stars have $d \lesssim$ 10--20~kpc.  

In both diagrams, the dynamical and supernova ejection scenarios produce an ensemble 
of stars at $d \approx$ 10~kpc with smaller proper motions than the locus of stars 
with solar reflex motion.  Within this group, stars with the smallest $\mu$ are
concentrated towards small $b$ at a variety of Galactic longitudes 
$l \approx \pm$100\deg--280\deg\ where the tangential velocity reaches a minimum.
Some of these stars have large ejection velocities parallel to the Sun's trajectory
(see Fig.~\ref{fig: v-rad50hel}). Others have modest ejection velocities perpendicular
the plane, which enable them to reach large $b$ but not escape the Galaxy
(see Fig.~\ref{fig: v-orbhel}). 

\subsection{Ejected Stars in the $d$-$v_r$ Plane}
\label{sec: res1-vrd}

To explore the variation of radial velocity with distance, we examine another
density diagram. As in the previous section, we (i) select stars with $|b| \ge$ 
30\deg, (ii) divide the log~$d$--$v_r$ plane into bins spaced by 0.01 in log~$d$ 
and 20~\kms\ in $v_r$, and (iii) count the number of stars in each bin.  In the 
diagrams, bright red represents the largest density; dark blue the smallest 
density. The range in density varies from a factor of 50 for 1~\msun\ runaways 
to a factor of 300 for 3~\msun\ HVSs and runaways.

\begin{figure}
\includegraphics[width=6.5in]{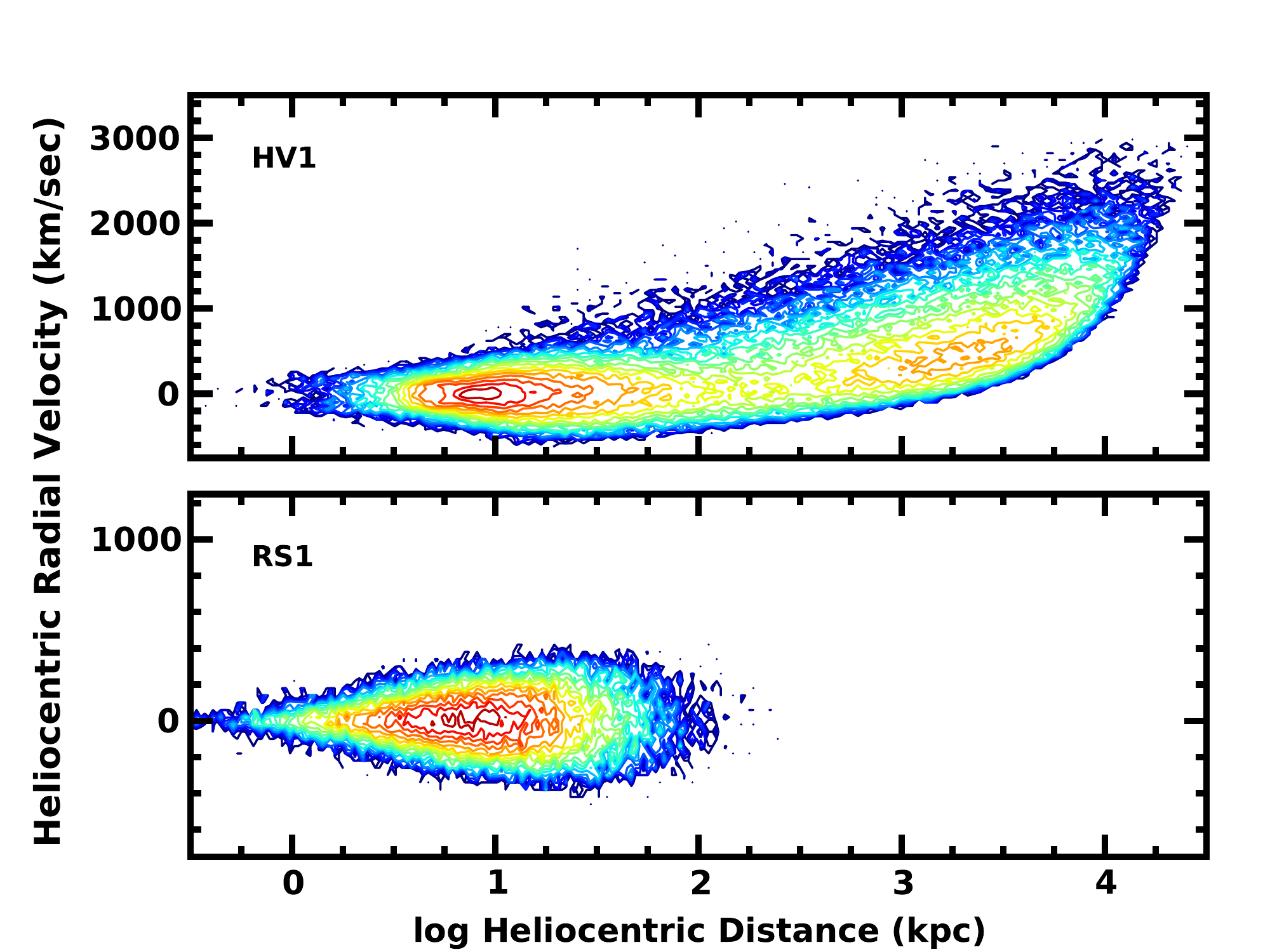}
\vskip 2ex
\caption{\label{fig: vrd1}
Predicted density distributions of 1~\msun\ HVSs and supernova-induced runaways 
with $|b| \ge$ 30\deg\ in the $v_r-d$ plane. In each panel, the density scale is 
logarithmic with a minimum of 0.0 (displayed as dark blue) and a maximum (displayed 
as bright red) of 2.1 (HV1) or 1.7 (RS1).
{\it Upper panel:} bound 1~\msun\ HVSs lie within a band from 5--50~kpc symmetric
about median $v_r$ = 0~\kms. Beyond this locus, unbound HVSs produce a secondary
peak at $d \approx$ 1--5~Mpc. 
{\it Lower panel:} compared to HVSs, 1~\msun\ runaways ejected during a supernova
have a much more symmetric distribution of $v_r$ with $d$. All supernova-induced
runaways are bound; most lie within 10--20~kpc and have small median $v_r$ and 
$\sigma_r$. None reach $d \gtrsim$ 100~kpc.
}
\end{figure}

HVSs and runaways from supernovae show a remarkable diversity in the relative 
density of 1~\msun\ stars as a function of $v_r$ and $d$ (Fig.~\ref{fig: vrd1}). 
Close to the Sun ($d \lesssim$ 1--3~kpc), the relatively few HVSs and runaways 
have fairly symmetric velocity distributions around a median $v_r \approx$ 0 \kms.  
At moderate distances ($d \approx$ 3--20~kpc), the spread in radial velocity grows 
smoothly with distance. Although HVSs have a much larger spread in radial velocity
(see also Table~1), both groups have a clear peak in the relative number of stars 
at $d \approx$ 10~kpc and $v_r \approx$ 0--100~\kms. 

For stars at large distances ($d \gtrsim$ 20~kpc), the velocity distributions
of HVSs and supernova-induced runaways differ dramatically. With their modest
maximum ejection velocities, few runaways reach the outer Galaxy (\S\ref{sec: halo-frac}; 
Fig.~\ref{fig: halo-frac}). For $d \gtrsim$ 20~kpc, the density of runaways drops 
significantly and falls very close to zero at $d \approx$ 100~kpc. Despite the steep 
fall in relative density, the median velocity and the spread in radial velocity are 
roughly constant with distance (Table~1).

The properties of distant HVSs provide a clear contrast with distant runaways.
For HVSs, the median radial velocity and the spread in the radial velocity grow 
with distance. The maximum radial velocity increases from roughly 1000~\kms\ at 
$d \approx$ 10~kpc to 3000~\kms\ at $d \approx$ 10--20 Mpc. Although the relative
density of HVSs falls from 20~kpc to 300~kpc, the relative density displays a
clear secondary peak at $d \approx$ 2~Mpc and $v_r \approx$ 400--500~\kms. 
Beyond 5~Mpc, the density slowly falls and reaches roughly zero at $d \approx$
20~Mpc. 

\begin{figure}
\includegraphics[width=6.5in]{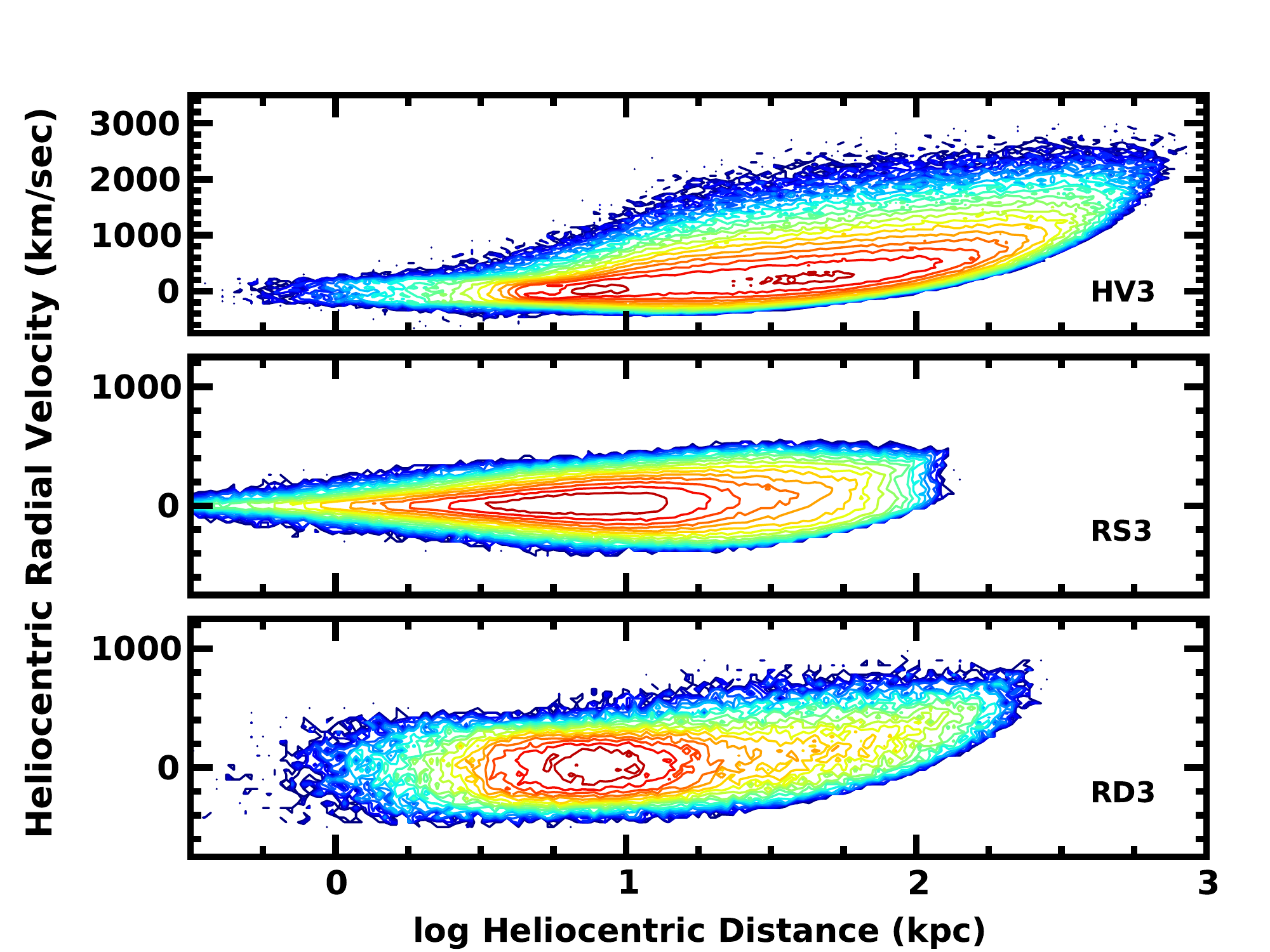}
\vskip 2ex
\caption{\label{fig: vrd3}
As in Fig.~\ref{fig: vrd1} for 3~\msun\ ejected stars. The maximum density is
2.5 (HV3), 2.6 (RS3), or 1.9 (RD3).
{\it Upper panel:} most bound 3~\msun\ HVSs lie within the narrow red band 
where the median $v_r$ increases with distance at $d \approx$ 10--100~kpc. 
Above this locus, unbound HVSs within the green and blue contours have large $v_r$.
{\it Middle panel:} compared to HVSs, 3~\msun\ runaways ejected during a supernova
have a much more symmetric distribution of $v_r$ with $d$. All supernova-induced
runaways are bound; most lie within 10--20~kpc and have small median $v_r$ and 
$\sigma_r$. None reach $d \gtrsim$ 100~kpc.
{\it Lower panel:} 3~\msun\ runaways ejected from dynamical interactions among
massive stars have median $v_r \approx$ 0~\kms\ and dispersion $\sigma_r \approx$
100--150~\kms\ at $d \lesssim$ 10--20~kpc. Nearly all of these runaways are bound.
At larger distances, a few unbound stars reach $d \gtrsim$ 50--100~kpc with large
$v_r$.
}
\end{figure}

The distributions of 3~\msun\ runaways are similar to those of 1~\msun\ runaways
(Fig.~\ref{fig: vrd3}). In the middle panel, supernova-induced runaways show a
clear increase in relative density from $d \approx$ 300~pc to $d \approx$ 10~kpc.
Within this range of distances, the spread in the radial velocity grows smoothly
with distance; the median $v_r$ is close to zero. Beyond this peak, the relative
density drops to zero at $d \approx$ 100~kpc. Among the more distant stars, the
maximum $v_r$ is roughly constant with distance; the minimum $v_r$ grows slowly
with distance. 

Dynamically-generated runaways yield similar results.  When we adopt a minimum 
$v_{ej}$ = 20~\kms\ for dynamically-generated runaways, the distribution is
nearly indistinguishable from the middle panel of Fig.~\ref{fig: vrd3}.  Increasing 
the minimum $v_{ej}$ to 50~\kms\ removes the ensemble of low velocity stars from 
the diagram, reducing the density and increasing the spread in $v_r$ for nearby 
stars (Fig.~\ref{fig: vrd3}, lower panel). Despite this difference, dynamically 
generated runaways still display a clear peak in relative density at $d \approx$
10~kpc. Around this peak, stars have a median $v_r$ close to zero and a spread of 
$\pm$500~\kms\ (Table~1). 

Because the dynamical model yields a maximum $v_{ej} \approx$ 800~\kms, a few 
runaways reach larger distances and have larger radial velocities than 
supernova-induced runaways. Despite this difference, very few runaways are 
unbound (\S\ref{sec: halo-frac}).

Although 3~\msun\ HVSs have a nearly identical distribution of ejection velocities
as 1~\msun\ HVSs, the density distributions in the $d$--$v_r$ plane show several clear 
differences. The primary peak in the density lies at somewhat smaller distances, at 
$\sim$ 8~kpc instead of $\sim$ 10~kpc. The secondary peak falls at much smaller
distances, $\sim$ 50~kpc instead of a few Mpc. The drop in density at large distances
is much more rapid, falling to zero just inside 1~Mpc instead of reaching to 30~Mpc.

These trends have simple physical explanations \citep{bromley2006,kenyon2008}.
For our adopted MW potential and HVS parameters, roughly 10\% of ejected stars
have velocities large enough to reach 10--20~kpc but too small to travel beyond 
60--100~kpc. Typical travel times of 100--400~Myr for these bound stars are a 
significant fraction of the main sequence lifetime of a 3~\msun\ star, but are much 
smaller than the lifetime of a 1~\msun\ star. Bound stars with long lifetimes
have median velocities close to zero and modest velocity dispersions of 
100--200~\kms. However, many bound stars with short lifetimes reach $d \approx$ 
30--60~kpc and evolve off the main sequence before returning to the solar circle. 
Thus, there is a large deficit of bound, massive stars with negative radial velocity.
For 3~\msun\ stars, this deficit leads to a median $v_r$ larger than zero and a
larger velocity dispersion than 1~\msun\ HVSs (see also Table~1).

Among unbound stars, finite stellar lifetimes are also responsible for trends in 
$v_r$ with distance. Stars with larger ejection velocities reach larger distances;
the median $v_r$ and dispersion in $v_r$ thus increase with $d$. Longer lifetimes
also enable stars to reach larger $d$. With a factor of 30 longer lifetime, the
1~\msun\ stars reach 30 times larger distances than 3~\msun\ stars (30~Mpc instead 
of 1~Mpc).

The initial velocity distributions of HVSs and runaways produce the stark differences 
in Figs.~\ref{fig: vrd1}--\ref{fig: vrd3}. Among runaways ejected from 3--30~kpc in 
the Galactic disk, nearly all have modest ejection velocities and remain
bound to the Galaxy (\S5). Bound stars with modest ejection velocities reach
maximum distances of roughly 100~kpc before falling back into the Galaxy.  
The small fraction of runaways which reach the halo have typical $d \approx$
10~kpc and $v_r \lesssim$ 100--150~\kms. 

\subsection{Radial Velocity and Proper Motion Histograms}
\label{sec: res1-vrmuhist}

The density plots in Figs.~\ref{fig: mud3}--\ref{fig: vrd3} demonstrate the rich
behavior in the predicted $\mu(d)$ and $v_r(d) $ as a function of initial 
conditions and stellar properties. To focus on predictions for large ensembles
of HVSs and runaways, we now consider the frequency distributions of $v_r$ and
$\mu$ for Galactic halo stars ($|b| \ge$ 30\deg) in a discrete set of distance 
bins. In these diagrams, color encodes distance (violet: $d \le$ 10~kpc, blue: 
10~kpc $< d \le$ 20~kpc, green: 20~kpc $< d \le$ 40~kpc, and orange: 40~kpc 
$< d \le$ 80~kpc). Tables~1--3 summarize statistics for $\mu$, $v_r$, and 
$v_t$ in each distance bin.

\begin{figure}
\includegraphics[width=6.5in]{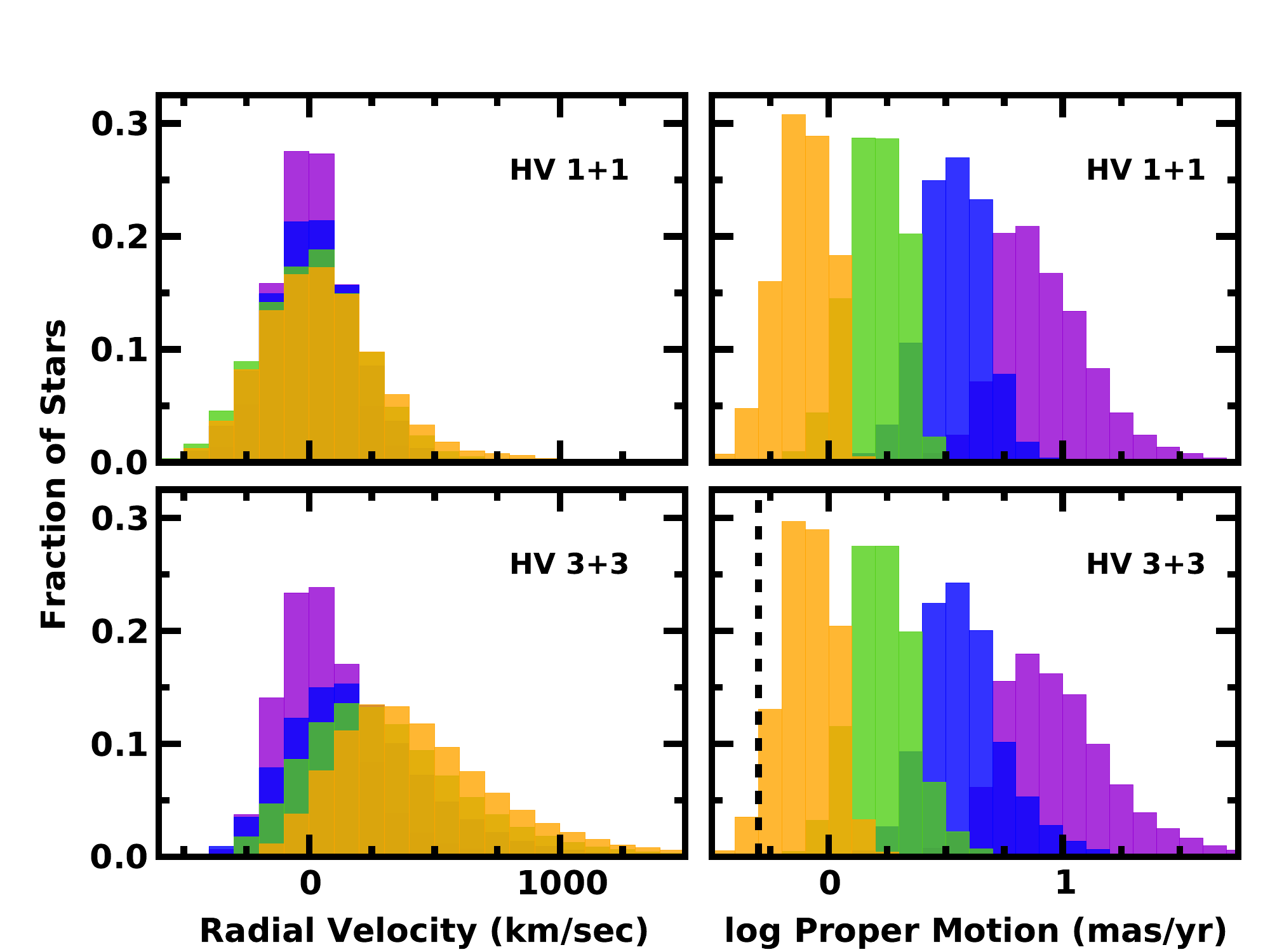}
\vskip 2ex
\caption{\label{fig: vhist-hvs}
Predicted distributions of radial velocity (left panels) and proper motion 
(right panels) velocity for HVSs produced from 1 \msun\ + 1 \msun\ binaries 
(upper panels) and from 3 \msun\ + 3 \msun\ binaries (lower panels) with 
Galactic latitude $|b| >$ 30\deg\ and distances $d <$ 10~kpc (violet histograms), 
10~kpc $< d <$ 20~kpc (blue histograms), 20~kpc $< d <$ 40~kpc (green histograms), 
and 40~kpc $< d <$ 80~kpc (orange histograms).  Dashed lines in the right panels
indicate the 3$\sigma$ GAIA detection limit for stars with $g \lesssim$ 20
\citep{linde2010}.
The median radial velocity (proper motion) grows (falls) with 
increasing distance. More massive HVSs have larger radial velocities.
}
\end{figure}

Fig.~\ref{fig: vhist-hvs} shows the distributions of $v_r$ (left panels) and
$\mu$ (right panels) for 1~\msun\ (upper panels) and 3~\msun\ (lower panels) 
HVSs.  The trends of radial velocity with stellar mass and distance follow the 
correlations in \S\ref{sec: res1-vrd} \citep[see also][]{bromley2006,kenyon2008}.
For 3~\msun\ stars, the median radial velocity grows with increasing distance. 
Among the more distant stars, there is a large tail of very high velocity stars 
with $v_r \gtrsim$ 1000~\kms. As distance decreases, a smaller and smaller 
fraction of stars have high velocities. In the nearby sample with $d \le$ 10~kpc, 
nearly all stars have $v_r \lesssim$ 500~\kms. 

For 1~\msun\ stars, the trend of increasing median velocity with increasing
distance is much weaker (Table~1). For all $d$, the velocity dispersion and 
inter-quartile range are smaller. Although the typical maximum velocity is 
similar, a much smaller fraction of stars has $v_r \gtrsim$ 1000~\kms.

The distributions of proper motion and tangential velocity reverse the trends 
of the radial velocity (Tables~2--3; Fig.~\ref{fig: vhist-hvs}, right panels). 
Distant HVSs moving radially away from the GC have small transverse components 
of their space motion, leading to small tangential velocities and small proper 
motions. As the distance decreases, the angle between the line-of-sight and the 
velocity vector for an HVS grows, leading to larger and larger tangential 
velocities. With $\mu \propto d^{-1}$, nearby HVSs have much larger proper
motions than more distant HVSs (see also Fig.~\ref{fig: v-rad-hd}). Although 
geometry requires that the maximum $v_r$ exceed the maximum $v_t$, some nearby 
HVSs have $v_t \approx$ 400--600~\kms\ and $\mu \approx$ 30~\masyr.

\begin{figure}
\includegraphics[width=6.5in]{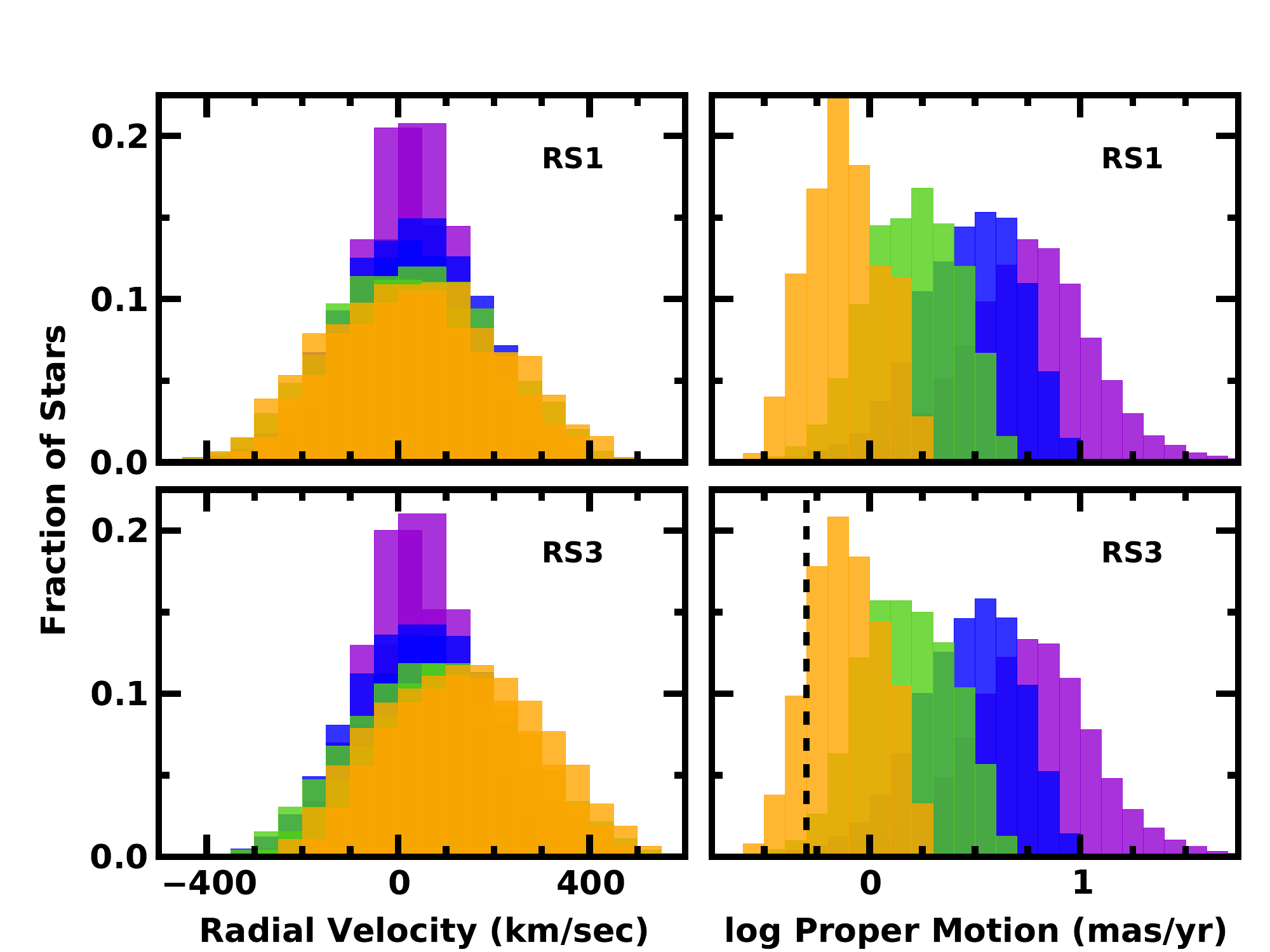}
\vskip 2ex
\caption{\label{fig: vhist-erun}
As in Fig.~\ref{fig: vhist-hvs} for 1 \msun\ runaway stars (upper panels) and 
3 \msun\ runaway stars.  The stars have an exponential distribution of ejection 
velocities (eq. [\ref{eq: pej-sn}]). As with HVSs, the median radial velocity 
(proper motion) grows (falls) with increasing distance. 
}
\end{figure}

Trends in the radial velocity distributions for supernova-induced runaways 
follow those of the HVSs (Fig.~\ref{fig: vhist-erun}; Table~1).  More distant 
runaways have a larger median radial velocity and a larger tail to very large 
radial velocity \citep[see also][]{bromley2009}.  At fixed distance, however, 
the average and median velocities of runaways are much smaller than those of 
HVSs. Typically, runaways are 100--500~\kms\ slower than HVSs, with velocity 
dispersions less than half the dispersions of HVSs. 

Differences between the radial velocity distributions for HVSs and runaways in
the snapshots
reflect the initial distribution of ejection velocities. As summarized in
\S\ref{sec: halo-frac}, more than 20\% of the HVSs ejected with initial velocities
exceeding 600~\kms\ reach the halo. Most HVSs that reach the halo are unbound. 
The fastest runaways receive a boost from Galactic rotation (eq.~[\ref{eq: vrun}]); 
they all lie in the Galactic plane \citep[see also][]{bromley2009}. Many fewer 
runaways reach the halo; nearly all of these have much smaller space velocities 
than HVSs. As a result, runaways in the halo have smaller median radial velocities 
and a larger fraction of bound stars than HVSs.

At similar distances, the proper motion distribution of both types of runaways is 
broader than that of HVSs (Fig.~\ref{fig: vhist-erun}, right panels).  As with HVSs, 
nearby runaways have larger tangential velocities than more distant runaways. 
However, galactic rotation produces a double-peaked distribution of proper motion 
for runaways at fixed distance (Fig.~\ref{fig: mud3}--\ref{fig: mud1}). In an
ensemble of stars with a broad range of distances, the double-peaked character of 
the distribution smears out into a single broad peak. Among stars with a smaller
range of distances, the double-peaked proper motion distribution is prominent.

\begin{figure}
\includegraphics[width=6.5in]{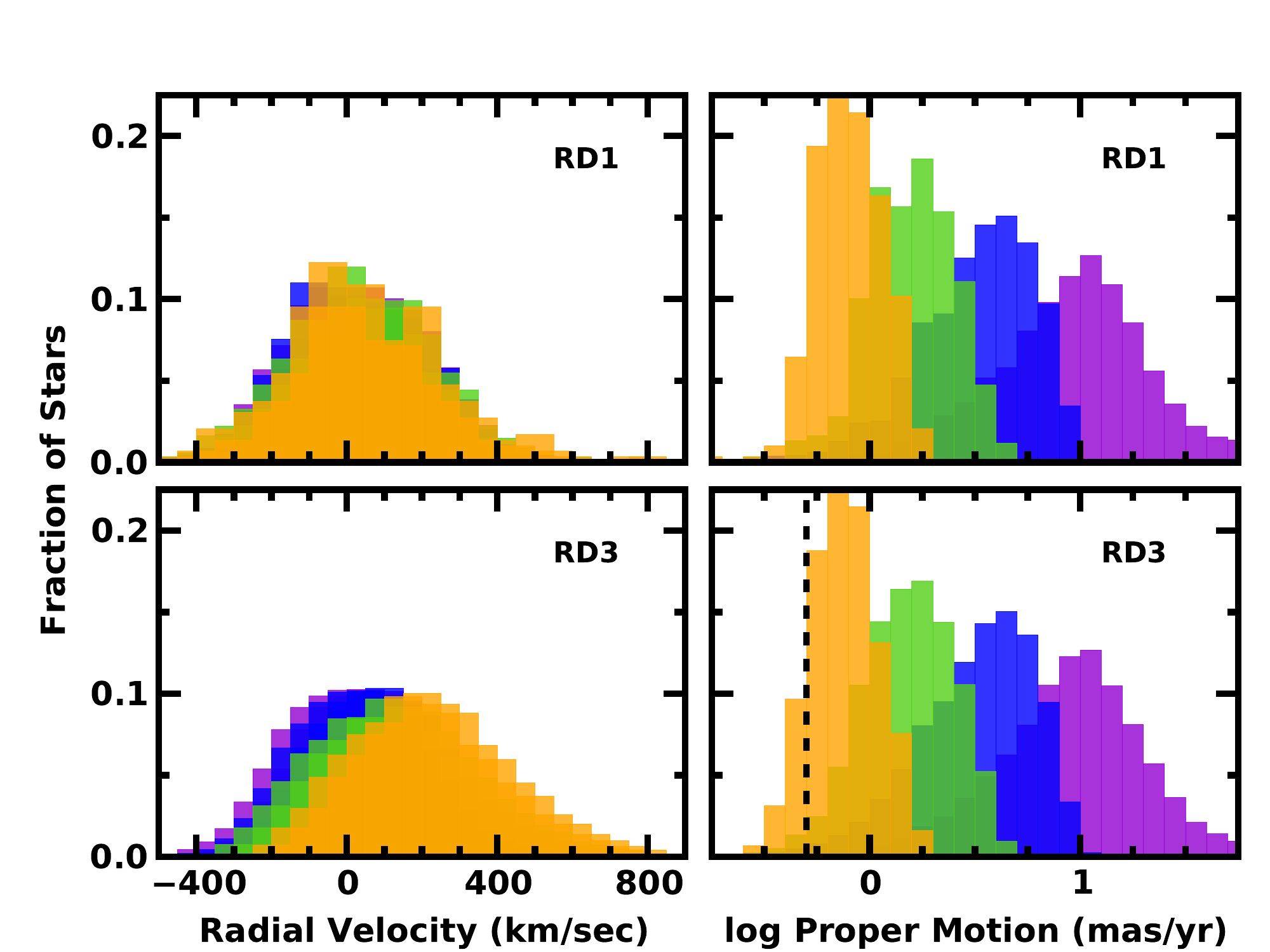}
\vskip 2ex
\caption{\label{fig: vhist-prun}
As in Fig.~\ref{fig: vhist-hvs} for runaways with a power-law distribution of 
ejection velocities (eq. [\ref{eq: pej-dy}]). 
}
\end{figure}

The distributions of $v_r$ and $\mu$ for runaways produced from dynamical ejections 
have the same features as supernova-induced runaways.  The median radial velocity 
grows with distance (Fig.~\ref{fig: vhist-prun}, left panels). Although dynamical 
ejections produce a smaller fraction of high velocity runaways, the largest ejection 
velocities exceed those produced from the supernova mechanism (Table~1). For 
calculations with a minimum ejection velocity of 20~\kms, dynamical ejections yield 
average and median velocities 5\%--10\% smaller than supernova-induced runaways. In 
simulations with a minimum ejection velocity of 50~\kms, the inter-quartile ranges 
and standard deviations for dynamical ejections lie between those of HVSs and runaways 
produced in supernovae.

The larger maximum velocities from dynamical ejections shift the peaks of the proper 
motion distributions to larger values (Fig.~\ref{fig: vhist-prun}, right panels; see
also Table~3). These peaks are also somewhat broader than those for other ejected stars.
As with HVSs and supernova-induced runaways, the width of the proper motion distribution 
narrows with increasing distance. 

For 3~\msun\ HVSs and runaways, GAIA can detect the typical proper motion in the $d$ = 
40--80~kpc bins (dashed lines in Figs.~\ref{fig: vhist-hvs}--\ref{fig: vhist-prun}).
To establish this conclusion, we use the predicted rms errors of roughly 0.16~\masyr\ for 
stars with $g \approx$ 20 \citep{linde2010}. Observed proper motions of 0.50~\masyr\ should
then be detectable at the 3$\sigma$ level. Although solar-type stars with $g \approx$ 20
have $d \approx$ 10~kpc, 3~\msun\ B-type stars with $g \approx$ 20 have $d \approx$ 100~kpc
(see also \S7). Thus, reliable distances and proper motions from GAIA can test these
predicted proper motion distributions.

As we described in \S\ref{sec: res1-mud}, proper motion distributions for HVSs and runaways 
are very sensitive to stellar lifetime. Long-lived unbound stars travel great distances
from the Galaxy; shorter-lived stars evolve off the main sequence before leaving the 
Galaxy. Long-lived bound stars generate fairly symmetric distributions around the GC; 
shorter-lived stars have more asymmetric distributions.

\begin{figure}
\includegraphics[width=6.5in]{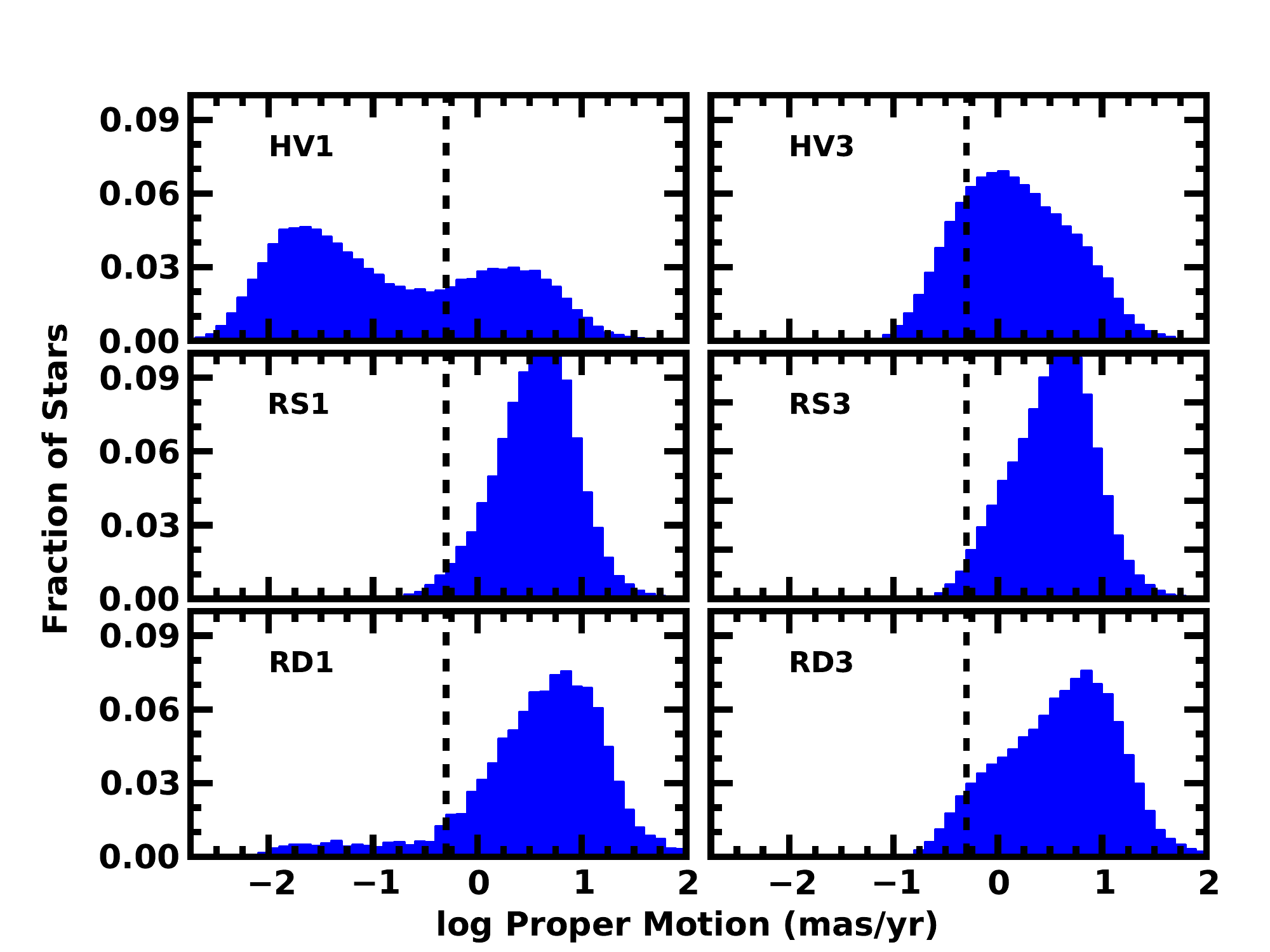}
\vskip 2ex
\caption{\label{fig: muhist}
Histograms of proper motions for HVSs (top panels), runaways produced during a
supernova (middle panels), and runaways from dynamical ejections (lower panels)
with $|b| \ge$ 30~\deg.
Left panels: histograms for 1~\msun\ stars; right panels: histograms for 
3~\msun\ stars; dashed lines indicate the 3$\sigma$ GAIA detection limit 
for stars with $g \lesssim$ 20 \citep{linde2010}.
Among long-lived 1~\msun\ stars, unbound stars at large distances
($d \gtrsim$ 100~kpc) have small proper motions $\mu \approx 10^{-2} - 10^{-1}$~\masyr.
Unbound higher mass stars do not live long enough to reach $d \gtrsim$ 100~kpc; 
they have proper motions $\mu \approx$ 0.1--1~\masyr. Bound stars of any mass
have small distances, $d \lesssim$ 20--30~kpc, and modest proper motions,
$\mu \approx$ 1--10~\masyr.
}
\end{figure}

At high galactic latitude ($|b| \ge$ 30\deg), HVSs provide the most extreme examples 
of this behavior (Fig.~\ref{fig: muhist}, top panels). With lifetimes of 10~Gyr, 
unbound 1~\msun\ stars reach maximum distances of roughly 30~Mpc from the GC. These 
unbound stars produce the prominent peak at $\mu \approx$ 0.01--0.05~\masyr\ in the 
upper left panel of Fig.~\ref{fig: muhist}. Bound 1~\msun\ stars have maximum distances
of roughly 60~kpc; they orbit the GC with periods of 700~Myr or less. Smaller
distances result in much larger proper motions.  These stars comprise the smaller
peak in the upper left histogram at $\mu \approx$ 1--10~\masyr.

Among all HVS ejections, bound stars outnumber unbound stars by roughly 4:1 (\S5). 
Outside the Galactic plane ($|b| \ge$~30\deg), however, unbound stars dominate.  
Thus, the peak of unbound stars at small $\mu$ is larger than the peak of bound 
stars at large $\mu$.

Despite having similar ejection velocities as low mass HVSs, massive unbound 
HVSs do not live long enough to reach large distances from the GC. With typical 
maximum distances of roughly 1~Mpc, the smallest proper motions of unbound 
3~\msun\ HVSs are roughly a factor of 30 larger than those of unbound 
1~\msun\ HVSs. Although there are a few massive HVSs with $\mu \approx$ 
0.03--0.10~\masyr, most have $\mu \gtrsim$ 0.1~\masyr. These stars lie within the 
peak at $\mu \approx$ 1~\masyr\ in the upper right panel of Fig.~\ref{fig: muhist}.

At $|b| \ge$ 30\deg, bound 3~\msun\ HVSs have a much larger range in proper 
motion. Marginally bound stars reach large distances from the GC, $r \approx$
40--60~kpc. Before they turn around and return to the GC, these stars evolve off 
the main sequence. This group has fairly small proper motion $\mu \approx$ 1~\masyr.
Among the much larger group of bound stars that reach small distances, 
$r \lesssim$ 10--20~kpc, some have $|b| \gtrsim$ 30\deg. These have large proper 
motions, $\mu \gtrsim$ 10~\masyr. In between, stars with $d \approx$ 20--40~kpc
fill in the histogram at $\mu \approx$ 1--10~\masyr. 

These general conclusions apply to both types of runaways. Among stars ejected by
dynamical processes, a few have ejection velocities of 600--900~\kms\ and can
reach large distances from the Galaxy. Low mass stars in this group are still
on the main sequence at $d \gtrsim$ 1~Mpc; these stars produce a long tail in
the proper motion distribution at $\mu \approx$ 0.01--0.5~\masyr\ 
(Fig.~\ref{fig: muhist}, lower left panel). More massive stars cannot reach 
these distances while on the main sequence; 3~\msun\ unbound stars produce a small 
shoulder in the distribution at $\mu \approx$ 0.1--0.3~\masyr\ (Fig.~\ref{fig: muhist},
upper right panel). 

For all masses, the dynamical ejection process yields a large population of bound,
relatively nearby stars. Typical proper motions are $\mu \approx$ 1--20~\masyr. 
Nearly all stars in the peaks of both histograms are bound stars.

With much lower maximum ejection velocities, stars ejected during a supernova 
are almost always bound to Galaxy. Among 1--3~\msun\ stars, most are nearby. 
Few have $\mu \lesssim$ 0.5~\masyr. These stars simply produce a single peak 
in the histogram at $\mu \approx$ 3--4~\masyr.

At large $\mu$, the shapes of all of the histograms are fairly similar. For
all of our snapshots, nearby stars with high velocity are rare. Most runaways
with $d \lesssim$ 1~kpc have small velocities and modest proper motions. HVSs
fill a much larger volume of the Galaxy and have much smaller space densities.
Thus, the frequency of ejected stars with $\mu \gtrsim$ 10~\masyr\ falls sharply
and reaches zero at $\mu \approx$ 100~\masyr. 

\subsection{Proper Motion in Galactic Coordinates}
\label{sec: res1-mul}

To conclude our analysis of complete snapshots of ejected stars, we focus on the
variation of proper motion with $l$ and $b$. After separating stars into four
galactic latitude bins equally spaced in $|{\rm sin}~b|$, we divide the $l$--$\mu$ 
plane into a set of bins spaced by 0.01 in log~$\mu$ and 1\deg\ in $l$. As with the 
$d$--$\mu$ and $d$--$v_r$ diagrams in \S\ref{sec: res1-mud}--\ref{sec: res1-vrd}, 
we plot the relative density of stars in each bin as a contour diagram where bright 
red represents the largest density and dark blue represents the smallest density.  
The range in density varies from a factor of 5--10 for 1~\msun\ runaways to a factor
of 50--500 for 1--3~\msun\ HVSs.

\begin{figure}
\includegraphics[width=6.5in]{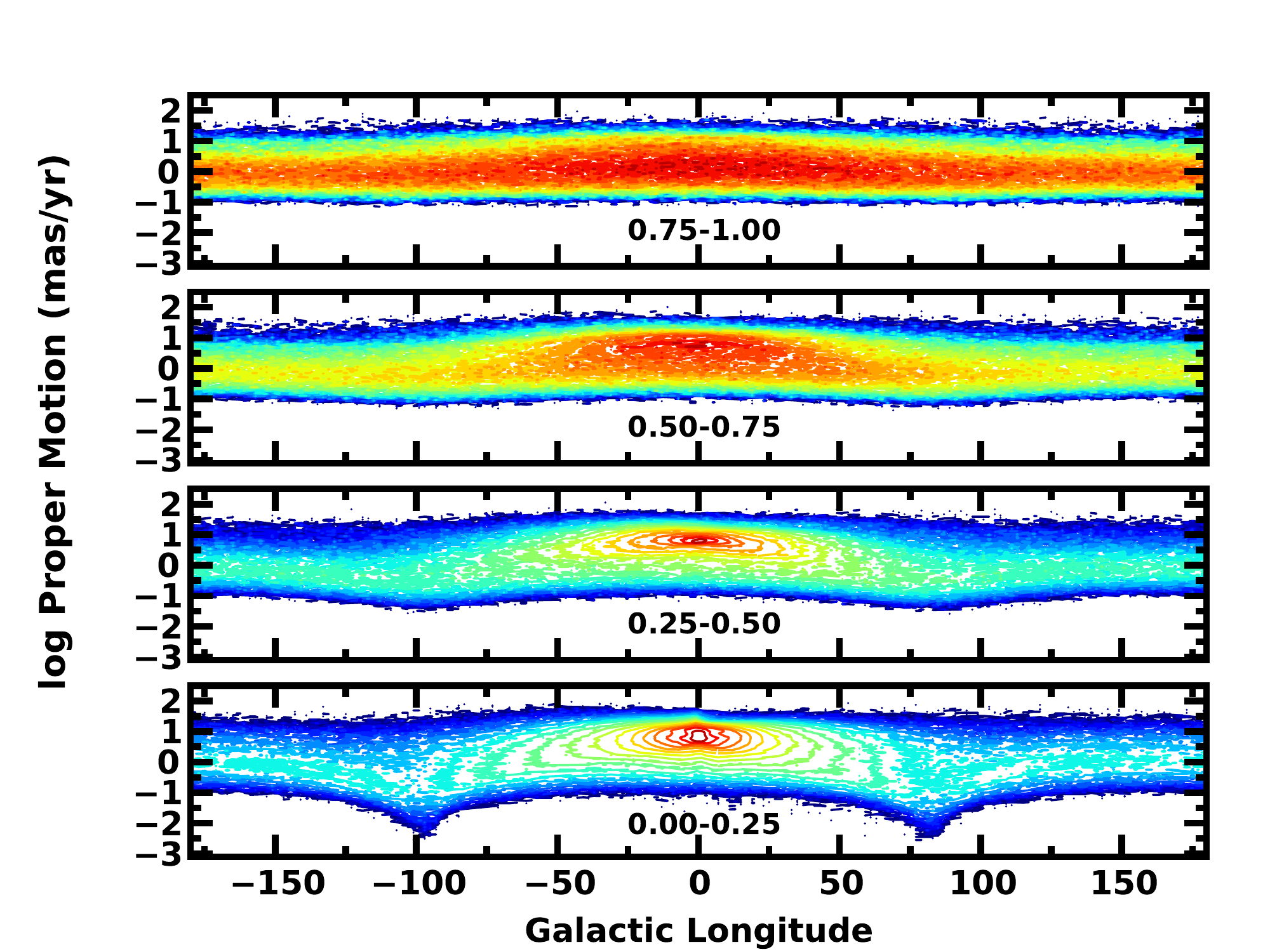}
\vskip 2ex
\caption{\label{fig: mul-hvs}
Predicted density distributions for 3~\msun\ HVSs as a function 
of $l$ and $b$. The range in $| {\rm sin}~b|$ is listed in each 
panel.  The density scale is logarithmic with a minimum of 0.0 
(displayed as dark blue). The maximum of the density (displayed 
as bright red) is 3.6 (lowermost panel), 2.6 (lower middle panel),
1.6 (upper middle panel), and 1.4 (uppermost panel).  At all $l$, 
there is at least a two order of magnitude range in $\mu$.  Towards 
the Galactic anti-center, the median proper motion is small, 
$\mu \approx$ 1~\masyr.  Ejections along the Galactic poles produce 
a dense concentration of HVSs with large $\mu$ towards the GC at all 
$b$. The concentration weakens with increasing $b$. Close to the 
Galactic plane, solar reflex motion produces clear minima in $\mu$ 
for radially outflowing stars at $l \approx$ $-$100\deg\ and $+80$\deg. 
}
\end{figure}

Fig.~\ref{fig: mul-hvs} shows a set of four contour diagrams for 3~\msun\ HVSs.
At all $b$, HVSs have a broad range of proper motion between 0.1~\masyr\ and 
30~\masyr. Outside the GC region, the typical proper motion is $\mu \approx$ 
1~\masyr.  Towards the GC, there is a strong concentration of stars with large 
proper motion, $\mu \approx$ 10--100~\masyr. This concentration is strongest
at low Galactic latitude and weakens considerably at larger $b$.  

In the Galactic plane (Fig.~\ref{fig: mul-hvs}, lowermost panel), the variation 
of $\mu$ with $l$ for HVSs shows a clear signature from the Sun's orbit around 
the GC (see eq.~[\ref{eq: vl-rad}]; compare with Fig.~\ref{fig: v-rad50hel}). 
The plot shows clear minima in $\mu$ at $l$ = $-$100\deg\ and at $l$ = $+$80\deg. 
The Sun's (i) spatial offset 
from the GC and (ii) orbit around the GC produces the lack of mirror symmetry in 
the minima (see also Fig.~\ref{fig: v-rad50hel}). At somewhat larger $b$ (middle 
two panels), the amplitude of the variation is visible but suppressed.  Towards 
the Galactic poles (uppermost panel), the Sun's position and orbital velocity
have no impact on the tangential velocity. Thus, the variation disappears. 

\begin{figure}
\includegraphics[width=6.5in]{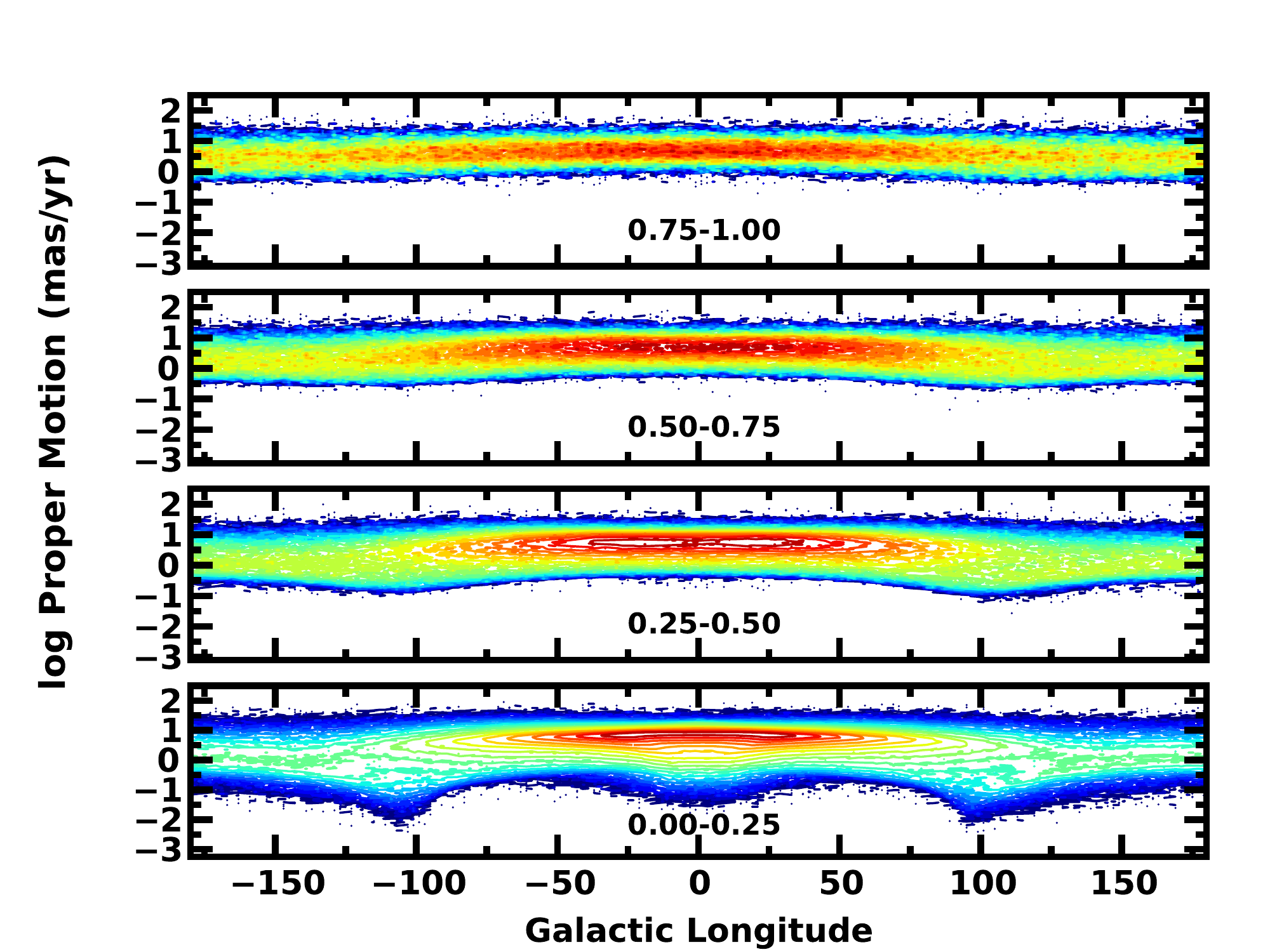}
\vskip 2ex
\caption{\label{fig: mul-erun}
As in Fig.~\ref{fig: mul-hvs} for supernova-induced runaways. The
maximum of the density is 3.6 (lowermost panel), 2.1 (lower middle 
panel), 1.6 (upper middle panel), and 1.2 (uppermost panel).  Stars
ejected from the disk are heavily concentrated to the GC, which 
yields runaways with large $\mu$ towards the GC at all $b$. As with
HVSs, the concentration weakens with increasing $b$.  Close to the 
Galactic plane, solar reflex motion produces clear minima in $\mu$ 
at $l \approx$ $\pm$100~\deg. Stars orbiting inside the solar circle
produce the distinct minimum in $\mu$ towards the GC.
}
\end{figure}

Contour diagrams for supernova-induced runaways display identical features
(Fig.~\ref{fig: mul-erun}). Runaways have somewhat larger proper motions
than HVSs, with a typical $\mu \approx$ 3~\masyr\ and a typical range of 
0.3--30~\masyr.  Although HVSs have larger space velocities, their much
larger distances result in smaller proper motions. 

As with HVSs, the contour diagrams change systematically with Galactic latitude.
In the Galactic plane (Fig.~\ref{fig: mul-erun}, lowermost panel), runaways 
display a large concentration of high proper motion stars towards the GC. 
At larger $b$, this concentration weakens and spreads to a broader range of
Galactic longitude. High velocity runaways outside the solar circle but close 
to the Galactic plane also show clear minima in $\mu$ at $l \approx$ $-$110\deg\ and 
$+$100\deg. As noted for high velocity HVS in Fig.~\ref{fig: mul-hvs}, these 
minima are a clear signature of solar rotation around the GC and the solar 
offset from the GC (see Fig.~\ref{fig: v-rad50hel}). 
This signal gradually diminishes with increasing $b$.

A clear minimum in $\mu$ at $l \approx$ 0 and $|{\rm sin}~b| <$ 0.25 
(Fig.~\ref{fig: mul-erun}, lowermost panel) distinguishes supernova-induced runaways
from HVSs. For stars inside the solar circle, this feature is the signature of stars
rotating in the Galactic disk (see Fig.~\ref{fig: v-orbhel}). Inside the solar
circle, the tangential velocities of stars orbiting the GC lie in an egg-shaped 
locus with $v_{t,max} \approx$ 2 \vsun\ and $v_{t,min} \approx$ 0. The lower edge 
of this egg produces the distinct minimum in $\mu$ at $l \approx$ 0.

Dynamical runaways with a minimum velocity of 20~\kms\ produce distributions of
$\mu(l,b)$ nearly identical to the distributions for supernova-induced runaways 
in Fig~\ref{fig: mul-erun}. Calculations with a minimum velocity of 50~\kms\ yield
dramatically different results (Fig.~\ref{fig: mul-prun}). Although (i) the typical 
range in $\mu(b)$, (ii) the heavy concentration of stars towards the GC, and (iii)
the clear minima in $\mu$ at $l \approx \pm$ 100\deg\ observed for dynamical runaways
are similar to results for supernova-induced runaways, there is 
(i) a clear lack of stars with very small $\mu$ at $l \approx$ 0\deg\ and 
(ii) a broad minimum of stars with very small $\mu$ at $l \approx \pm$150--180\deg. 

The higher minimum ejection velocity in these calculations produces both features.
Stars on unperturbed orbits around the GC produce the distinct minimum in $\mu$ 
at $l \approx$ 0\deg. Setting a high minimum ejection velocity in our calculations
produces ensembles of stars with modest tangential velocities and proper motions at 
$l \approx$ 0\deg, eliminating the pronounced minimum in $\mu$ at $l$ = 0\deg\ in
Fig.~\ref{fig: mul-erun}. This high minimum ejection velocity also tends to place
stars onto orbits with modest eccentricity. Stars originating inside the solar 
circle -- where the stellar density is large -- then spend some time 
outside the solar circle -- where the stellar density is small. This behavior increases
the density of stars with small proper motion in the direction of the Galactic 
anti-center, where the tangential velocity is very small (see Fig.~\ref{fig: v-orbgc}).

\subsection{Summary}
\label{sec: res1-summ} 

Analyzing the complete sample of stars in simulations of HVSs and runaways leads to
several clear results.

\begin{figure}
\includegraphics[width=6.5in]{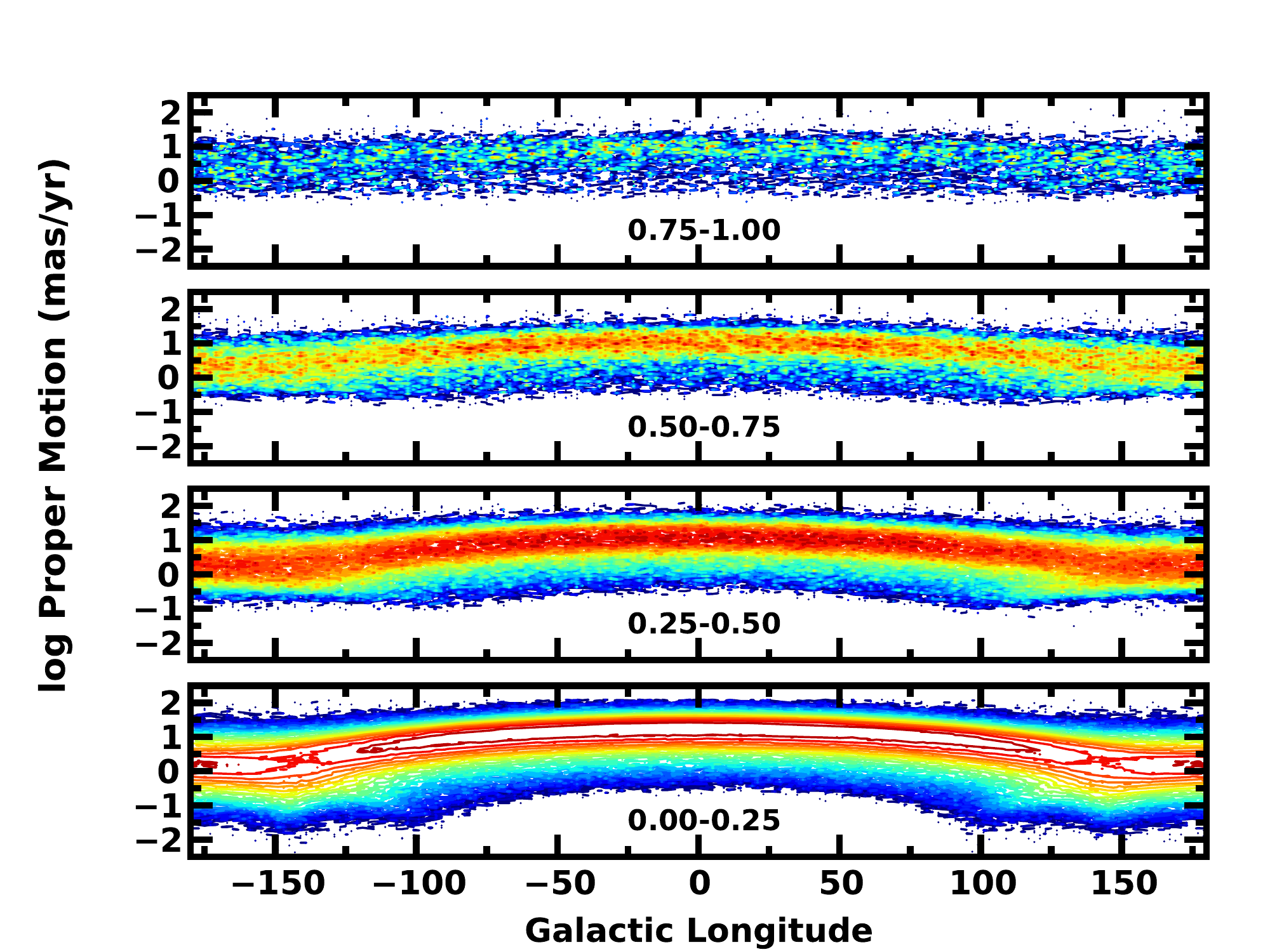}
\vskip 2ex
\caption{\label{fig: mul-prun}
As in Fig.~\ref{fig: mul-hvs} for dynamically-produced runaways with
a minimum ejection velocity of 50~\kms.  The maximum of the density is 
2.5 (lowermost panel), 1.4 (lower middle panel), 1.0 (upper middle panel), 
and 0.8 (uppermost panel).  As with supernova induced runaways, runaways 
with large $\mu$ are heavily concentrated towards the GC at all $b$. The 
concentration weakens with increasing $b$.  Close to the Galactic plane, 
solar reflex motion produces clear minima in $\mu$ at $l \approx$ $\pm$100~\deg.  
With a minimum ejection velocity of 50~\kms, stars orbiting inside the
solar circle at $l \approx$ 0\deg\ never have a distinct minimum in $\mu$
towards the GC. However, small ejection velocities produce a deeper minimum
in $\mu$ towards the Galactic anti-center.
}
\end{figure}

\begin{itemize}

\item Ejections produce a broad range in $v_r$ (0--1000~\kms), $v_t$ (0--600~\kms),
and $\mu$ (0.01--100~\masyr). HVSs have the largest velocities and proper motions
(Figs.~\ref{fig: mud3}--\ref{fig: vrd3}).

\item For distant stars, radial velocities separate unbound stars from bound stars
(Figs.~\ref{fig: vrd1}--\ref{fig: vrd3} and Figs.~\ref{fig: vhist-hvs}--\ref{fig: vhist-prun}).

\item For nearby stars, proper motions distinguish between unbound stars and bound stars
on radial or circular orbits. However, nearby unbound stars are relatively rare compared
to nearby bound stars (Figs.~\ref{fig: mud3}--\ref{fig: mud1} and Fig.~\ref{fig: muhist}).

\item Unbound runaways retain memory of their original orbital velocity around the GC.
At large distances and high Galactic latitudes, the double-peaked proper motion
distribution of runaways distinguishes them from stars on radial orbits
(Figs.~\ref{fig: mud3}--\ref{fig: mud1}).

\item In the Galactic plane, HVSs and runaways have clear, distinctive minima in $\mu(l)$ 
at $l \approx \pm$100\deg\ (Figs.~\ref{fig: mul-hvs}--\ref{fig: mul-prun}). 

\item Concentrations of high proper motion stars towards the GC are a unique signature 
of HVSs ejected from the GC or runaways ejected from the inner galaxy
(Figs.~\ref{fig: mul-hvs}--\ref{fig: mul-prun}).

\item GAIA can detect predicted proper motions of B-type HVSs and runaways with 
$d \lesssim$ 100~kpc (Figs.~\ref{fig: vhist-hvs}--\ref{fig: vhist-prun}).

\end{itemize}

These results clearly demonstrate the ability of radial velocity measurements to
distinguish the highest velocity HVSs and runaways from indigenous halo stars
\citep[see also][and references therein]{brown2005,brown2006,bromley2006,kenyon2008,silva2011,brown2014}.
For HVSs and runaways with $d \gtrsim$ 50--100~kpc, the median radial velocity,
500--600~\kms, and radial velocity dispersion, $\sigma_r \approx$ 200--400~\kms, are
very different from the population of halo stars with median $v_r$ close to zero
and $\sigma_r \approx$ 100--110~\kms. Thus, radial velocity surveys easily separate 
the highest velocity HVSs and runaways from indigenous halo stars.

Among stars with intermediate distances, $d \approx$ 20--40~kpc, proper motions 
can distinguish runaways from indigenous halo stars. For runaway stars with a 
small range of distances, galactic rotation produces a clearly double-peaked
distribution of proper motions. With little or no rotation about the GC 
\citep[e.g.,][]{bond2010}, halo stars should have a broad, single-peaked distribution. 
Because HVSs are ejected on purely radial orbits, their distribution of
proper motions should resemble the halo distribution.

Among nearby halo stars with $d \lesssim$ 10~kpc and $|b| \gtrsim$ 30\deg, 
kinematic data do not offer a simple path for identifying ejected stars.  
The velocity dispersions, $\sigma_r \approx$ 100--175~\kms, of nearby HVSs 
and runaways are comparable to the typical velocity dispersion of halo stars 
\citep[e.g.,][and references therein]{brown2014}.  Thus, it is not possible 
to use $v_r$ to distinguish nearby HVSs and runaways from halo stars.  Although 
the dispersions in $v_t$ for nearby HVSs and runaways are small, the median 
$v_t \approx$ 250~\kms\ for HVSs and dynamically generated runaways is much 
larger than the roughly 150~\kms\ dispersion in $v_t$ expected for indigenous 
halo stars. Because nearby HVSs and runaways are rare, their proper motions 
are fairly similar to the proper motions of nearby halo stars.

As with radial velocities, obvious outliers in proper motion are promising 
candidates for ejected stars. Our simulations yield maximum proper motions 
of 100~\masyr\ for stars with $d \lesssim$ 1--2~kpc. Along any line-of-sight,
however, such high proper motion stars comprise only 0.01--0.1\% of the 
complete population of ejected stars. Thus, high proper motion outliers
should be very rare.

\section{DISTANCE LIMITED SAMPLES OF STARS}
\label{sec: res2}

To construct a clear set of testable predictions from the simulations, we now focus on
distance-limited (magnitude-limited) samples of 1~\msun\ and 3~\msun\ HVSs and runaways. 
Based on the expected sensitivity of GAIA, we establish a magnitude limit.  For 
convenience, we base this limit on the SDSS $g$ magnitude \citep[e.g.,][]{brown2014}.  
Using stellar evolution models, we convert the magnitude limit into a distance limit 
$d_{max}$.  Finally, we draw stars with $d \le d_{max}$ from the complete simulations 
of HVSs and runaways. These catalogs allow us to predict distributions of $v_r$ and 
$\mu$ for comparison with observations.

For 1--3~\msun\ stars with $g \lesssim$ 20, GAIA observations should yield proper 
motions with rms errors of 0.16~\masyr\ \citep[e.g.,][]{linde2010,linde2012}. Brighter
stars have much smaller errors, roughly 0.08~\masyr\ for $g \approx$ 19 and 0.05~\masyr\ for 
$g \approx$ 18. Because GAIA should detect proper motions of roughly 0.15--0.5~\masyr\ for
stars with $g \approx$ 18--20, we set a magnitude limit of $g$ = 20.

To convert this magnitude limit to a distance limit, we derive absolute magnitudes 
$M_g$ from the Padova stellar evolution models.  For solar metallicity 
\citep[Z = 0.019; e.g.,][]{bress2012}, $M_g$(1~\msun) = 5.1 at $t$ = 5~Gyr and 
$M_g$(3~\msun) = 0.17 at $t$ = 172.5~Myr.  These ages are half of the adopted main-sequence 
lifetimes of 10~Gyr and 345~Myr for these stars \citep[e.g.,][]{bress2012}.  These 
magnitudes then yield distance limits of $d_{max}$ = 9.55~kpc (92.5~kpc) for an ensemble of 
middle-aged 1~\msun\ (3~\msun) stars.  We round these limits to 10~kpc for 
1~\msun\ stars and 100~kpc for 3~\msun\ stars. 

In this first exploration of distance-limited samples of HVSs and runaways
from our simulations, we ignore
several aspects of stellar evolution and Galactic structure. We assume stars are 
unreddened, which is reasonable for halo stars at high galactic latitude
\citep[e.g.,][]{schl1998}.  Although metallicity has a small impact ($\sim$ 0.1 mag) 
on $M_g$ for 3~\msun\ stars, these stars brighten by roughly 1 mag during their main
sequence lifetime \citep{bress2012}. For ensembles of 1~\msun\ stars, metallicity and 
stellar evolution produce a 0.5--0.75 mag spread in $M_g$ \citep{bress2012}. Thus, 
$g$ = 20 3~\msun\ (1~\msun) main sequence stars with a range of ages and metallicities 
have a 10\% to 20\% range in distances.  This range is small compared to the obvious 
trends in $v_r$, $v_t$, and $\mu$ with $d$, $l$, and $b$ derived from our simulations. 
Thus, we can safely adopt uniform samples of identical stars with no range in age or 
metallicity.

\begin{figure}
\includegraphics[width=6.5in]{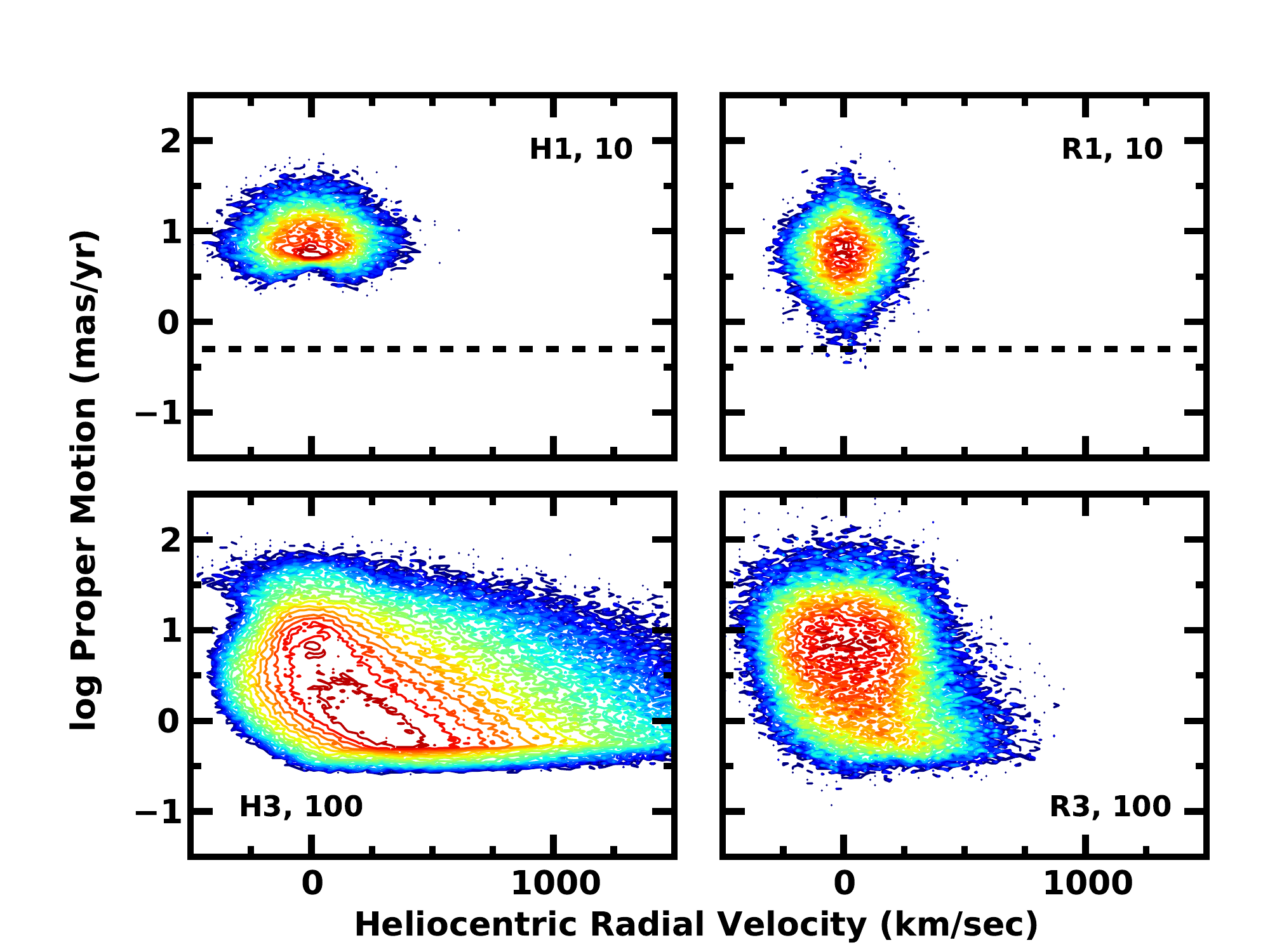}
\vskip 2ex
\caption{\label{fig: muvr}
Predicted density plots in the $v_r$-$\mu$ plane for {\it distance-limited}
samples of HVSs (left panels) and runaways (right panels). The legend
codifies the stellar mass (1~\msun\ or 3~\msun) and the distance
(10~kpc or 100~kpc). Magnitude-limited samples of 1~\msun\ stars (top
panels) have a small predicted range for $v_r$ and $\mu$. All predicted
proper motions lie above the 3$\sigma$ GAIA detection limit for stars u
with $g \lesssim$ 20 \citep[dashed line;][]{linde2010}.
Magnitude-limited samples of 3~\msun\ stars (lower panels) cover a 
broader range in $v_r$ and $\mu$.
}
\end{figure}

Fig.~\ref{fig: muvr} compares predicted density distributions in the $v_r$--$\mu$ 
plane for distance-limited samples of HVSs (left panels) and runaways (right panels).  
Contours for 1~\msun\ (3~\msun) stars are in the upper (lower) panels. For 
simplicity, we show results for 1~\msun\ supernova-induced runaways and for
3~\msun\ dynamically generated runaways. Density distributions for other runaway
models have similar morphology to those shown in this diagram.

The loci for 1~\msun\ HVSs and runaways are fairly similar. Compared to the set
of contours for 3~\msun\ stars in the lower panels, both ensembles have a limited 
extent in radial velocity and proper motion, with $v_r \approx$ $-$250~\kms\ to 
$+$250~\kms\ and $\mu \approx$ 1--30~\masyr. 
The HVS contours have a broader extent in $v_r$ and a much narrower extent in $\mu$ 
than the contours for the runaways. The median proper motion of roughly 6 \masyr\ for 
1~\msun\ HVSs is somewhat larger than the median proper motion of 5~\masyr\ for 
1~\msun\ runaways. 

The ensemble of 3~\msun\ ejected stars fills a much larger portion of $v_r$--$\mu$ 
space. HVSs and runaways have a large concentration of bound stars with median 
$v_r$ close to 0~\kms\ and median proper motion of roughly 3--10~\masyr. Both
populations contain a group of unbound stars with larger $v_r$ and smaller $\mu$. 
Among the runaways, this group produces a modest `tail' in the distribution which 
comprises less than 0.1\% of the entire population. For HVSs, however, the tail 
extends to $v_r \approx$ 1000--1500~\kms\ and contains more than half of the 
ensemble. Nearly all of the unbound HVSs have small proper motion, $\mu \lesssim$
1~\masyr. 

\begin{figure}
\includegraphics[width=6.5in]{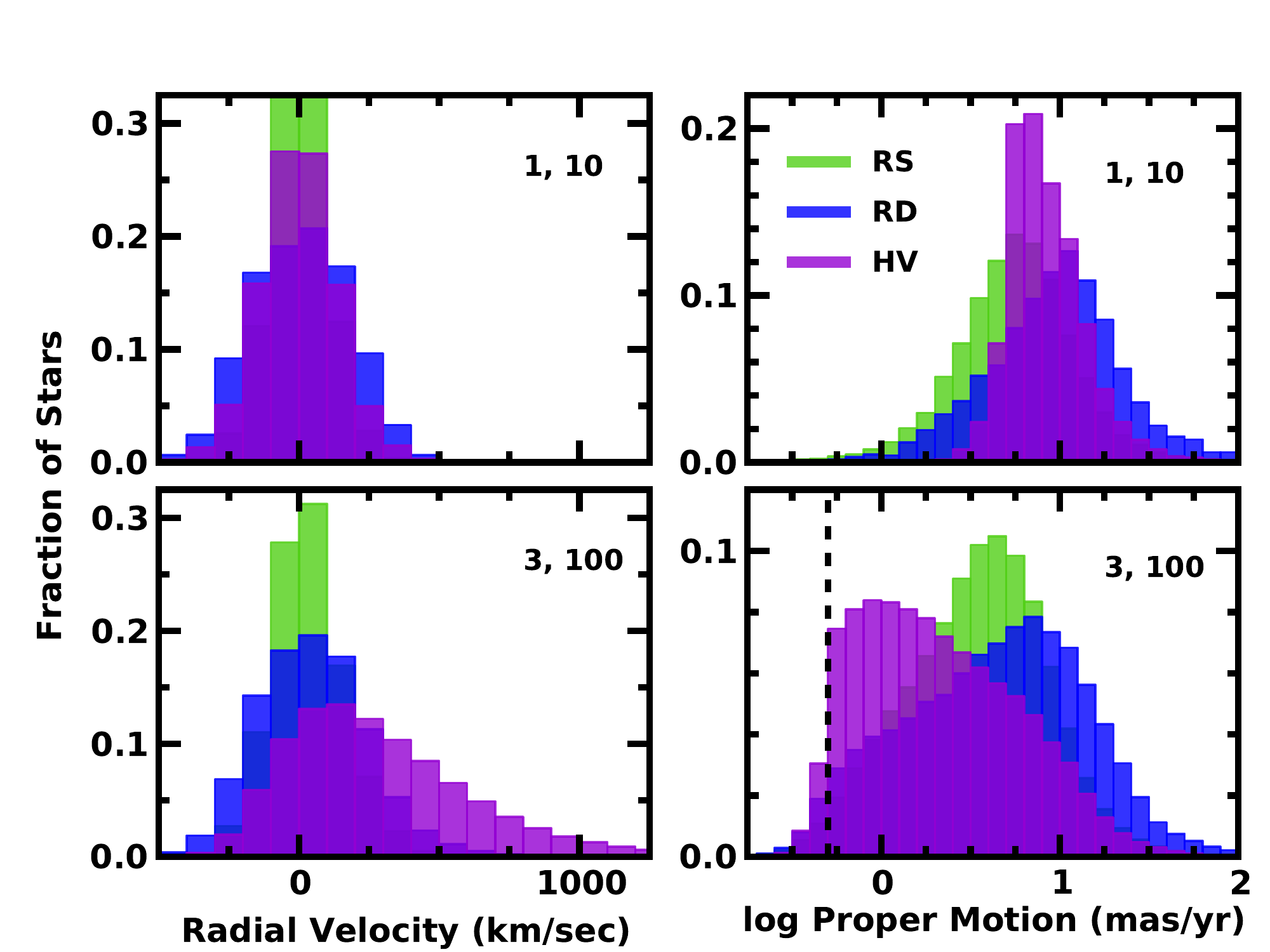}
\vskip 2ex
\caption{\label{fig: vrmuhist}
Predicted histograms of radial velocity (left panels) and proper motion
(right panels) for distance-limited samples of HVSs (violet), 
supernova-induced runaways (green), and dynamically generated runaways 
(blue). The legend codifies the stellar mass (1~\msun\ or 3~\msun) and
the distance (10~kpc or 100~kpc).  Dashed line in the lower right panel
indicates the 3$\sigma$ GAIA detection limit for stars with $g \lesssim$ 20
\citep{linde2010}.
}
\end{figure}

To compare the distributions of proper motion and radial velocity in more detail,
Fig.~\ref{fig: vrmuhist} shows histograms of radial velocity (left panels) and
proper motion (right panels) for 1~\msun\ (upper panels) and 3~\msun\ (lower panels)
ejected stars. The radial velocity histograms for 1~\msun\ stars in the upper left 
panel are amazingly similar with clear peaks at roughly $-$50~\kms\ and modest 
velocity dispersions. In this group, supernova-induced runaways have the smallest 
velocity dispersion, $\sim$ 103~\kms. HVSs have a somewhat smaller velocity 
dispersion, $\sim$ 138~\kms, than the dynamically generated runaways, $\sim$ 
172~\kms\ (Table~1). 

Proper motion distributions for different types of 1~\msun\ ejected stars are also 
very similar (Fig.~\ref{fig: vrmuhist}, upper right panel). The HVSs have a sharp peak 
at 5--10~\masyr; nearly all 1~\msun\ HVSs have $\mu \approx$ 3--30~\masyr. Runaways
produced during a supernova have a broader distribution displaced to smaller $\mu$.
The median $\mu$ is roughly 50\% smaller; the dispersion is roughly 25\% larger (Table~3). 
Dynamically generated runaways have median $\mu$ comparable to the HVSs and a 40\%
larger dispersion. Thus, the distribution of dynamically generated runaways extends
to much larger $\mu$ than the HVSs.

The $v_r$ distributions for 3~\msun\ stars are much easier to distinguish 
(Fig.~\ref{fig: vrmuhist}, lower left panel).
Supernova-induced runaways have a very narrow radial velocity distribution with a
median near zero velocity and a dispersion of roughly 100~\kms\ (see also Table~1).
Dynamically generated runaways also have a median velocity near zero and a larger
dispersion of 170~\kms. In contrast, 3~\msun\ HVSs have a much larger median, 
$\sim$ 200~\kms, and dispersion, $\sim$ 300--350~\kms. More than 1\% of the HVSs have 
radial velocities exceeding 1000~\kms, compared to 0\% for both types of runaways.

The proper motion distributions of ejected 3~\msun\ stars also show a clear 
separation (Fig.~\ref{fig: vrmuhist}, lower right panel).  Most high velocity HVSs
lie at large distances and have small proper motions, producing a clear peak in
the proper motion histogram at roughly 1~\masyr.  The distance limit establishes
the sharp drop in the population at smaller $\mu$.  A few nearby HVSs have maximum 
proper motions of 30--50~\masyr. 

Most 3~\msun\ runaways have much larger proper motions than 3~\msun\ HVSs. 
Stars ejected during a supernova have a fairly symmetric distribution of $\mu$,
with a median at 3--5~\masyr\ and a dispersion of roughly 3--4~\masyr. The
dynamical ejections produce a broader peak with a larger median at roughly 
8--9~\masyr. Compared to the supernova-induced runaways, there are fewer 
dynamically generated runaways with $\mu \approx$ 3--5~\masyr\ and more with
$\mu \approx$ 1--2~\masyr.

\section{OBSERVATIONAL TESTS}
\label{sec: obs}

The distance-limited samples suggest several clear tests of the models based 
on existing samples of ejected stars. Among 3~\msun\ stars, HVS models predict
a much larger group of stars with large $v_r$ compared to either model for
runaways. Because these stars lie at larger distances than runaways, HVSs 
should also have much smaller proper motions.

Among 1~\msun\ stars, the models predict a large overlap in the observed
$v_r$ and $\mu$. Despite this large overlap, it might be possible to isolate 
HVSs and dynamically generated runaways within a large sample of halo stars.
The observed radial velocity dispersion of halo stars 
\citep[100--110~km~s$^{-1}$; e.g.,][]{xue2008,brown2010a,gnedin2010,piffl2014,brown2014} 
is somewhat smaller than the predicted velocity dispersion -- 140--170 ~\kms\ -- of 
HVSs and dynamically generated runaways. Large samples of stars might also provide 
a distinction between the narrow proper motion distribution predicted for HVSs 
from the broader distribution predicted for runaways.

To begin to investigate these possibilities, we consider several sets of ejected 
stars derived from the SDSS.  For 1~\msun\ stars, we examine candidates drawn from 
the G--K dwarfs in SEGUE \citep{yanny2009}. The candidates have a broad range in 
brightness $g \approx$ 14--20. The large surface density of G--K dwarfs results in 
a sample of more than 28,000 stars with moderate resolution spectroscopy and high 
quality radial velocities and atmospheric parameters.  From this ensemble, 
\citet{palla2014} use SDSS proper motion data to select 20 stars with 
$d \approx$ 1--6~kpc, $\mu \approx$ 10--100~\masyr, and $v_r \approx$ 
$-200$~\kms\ to $+130$~\kms. Most of these candidates are metal-poor, with
$[{\rm Fe/H} ] \approx$ $-1.27$ to $-0.06$ and modest enhancements of $\alpha$
nuclei relative to Fe.

To compare with models for 3~\msun\ stars, we focus on the targeted search for HVSs 
from \citet{brown2014}. This set of $\sim$ 20 HVS candidates derives from a nearly 
complete spectroscopic survey of 1126 candidate B-type main sequence stars with 
$g \approx$ 17--20.25 selected from the SDSS \citep{brown2012a}.  Moderate resolution 
MMT spectra yield high quality radial velocities ($v_r \approx$ 250--800~\kms),
atmospheric parameters (log~$g \approx$ 3.75--4.6 and $T_{eff} \approx$ 10000--14000~K),
and distances ($d \approx$ 40--100~kpc).  For a few candidates, high resolution spectra 
confirm their main sequence nature and yield stronger constraints on the atmospheric
parameters \citep{lopez2008,przy2008,brown2012b,brown2013}. 

To compare simulations with a sample of likely runaway stars, we select 16 runaway 
main sequence stars with high Galactic latitude and masses of 2.5--4.0~\msun\ \citep{silva2011}.  
Although derived from several surveys, this set of stars has reliable distances, 
radial velocities, proper motions, and atmospheric parameters. 
With $|v_r| \lesssim$ 250~\kms\ and $d \lesssim$ 5~kpc, these runaways are closer
and have much smaller space velocities than the HVS candidates. The accurate proper 
motions allow us to test whether the lack of proper motion information for the HVS 
candidates limits our ability to compare their radial velocity distributuon with
our simulations.

To test the simulations in more detail, we examine a handful of miscellaneous 
high velocity main sequence stars identified in other surveys.  In order of 
increasing mass, these stars are:
SDSS J013655.91+242546.0, a 2.5~\msun\ A-type sequence star with $d \approx$ 11~kpc 
and $v_r \approx$ 325~\kms\ \citep{till2009};
HIP 60350, a 5~\msun\ B-type main sequence star with $d \approx$ 3 kpc and 
$v_r \approx$ 260~\kms\ \citep{irrg2010};
HE0437-5439, a 9~\msun\ B-type star close to the LMC with $d \approx$ 61~kpc 
and $v_r \approx$ 725~\kms\ \citep{edel2005};
HD 271791, an 11~\msun\ B-type main sequence star with $d \approx$ 21~kpc and 
$v_r \approx$ 440~\kms\ \citep{heber2008}; and 
J091206.52+091621.8, another 11~\msun\ B-type main sequence star with 
$d \approx$ 13~kpc and $v_r \approx$ 620~\kms\ \citep{zheng2014}. Selected in
a variety of ways, this group of stars has reliable distances, space motions,
and atmospheric properties. As with the runaways, the additional proper motion 
information yields a good test of our calculations.

Finally, we consider how well proper motion and radial velocity data isolate
HVSs and runaways from samples of indigenous halo stars. For this study, we
include data from recent surveys of halo stars towards the north Galactic 
pole \citep{kinman2007} and the Galactic anti-center \citep{kinman2012}. 
Selected from several surveys, the RR Lyr stars have measured pulsation
periods and metallicities; the blue horizontal branch (BHB) stars have no
metallicity data. Both groups have proper motion and radial velocity 
measurements with typical uncertainties of a few \masyr\ and 10~\kms. 
With $\sim$ 100--125 confirmed halo stars in each sample, these data enable
a first comparison between observations of halo stars and our calculations.

\subsection{Surveys of solar-type stars}
\label{sec: obs1}

\begin{figure}
\includegraphics[width=6.5in]{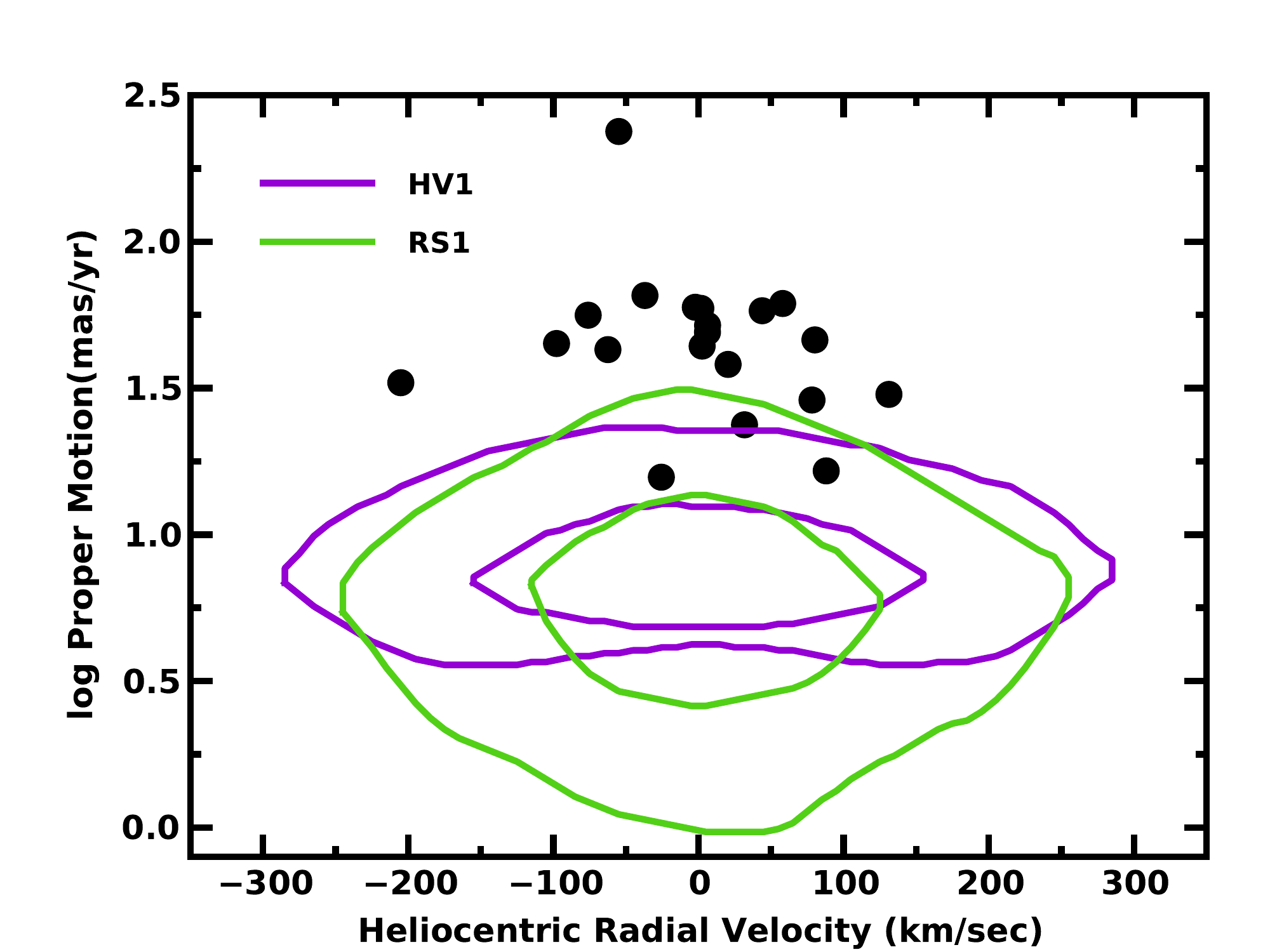}
\vskip 2ex
\caption{\label{fig: con1-obs}
Comparison of observations for 1~\msun\ candidates HVSs from \citet{palla2014}
(filled circles) with predicted density contours in the $v_r$--$\mu$ plane for 
a distance-limited ($d$ = 10~kpc) sample of 1~\msun\ HVSs (`HV1'; violet curves) 
and 1~\msun\ runaways (`RS1'; green curves).  For each model, the inner (outer) 
contours include 50\% (90\%) of stars in the the distance-limited samples.  
Aside from a few stars with $\mu \lesssim$ 30~\masyr, the data fall well above 
predictions for either model.
}
\end{figure}

Fig.~\ref{fig: con1-obs} compares the HVS candidates from \citet{palla2014}
with predictions from the HVS and supernova-induced runaway models. Solid 
lines show contours of constant stellar density which contain 50\% (inner) 
and 90\% (outer) of the stars in the complete samples of simulated stars. 
Contours for dynamically generated runaways are nearly identical to those
for the supernova-induced runaways.  Filled circles indicate the measured 
($d, v_r$) for the HVS candidates.

Based on this comparison, the HVS model and both runaway models are wildly
inconsistent with the observations. Although the observed radial velocities fall
within the bounds predicted for all three models, the observed proper motions 
lie well above model predictions.  None of the survey stars fall within the 
50\% contours of any model; only 3 fall within the 90\% contours. Using the
95\% contours (not shown on the Fig.), the runaway models fare a little better
than the HVS model, with 5 (instead of 3) stars lying within the contours. 
Despite this marginally better success, all models are excluded at better than
3$\sigma$ confidence.

There are two possible origins of the mismatch between the models and the 
observations in Fig.~\ref{fig: con1-obs}. Reducing the proper motions by 
a factor of 5 would place the data within the 50\% contours of all models.
Although \citet{palla2014} derive high reliabilities for their proper motion 
measurements, the ratio of the transverse to radial velocity for this sample 
is much larger than expected for a random selection of high velocity stars.
Thus, there is a reasonable probability that at least some of the large 
proper motions are spurious \citep{palla2014}. Testing this hypothesis with
another epoch of imaging data from the ground or with GAIA is straightforward.

The only alternative to explain the large proper motions of these stars is to
develop another model which produces high velocity stars. Modifying the existing HVS 
or runaway models is unlikely to produce a better match: the geometry of ejections
from the GC or the Galactic disk simply precludes a large population of nearby
1~\msun\ stars with modest radial velocity and very large proper motions. Supernova-induced
runaways from binaries in the halo might allow a better match; however, supernova
rates for halo binaries are probably much smaller than those in the disk.  The
\citet{abadi2009} proposal of ejections from disrupted dwarf galaxies requires a
somewhat massive Milky Way, $\sim$ 1.5--2 $\times 10^{12}$~\msun, but seems 
otherwise plausible. Numerical simulations of the velocity distribution of stars
from disrupted dwarfs are required to test this interesting idea in more detail but
are beyond the scope of this paper.

\subsection{Surveys of B-type stars}
\label{sec: obs2}
 
\begin{figure}
\includegraphics[width=6.5in]{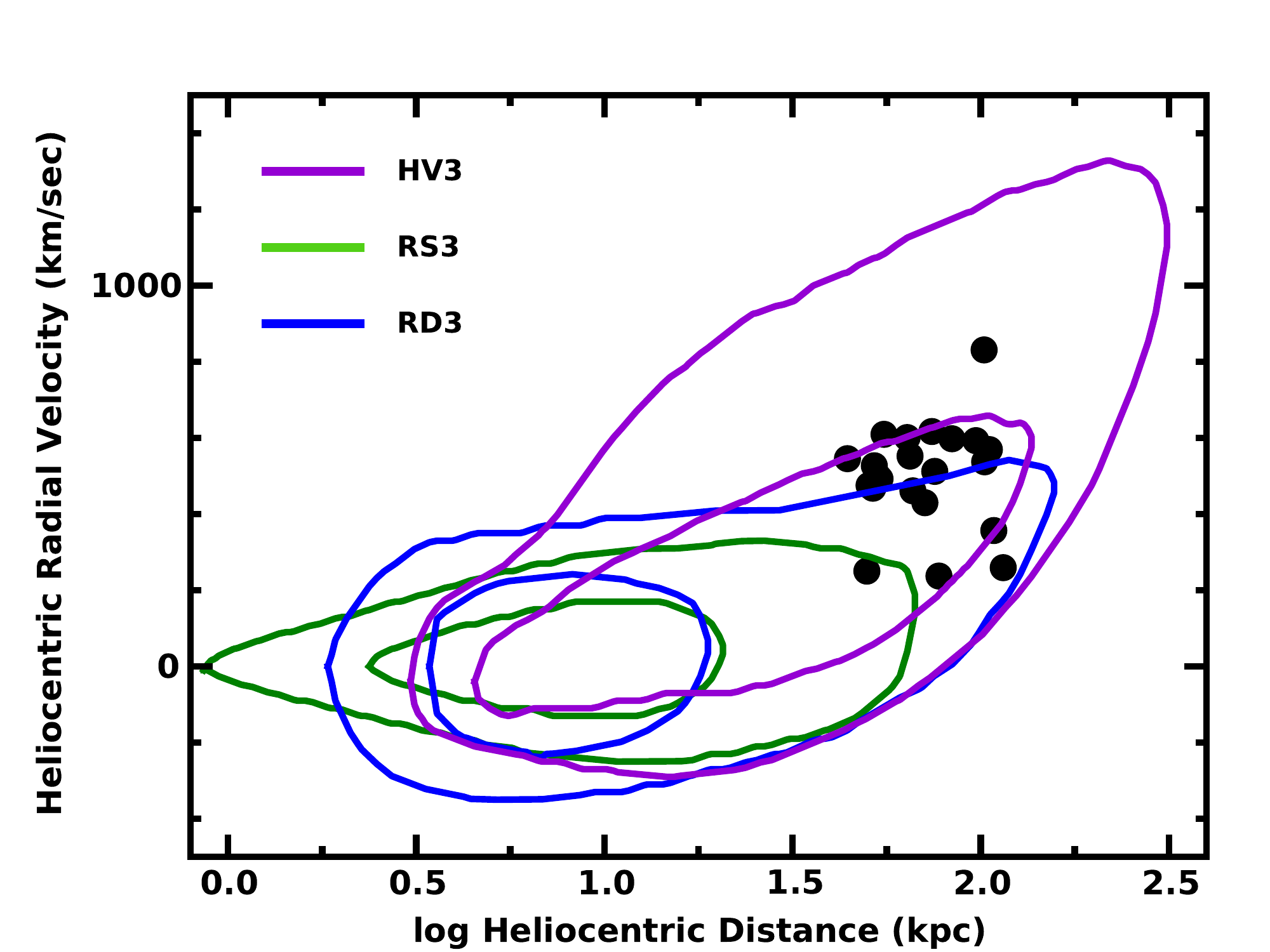}
\vskip 2ex
\caption{\label{fig: con3h-obs}
Comparison of observations for 3~\msun\ candidates HVSs from \citet{brown2014}
(filled circles) with predicted density contours in the $d$--$v_r$ plane for 
a distance-limited ($d$ = 100~kpc) sample of 3~\msun\ HVSs (`HV3'; violet curves), 
3~\msun\ supernova-induced runaways (`RS3'; green curves), and 3~\msun\ dynamically
generated runaways (`RD3'; blue curves).  For each model, the inner (outer) contours 
include 50\% (90\%) of stars in the the distance-limited samples.  Although a few 
observations fall within the 90\% contour for the dynamically generated runaway model, 
most (all) of the data fall within the 50\% (90\%) contours for the HVS model.
}
\end{figure}

Fig.~\ref{fig: con3h-obs} compares the HVS candidates from \citet{brown2014}
with predictions from the HVS and runaway models. As in Fig.~\ref{fig: con1-obs},
solid lines show contours of stellar density which contain 50\% (inner) and 90\% 
(outer) of the stars in each simulation. Filled circles indicate the measured 
($d, v_r$) for the HVS candidates. Unlike the 1~\msun\ targets, these candidates
lie close to or within the 90\% contours for all three models.

This comparison strongly favors the HVS model for the origin of these high
velocity stars. Only one (four) of the candidates lies within the 90\% contour 
for the supernova-induced (dynamically generated) runaway model. Nearly all 
lie beyond the 95\% contours (not shown on the Fig.). In contrast, more than
half of the candidates lie within the 50\% contour for the HVS model; all lie
within the 90\% contour. Taking the results at face value, this comparison
rules out the runaway models at better than 3$\sigma$ confidence. 

To test the models in another way, we attempt to match the observed $d$ and 
$v_r$ of the candidates from the complete ensemble of stars in the HVS and
runaway models. For each HVS candidate with Galactic coordinates ($l, b$), 
we select all model stars within a 20\deg\ $\times$ 20\deg\ window centered 
on the measured coordinates. Among this group, we count the number of model 
stars with distances and radial velocities within 10\% of the measurements
and tabulate the number of `matches' $N_i$ for each candidate.  

\begin{figure}
\includegraphics[width=6.5in]{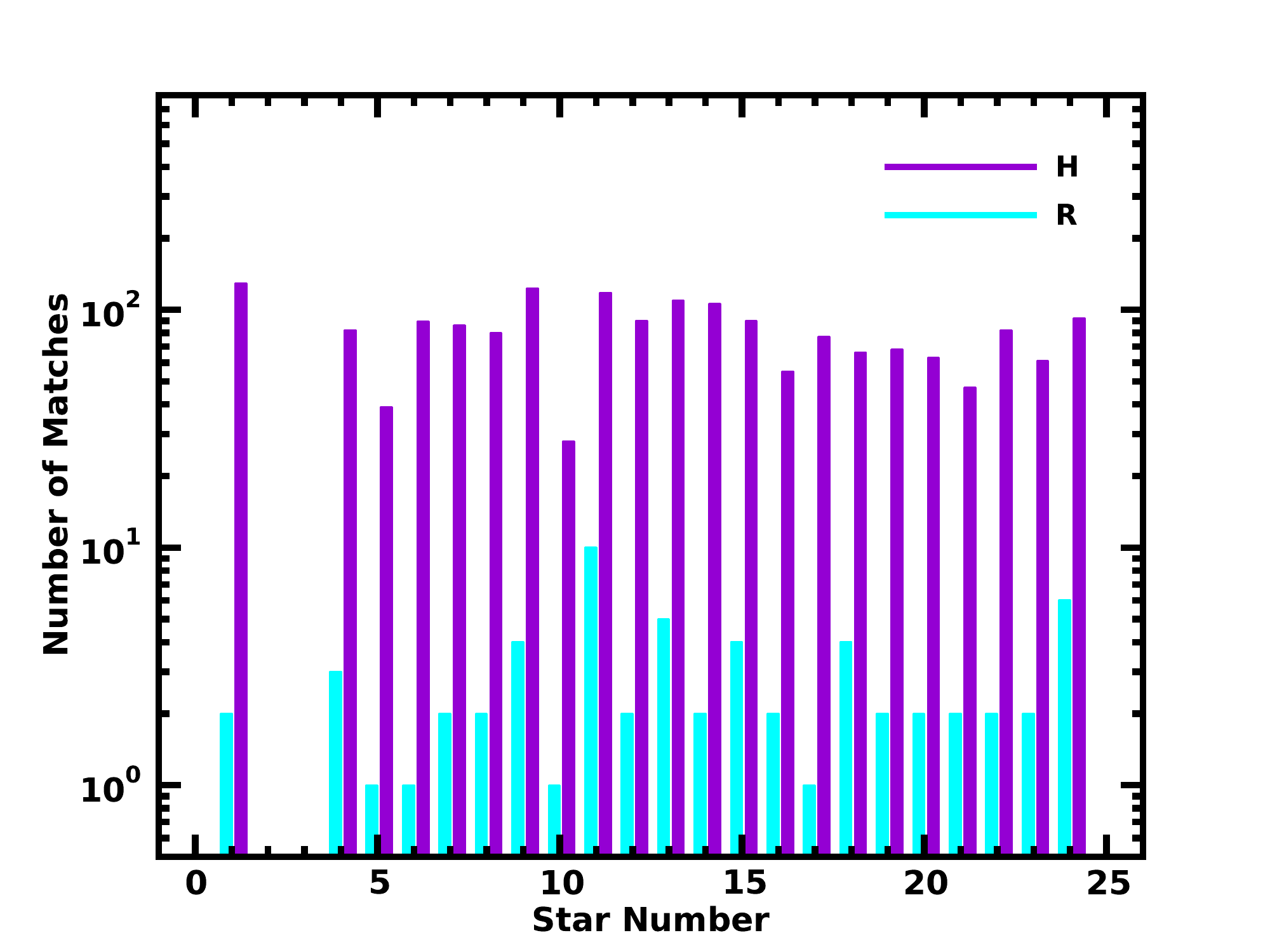}
\vskip 2ex
\caption{\label{fig: match}
Distribution of matches between the 3~\msun\ HVS (`H'; violet) and 
runaway (`R'; cyan) models and observations of HVSs candidates from
\citet{brown2014}. For most known HVSs, the HVS model provides
a better match to the data than the runaway model.
}
\end{figure}

Fig.~\ref{fig: match} summarizes the results of this matching exercise. For
each survey star, the bars show $N_i$ for the HVS (violet) and dynamically 
generated runaway (cyan) models. With only one match among all HVS candidates, 
the supernova-induced runaway model fails the exercise.  The HVS and dynamical 
runaway models fare much better, with $N_i \approx$ 30--100 for the HVS model 
and $N_i \approx$ 1--10 for the runaway model.

Based on this approach, we conclude with strong confidence that the candidate
HVSs from the \citet{brown2014} survey are much more likely to be HVSs ejected 
from the GC than runaway stars ejected from the Galactic disk.  For every survey 
star, the HVS model yields a larger $N_i$ with $N_i({\rm HVS}) \approx$ 
10--100 $N_i({\rm runaway})$.  Factor of two changes in the size of the windows 
for the distance, galactic coordinates, and radial velocity in the matching 
algorithm yield indistinguishable results. Among all HVS candidates, the
supernova-induced runaway model yields 0--1 matches. Analyzing each candidate
separately, an ensemble of HVSs ejected from the GC always produces a factor of 
10--100 more matches than an ensemble of runaways generated from dynamical
interactions.

Modest changes to the probability distribution for the ejection velocity of
runaways cannot change this conclusion. Matching the observed radial velocities
of the 3~\msun\ HVS candidates requires much larger maximum ejection velocities
(e.g., 800~\kms\ instead of our adopted 400~\kms) for supernova-induced ejections 
or a much shallower power-law distribution of ejection velocities (e.g.,
$p_D(v_{ej}) \propto v_{ej}^{-n}$, with $n \lesssim$ 2 instead of our adopted 8/3)
for dynamically generated runaways.  Although doubling the maximum velocity from a 
supernova ejection is possible in rare circumstances \citep[e.g.,][]{port2000}, 
observations suggest a maximum ejection velocity of 400--450~\kms, close to our 
adopted value \citep[e.g.,][]{silva2011}. Observations also appear to preclude 
placing a larger fraction of runaways at the highest velocities \citep{silva2011}. 

\subsection{Runaway B-type Stars}
\label{sec: obs3}

\begin{figure}
\includegraphics[width=6.5in]{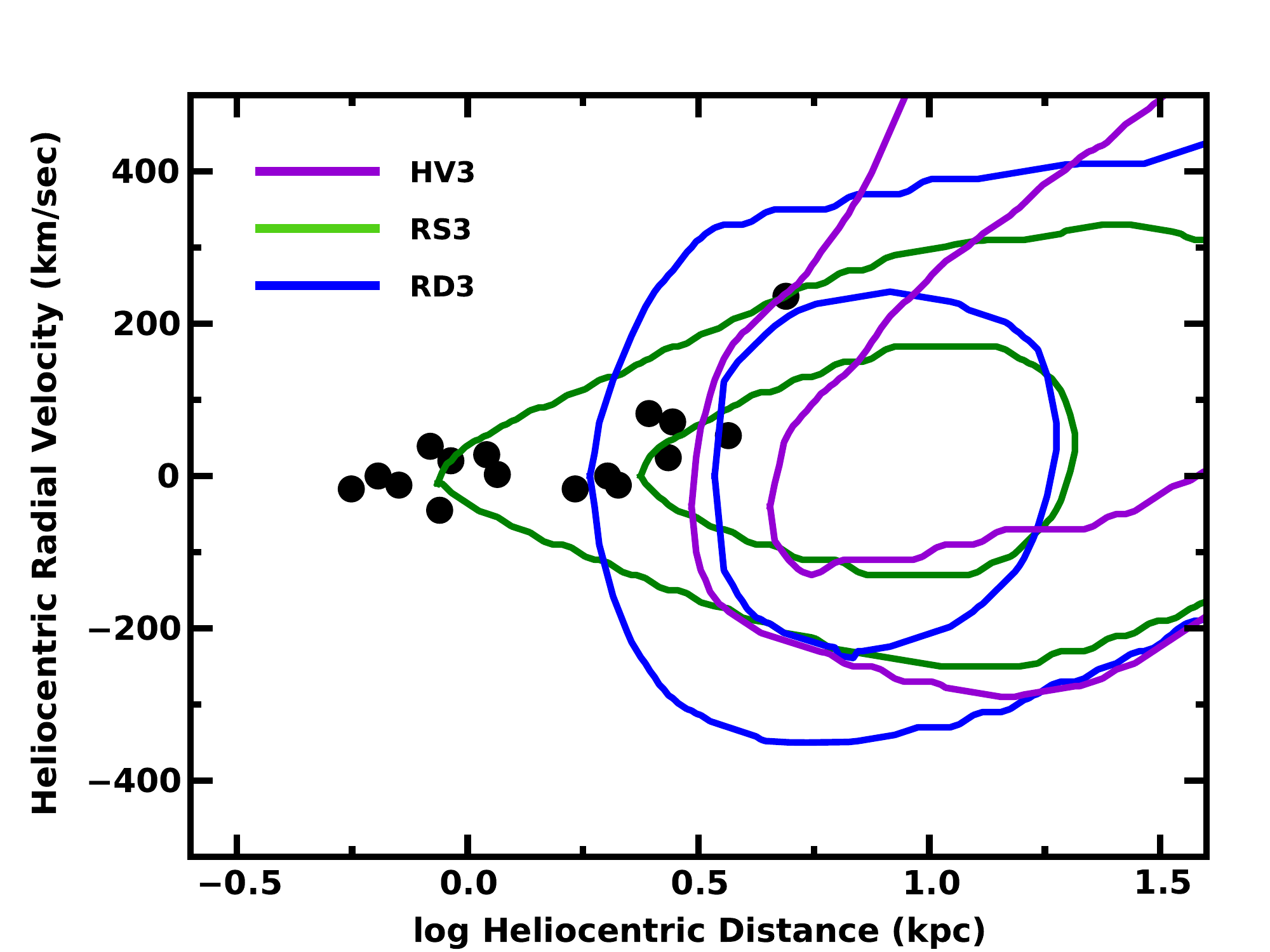}
\vskip 2ex
\caption{\label{fig: con3r-obs}
Comparison of observations (filled circles) for 3~\msun\ runaway stars 
from \citet{silva2011} with predicted density contours in 
the $d$--$v_r$ plane for a distance-limited ($d$ = 100~kpc) sample of 
3~\msun\ HVSs (`HV3'; violet curves), 3~\msun\ supernova-induced runaways 
(`RS3'; green curves), and 3~\msun\ dynamically generated runaways 
(`RD3'; blue curves).  
For each model, the inner (outer) contours include 50\% (90\%) of stars 
in the the distance-limited samples.  Only two observations lie within the 
90\% contours for the HVS model.  Although roughly half of the sample falls 
within the 90\% contour for dynamically generated runaways, nearly all of the
data fall within the 50\%--90\% contours for supernova-induced runaways.
}
\end{figure}

Fig.~\ref{fig: con3r-obs} compares the runaways from \citet{silva2011} with 
predictions from the HVS and runaway models. As in Fig.~\ref{fig: con1-obs},
solid lines show contours of stellar density which contain 50\% (inner) and 90\% 
(outer) of the stars in each simulation. Filled circles indicate the measured 
($d, v_r$) for the runaway candidates; these stars lie close to or within the 
90\% contours for all three models.

This comparison strongly favors the runaway model.  Only two candidates lie 
within the 90\% contour for the HVSs model; none lie within the 50\% contour.  
Many of the candidates are close to or within contours from dynamically generated 
runaways with a minimum ejection velocity of 50~\kms.  More than half of the 
candidates fall within the 90\% contours of the supernova-induced runaway model. 
Contours generated from a set of dynamically generated runaways with a minimum 
ejection velocity of 20~\kms\ show a similarly good agreement with the observations. 

Our matching algorithm also favors the runaway model for these stars. Only one
star -- EC 09452-1403 with $v_r$ = 236~\kms\ -- yields any matches (three) to
the HVS model. The supernova-induced runaway model typically yields 10--100
matches for each target.

To add a little more realism to the matching algorithm, we add the measured
proper motions. Requiring the proper motion in $l$ and $b$ to match within
$\pm$10\% eliminates all matches for the HVS model to data for EC 09452-1403.
Although including proper motion data also reduces the number of matches for 
the supernova-induced runaway model, the typical number of matches is still
3--20 per star. 

These tests demonstrate our ability to explain observations of {\it bona-fide}
runaways from a set of numerical simulations and to discriminate between HVSs 
and runaways. The positions of these stars in the $d-v_r$ diagram and the number 
of matches in $(d, \mu, v_r)$ space clearly favor an identification as runaways 
rather than HVSs.

\subsection{Miscellaneous High Velocity Stars}
\label{sec: obsm}

We now consider five miscellaneous high velocity stars identified as possible 
HVSs or `hyper-runaways.'  Here, we focus on two basic predictions from ejected 
star models: (i) HVSs should have radial orbits from the GC and
(ii) runaways should have a significant non-radial component of motion 
consistent with ejection from the disk. 

From the observed $l$, $b$, and $v_r$, we derive the radial velocity in the GC 
frame and the predicted proper motion for a purely radial orbit (e.g., 
eq.~[\ref{eq: pm}] and eqs.~[\ref{eq: vr-rad}--\ref{eq: vb-rad}]).  For nearby 
HVS candidates, the tangential component of the velocity dominates the space 
motion; for more distant HVSs, the radial component dominates. Among runaways,
the tangential component of the motion is smaller than HVS for nearby stars and 
larger than HVS for more distant stars. If a nearby (distant) candidate is a 
runaway, we expect the analytic proper motion to exceed (fall below) the observed 
proper motion.  In addition to yielding a reasonably simple way to distinguish
between HVSs and runaways, this approach avoids cpu-intensive calculations 
for stars with masses of 2--11~\msun\ which are not included in our suite of
simulations. 

To test this approach, we derive predicted proper motions for HVS candidates
from \citet{brown2014}. Our analysis yields predictions of 0.3--1.2~\masyr.
For each HVS candidate, we then compare this analytic proper motion with the 
average proper motion from the simulated stars selected with the matching 
algorithm outlined in \S\ref{sec: obs3}. The typical difference between the 
analytic and numerical results is small, $\sim$ 0.1--0.2~\masyr, with a
dispersion of 0.5~\masyr. Thus, the analytic approach works reasonably well.

This analysis strongly favors runaway models for SDSS J013655.91+242546.0, 
HIP 60350, and HD 271791.  For SDSS J013655.91+242546.0 and HD 271791, the 
observed magnitude of the proper motion is much larger than predicted from
the simple HVS model.  Coupled with the observed $v_r$, proper motions in 
$l$ and $b$ strongly favor ejection from the disk instead of the GC 
\citep[see also][]{heber2008,till2009}.  For HIP 60350, the HVS model 
predicts much larger proper motion than observed.  Proper motions in $l$ and 
$b$ also favor a disk ejection \citep{irrg2010}.

For HE0437-5439 and J091206.52+091621.8, the small proper motions favor
HVS models. HE0437-5439 has a distance and radial velocity similar to
HVS-1 \citep{brown2005,edel2005}. HST proper motion data suggest origin
in the GC \citep{brown2010b}; ejection from the LMC is also possible
\citep{edel2005,brown2010b}. For J091206.52+091621.8, the observed proper 
motion is comparable to or less than the predicted proper motion of roughly 
3~\masyr\ \citep{zheng2014}.  However, the relatively large errors in the 
proper motion prevent isolating the ejection solely from the GC \citep{zheng2014}.

\subsection{Surveys of Halo Stars}
\label{sec: obs-halo}

\begin{figure}
\includegraphics[width=6.5in]{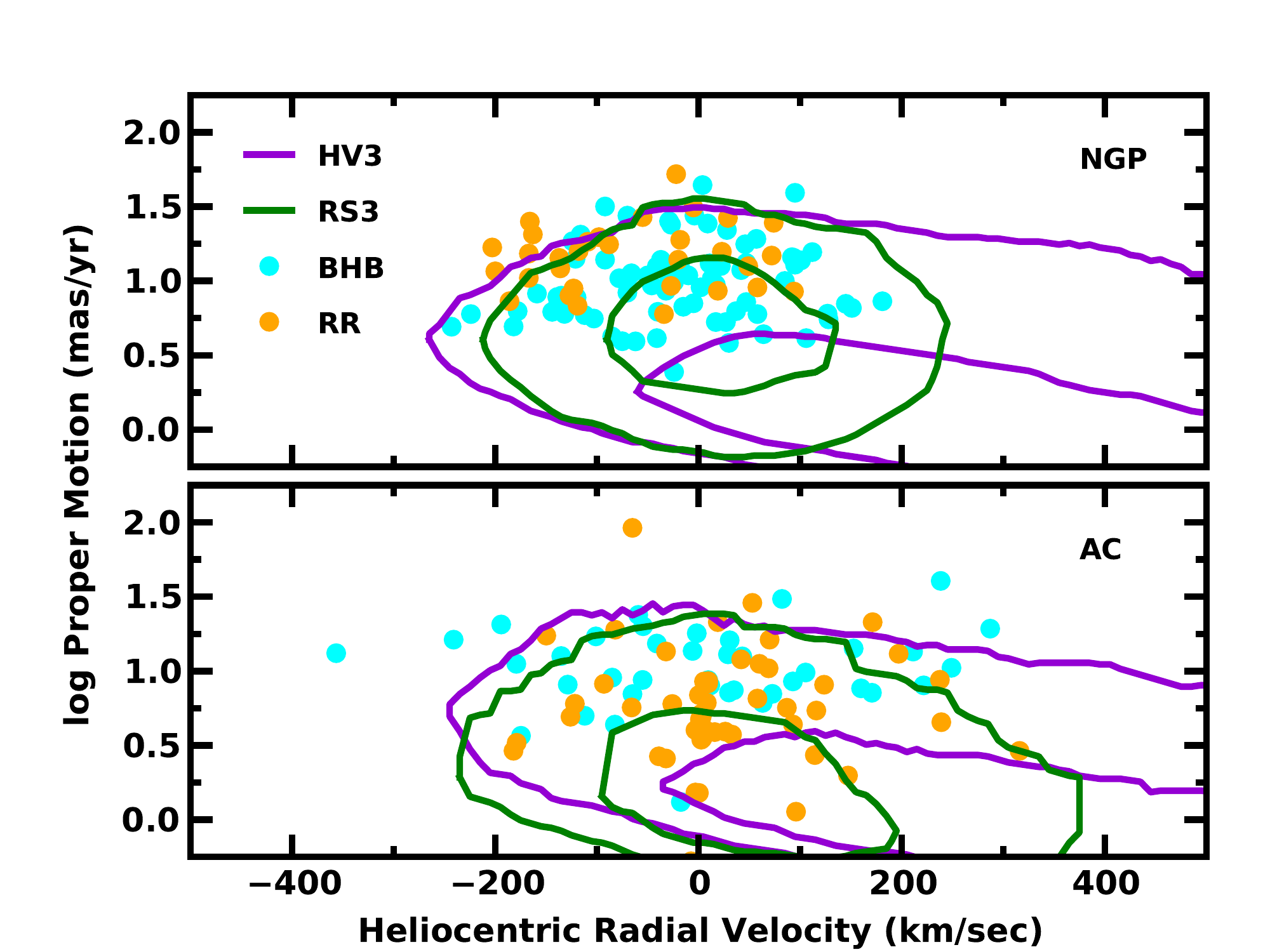}
\vskip 2ex
\caption{\label{fig: con4r-obs}
Comparison of observations (filled circles) for RR Lyr (`RR'; orange points)
and BHB (`BHB'; cyan points) stars from \citet{kinman2007,kinman2012} with
predicted density contours in the $v_r$--$\mu$ plane for a distance-limited 
($d$ = 100~kpc) sample of 3~\msun\ HVSs (`HV3'; violet curves) and 
3~\msun\ supernova-induced runaways (`RS3'; green curves).  Typical errors
are $\pm$1--2~\masyr\ for $\mu$ and $\pm$10~\kms\ for $v_r$. Survey limits 
and measurement errors preclude stars with $\mu \lesssim$ 1--3~\masyr\ in 
each panel.  For each model, the inner (outer) contours include 50\% (95\%) 
of stars in the the distance-limited samples.  
{\it Upper panel:} stars in the direction of the north Galactic pole (NGP).
{\it Lower panel:} stars in the direction of the Galactic anti-center (AC).
The halo star observations uniformly fill the upper half of the contours for 
runaway stars and the low velocity portion of the HVS model contours.
}
\end{figure}

To consider whether observations can distinguish HVSs and runaways from halo 
stars, we examine kinematic measurements from \citet{kinman2007,kinman2012}. 
Fig.~\ref{fig: con4r-obs} compares data for blue horizontal branch stars 
(`BHB'; cyan points) and RR Lyr stars (`RR'; orange points) with model 
contours for distance-limited samples of 3~\msun\ HVSs (`HV3'; violet curves) 
and runaways (`RS3'; green curves).  The contours enclose 50\% (inner) 
and 95\% (outer) of the stars in each simulation. To provide an approximate 
match to the Galactic coordinates for halo stars in the upper (lower) panel, 
we select HVSs and runaways with $b \ge$ 75\deg\ ($l =$ 160\deg\ to $l$ = 
200\deg\ and $b$ = 25\deg\ to 50\deg). 

This comparison confirms our previous conclusion that radial velocity 
measurements directly discriminate between HVSs and indigenous halo stars
\citep[e.g.,][]{brown2006,bromley2006,brown2007,kenyon2008,brown2009,
bromley2009,brown2014}.  For $v_r \approx$ $-$200~\kms\ to $+$300~\kms, 
the distributions of halo stars and HVSs overlap. Beyond $v_r \approx$ 
300~\kms, however, HVSs dominate.  This separation has a simple origin: 
HVSs and halo stars with $v_r \lesssim$ 300~\kms\ are bound to the 
galaxy. HVSs with larger $v_r$ are unbound.

Although the radial velocity distributions for halo stars and runaways 
overlap in the direction of the north Galactic pole (NGP), the highest 
velocity runaways are easily distinguished from halo stars towards the 
Galactic anti-center. At the NGP, nearly all runaways are bound to the 
Galaxy. Thus, they have radial velocities similar to indigenous halo 
stars. Towards the anti-center, the velocity from Galactic rotation
enables a significant population of unbound runaways on outbound 
trajectories. Bound halo stars never reach the large $v_r$ of these
unbound runaways.

Proper motion data alone do not easily discriminate HVSs and runaways
from indigenous halo stars. The maximum proper motion of halo stars
in \citet{kinman2007,kinman2012}, $\sim$ 30--50~\masyr, exceeds the 
proper motions of 99\% ($\sim$ 100\%) of HVSs and runaways towards 
the NGP (anti-center). Towards the anti-center, the number of high 
proper motion outliers, $\mu \gtrsim$ 100~\masyr, is nearly zero. 
Towards the NGP, however, 0.1--1\% of HVSs and runaways have proper 
motions larger than 100~\masyr. Thus, occasional HVSs and runaways 
can be identified as proper motion outliers \citep[e.g.,][]{heber2008,
till2009,irrg2010}.

For large samples of halo stars, accurate distances might provide a 
way to isolate runaways from indigenous stars (Fig.~\ref{fig: mud3}). 
Because runaways share the rotation of stars in the disk, they should
exhibit a double-peaked distribution of proper motion. With little or 
no rotation \citep[e.g.,][]{bond2010}, indigenous halo stars should
have a single-peaked distribution. Although synthesizing the expected 
distribution for indigenous halo stars is beyond our scope, isolating 
the runaways probably requires a significant population of ejected 
stars within the halo.

\subsection{Ejection Rates}
\label{sec: obs-rates}

To complete our comparisons between the models and available data, we now estimate
the relative production rates for HVSs and runaways. Accurate rate estimates
allow us to predict the relative space density of ejected stars as a function
of distance and Galactic latitude \citep[e.g.,][]{brown2009,bromley2009}. 
Because the matching algorithm and the contour maps draw from equal numbers 
of simulated HVSs and runaways, production rates allow us to normalize the
number of matches to the expected space density.

Predictions for HVSs depend on the rate binaries encounter the black hole 
at the GC \citep[e.g.,][]{yu2003}. The time variation of the population of 
binaries within the `loss cone' -- the ensemble of orbits which pass within 
the black hole's tidal radius -- is an important issue in these derivations.
Binary encounters with the black hole empty the loss cone; encounters between
binaries and molecular clouds or other field stars fill the loss cone
\citep[e.g.,][]{yu2003,perets2010,zhang2013,vas2013,mad2014}.  Rates with a 
`full' loss cone are larger than those with an `empty' loss cone; typical
estimates are $10^{-5} - 10^{-3}$~yr$^{-1}$ for binaries of all types
\citep[e.g.,][]{yu2003,bromley2012,zhang2013}. For reasonable initial mass
functions, limiting the binary population to 2.5--3.5~\msun\ B-type stars 
(0.8--1.2~\msun\ G-type stars) implies rates 40 times (10 times) smaller,
2.5--250~$\times 10^{-7}$ yr$^{-1}$ for B-type stars and 
1--100~$\times 10^{-6}$ yr$^{-1}$ for G-type stars.

To infer empirical HVS rates, we focus on the S stars at the GC 
\citep{eck1997,ghez1998} and HVSs in the Galactic halo \citep{brown2014}.
The S stars are luminous B-type stars orbiting the GC which are the
plausible captured partners of HVSs ejected into the outer Galaxy 
\citep{gould2003,oleary2008,perets2009,mad2014}.  The population of S stars 
implies a capture rate of roughly $2 \times 10^{-7}$~yr$^{-1}$ for stars
with masses exceeding 5~\msun\ \citep[see also][]{bromley2012}. Using their
complete spectroscopic sample of B-type stars in the outer halo, \citet{brown2014}
estimate a production rate of $1.5 \times 10^{-6}$~yr$^{-1}$ for unbound HVSs
with masses of 2.5--4~\msun.  For an ensemble of stars selected from a Salpeter 
IMF, these rates agree with theoretical predictions and with each other to 
within a factor of two.

Predictions rates for runaways from supernovae depend on the local star formation 
rate, the mass range adopted for B-type stars, the fraction of stars in binaries, 
and the fraction of binaries with mass ratios and orbital periods capable of ejecting 
a star with a velocity exceeding 10--20~\kms\ \citep[e.g.,][]{brown2009,bromley2009}. 
For a star formation rate of 0.5~\msunyr\ \citep{lada2003} and for the observed 
properties of binaries composed of B-type stars with masses of 
2.5--4~\msun\ \citep{kobul2007}, supernovae produce runaways with minimum ejection 
velocities of 10~\kms\ at a rate of roughly $3 \times 10^{-6}$~yr$^{-1}$ 
\citep[see also][]{brown2009,bromley2009}. 

Predicting ejection rates for the dynamical runaway mechanism is more challenging. 
From numerical simulations of dense clusters, \citet{perets2012} estimate an ejection 
rate for hyper-runaway B-type stars with $v_{esc} \gtrsim$ 450~\kms\ of 
$1 - 2 \times 10^{-8}$~yr$^{-1}$. For a power-law probability of the ejection velocity 
(eq.~[\ref{eq: pej-dy}]), the total ejection rate for B-type runaway stars with 
$v_{esc} \gtrsim$ 10~\kms, is roughly $1 - 2 \times 10^{-5}$~yr$^{-1}$.  This rate 
suggests that dynamically generated runaways are somewhat more common
than supernova-induced runaways.

Empirical rates for runaways generally agree with these estimates. To derive these rates, 
we estimate the formation rate of 2.5--3.5~\msun\ stars from the observed star formation
rates and a Salpeter initial mass function.  For the \citet{lada2003} star formation rate 
of 0.5~\msunyr, one star with a mass of 2.5--3.5~\msun\ is born every 300--350 years. 
To infer the fraction of runaway stars in this group, we rely on the observed 
frequencies of less than 1\% for A-type stars \citep{stet1981,bromley2009}, 5\% for 
B-type stars \citep{gies1986}, and more than 20\% for O-type stars \citep{tetz2011}. 
Adopting a 1--2\% runaway frequency among 2.5--3.5~\msun\ stars yields a production 
rate of $3 - 6 \times 10^{-5}$~yr$^{-1}$. This rate is nearly identical to our purely
theoretical rate estimate.

Altogether, these estimates suggest our matching algorithm selects stars from 
ensembles with similar formation rates. If we assume that bound HVSs have 
a comparable frequency to unbound HVSs in the \citet{brown2014} sample, the 
combined HVS production rate of $3 \times 10^{-6}$~yr$^{-1}$ is identical
to the production rate of supernova-induced runaways. For the dynamically
generated runaways, we correct the complete ensemble for the fraction of
runaways with $v_{ej} \gtrsim$ 50~\kms. This correction yields an expected
rate of $3 - 6 \times 10^{-6}$~yr$^{-1}$, very close to the rates for HVSs
and supernova-induced runaways.

Although these estimates yield roughly similar production rates for HVSs 
and both types of runaways, we expect unbound runaways to be much less 
frequent than unbound HVSs. More than 25\% of HVSs ejected from the GC to 
distances $\ge$ 10~kpc are unbound (\S5). In contrast, $\le$ 1\% of runaways 
have ejection velocities larger than the local escape velocity.  Thus,
unbound HVSs should greatly outnumber unbound runaways in the halo. 

Overall, observations confirm the expectation that the number of unbound 
runaways is smaller then the number of unbound HVSs 
\citep[see also][]{brown2009,bromley2009,perets2012}.  For every unbound 
runaway, theory predicts 10--30 HVSs. Among known unbound stars, the vast
majority are HVSs \citep[\S\ref{sec: obs3}; see also][]{brown2009}.

\subsection{Summary}
\label{sec: obs-summ}

Observations of HVS candidates and known runaways yield good tests of the
numerical simulations. The comparisons in the preceding subsections lead
to several clear conclusions.

\begin{itemize}

\item Runaway models provide an excellent match to observations of known 
runaway stars with modest radial velocities, 
$v_r \lesssim$ 250~\kms\ \citep{silva2011}.  

\item The HVS models match observations of HVS candidates from \citet{brown2014}.

\item HVS and runaway star models fail to match observations of HVS candidates
from SEGUE \citep{palla2014}. If the proper motions are correct, some other 
model for ejected stars is required to match the observations.

\item Among a few miscellaneous high velocity stars, at least three are 
runaways \citep{heber2008,till2009,irrg2010}. Two others are more likely
HVSs \citep{edel2005,brown2010b,zheng2014}.

\item Cleanly isolating unbound ejected stars from indigenous halo stars 
requires radial velocities. If the halo contains a large population of 
runaways ejected from the disk, these stars can be identified as high 
proper motion outliers or by their rotational motion about the GC.

\item Observed and theoretical formation rates for HVSs and runaways suggest
that most unbound stars in the halo are HVSs.

\end{itemize}

\section{DISCUSSION AND SUMMARY}
\label{sec: disc-summ}

We have explored analytic treatments and have developed numerical calculations of 
HVSs and runaway stars moving through the Galaxy. The simulations use a realistic 
Galactic potential which matches observations in the GC, the bulge, the disk, and 
the halo.  Algorithms for the velocities of ejected stars are based on detailed 
analytic and numerical calculations of 
(i) binaries interacting with the black hole in the GC,
(ii) binaries where one component undergoes a supernova explosion, and
(iii) single and multiple stars interacting in a massive star cluster. 
Realistic main sequence lifetimes allow snapshots of the positions and space motions 
for ejected stars as a function of time.

The following theoretical results serve as a guide for interpreting data from 
large-scale surveys with GAIA and ground-based telescopes.

\begin{itemize}

\item Ejected stars have a broad range in 
radial velocity ($v_r \approx$ 0--1000~\kms) and
proper motion ($\mu \approx$ 0.01--100~\masyr). 

\item At all distances, HVSs have larger space velocities than runaways. 
For $d \approx$ 50--150~kpc, unbound HVSs dominate the population. Nearby
($d \lesssim$ 10~kpc), unbound stars are rare; bound HVSs and runaways 
are common.

\item For nearby stars with $d \lesssim$ 10~kpc, proper motions cannot 
isolate high velocity ejected stars from horizontal branch stars in the halo. 

\item At larger distances, radial velocities excel at distinguishing
ejected stars from indigenous halo stars.

\item For runaway stars ejected into the halo at $d \approx$ 20--50~kpc, 
a double-peaked proper motion distribution results from orbital motion 
around the GC. This distribution cleanly separates runaways from either 
HVSs or halo stars on radial orbits.

\item Concentrations of high proper motion stars near the GC are a 
unique signature of HVSs ejected from the GC or runaways ejected from 
the inner galaxy.

\end{itemize}

These general conclusions are fairly independent of the modeling approach. The main 
trends for bound and unbound stars depend on the point of origin (GC for HVSs or disk
for runaways) and the total mass of the Galaxy within roughly 10~kpc 
\citep[e.g.,][]{kenyon2008}. Thus, modest changes to the mass of the Galaxy, the 
mass in the bulge, disk, or halo, or to the velocity distributions of ejected stars
cannot modify our main conclusions.  Adopting a triaxial potential for the bulge 
or a more realistic disk-like potential for the GC can modify the median velocities,
the velocity dispersions, and the relative densities of HVSs with Galactic latitude 
and longitude \citep[e.g.,][]{gnedin2005,yu2007,lu2010,zhang2010,rossi2013}.
However, tests with realistic anisotropic potentials show that the general trends 
do not depend on the form of the Galactic potential.  Similarly, modest changes in
the adopted velocity distributions of ejected stars yield modest differences in
the shape of the predicted distributions of $v_r$ and $\mu$ with distance without
changing the nature of systematic variations with distance and Galactic coordinates.

Comparisons between these predictions and existing observations of ejected stars 
are encouraging. Several large surveys of HVSs and runaways demonstrate several
unambiguous conclusions.

\begin{itemize}

\item Simulations of supernova-induced and dynamically generated runaways 
match observations of known runaway stars with modest radial velocities, 
$v_r \lesssim$ 250~\kms\ \citep{silva2011}.  HVS models fail to match these data.

\item HVS models explain observations of HVS candidates from \citet{brown2014}.
Runaway star models cannot explain the data.

\item HVS and runaway star models fail to account for observations of HVS candidates
from SEGUE \citep{palla2014}. If the proper motions of the SEGUE candidates are 
correct, another model for ejected stars is required to match the observations.

\item Among a few miscellaneous high velocity stars, at least three are 
runaways \citep{heber2008,till2009,irrg2010}. Two others are more likely
HVSs \citep{brown2010b,zheng2014}.

\item Observed and theoretical formation rates for HVSs and runaways suggest
that most unbound stars in the halo are HVSs.

\end{itemize}

GAIA data will provide clear tests of the theoretical predictions and observational
comparisons outlined above. For solar-type stars with $g \lesssim$ 20, GAIA data will 
yield robust samples of high velocity stars with $\mu \gtrsim$ 0.5~\masyr\ and 
$d \lesssim$ 10~kpc. Within this group, quantifying the (probably very low) fraction 
of unbound stars can place useful limits on the velocity distributions of ejected stars. 
More likely, identifying outliers in the distributions of proper motion and radial 
velocity will yield a population of bound HVSs and runaways. Aside from the properties 
of ejected stars, comparisons with theoretical models can yield information on the 
potential of the bulge and inner disk.

Accurate GAIA proper motions ($\delta \mu \lesssim$ 0.16~\masyr) for stars with 
$g \lesssim$ 20 \citep{linde2010,linde2012}, high quality radial velocities from 
large ground-based telescopes \citep[e.g.,][]{brown2014}, and good spectroscopic 
parallaxes will probe the properties of unbound B-type stars 
out to $d \approx$ 100~kpc. These data will yield improved constraints on 
(i) the production rates for HVSs and runaways and 
(ii) the relative populations of bound and unbound stars.

Although GAIA can identify stars with masses of roughly 10~\msun\ to much larger 
distances, the shorter main sequence lifetimes of these stars preclude large 
populations beyond 20--30 kpc \citep[see also][]{bromley2006,bromley2009}. If
ejections of very massive HVSs or runaways are more frequent than those of 
3~\msun\ stars, GAIA could discover a few massive, very high velocity ejected 
stars beyond 100~kpc.

As dynamical models for ejected stars improve, GAIA observations of the complete 
sample of 1--10~\msun\ HVSs and runaways will enable new measurements of 
anisotropies in the shape of the Galactic potential \citep{gnedin2005,yu2007}. 
HVSs and runaways near the GC constrain the shape of the bulge.  More distant 
stars probe the shape of the halo. Together, HVSs and runaways in the bulge 
and halo may yield new insights into the distribution of dark matter throughout
the Galaxy.

\vskip 6ex
We acknowledge generous allotments of computer time on the NASA `discover' cluster.
Clear and respectful comments from an anonymous referee improved our
presentation.

\bibliography{ms.bbl}

\begin{deluxetable}{lccccccc}
\tablecolumns{8}
\tablecaption{Radial Velocity Statistics for Runaway and Hypervelocity Stars}
\tabletypesize{\scriptsize}
\tablehead{
{Model} &
{~~d (kpc)~~} &
{~~~~$f_d$ (\%)~~~~} &
{$v_{med}$ (km~s$^{-1}$) } &
{$v_{q1}$ (km~s$^{-1}$) } &
{$v_{q3}$ (km~s$^{-1}$) } &
{$v_{avg}$ (km~s$^{-1}$) } &
{$\sigma_r$ (km~s$^{-1}$) }
}
\startdata
HV1 & ~~0--10  & 2.96 & $-$0.3~ & ~$-$89.3~ & ~89.7 & ~~0.9 & 138.3 \\
HV1 & ~10--20  & 2.54 &  ~~5.1~ & $-$113.6~ & 126.0 & ~~9.2 & 188.7 \\
HV1 & ~20--40  & 1.73 &  ~15.9~ & $-$128.5~ & 158.7 & ~25.2 & 233.9 \\
HV1 & ~40--80  & 1.12 &  ~39.5~ & $-$110.4~ & 197.5 & ~67.9 & 275.3 \\
HV1 &  80--160 & 0.86 &  ~90.8~ & ~$-$71.5~ & 292.8 & 152.7 & 337.3 \\
HV3 & ~~0--10  & 2.80 & ~31.6 & $-$70.7~ &  146.6 &  ~55.8 & 190.4 \\
HV3 & ~10--20  & 3.10 & 167.3 &   ~~2.8~ &  362.5 &  208.5 & 297.0 \\
HV3 & ~20--40  & 3.74 & 268.8 &   ~83.2~ &  498.3 &  321.5 & 339.7 \\
HV3 & ~40--80  & 3.77 & 395.1 &   209.1~ &  635.9 &  457.3 & 355.6 \\
HV3 &  80--160 & 2.77 & 605.9 &   414.5~ &  855.6 &  672.1 & 367.9 \\
\\
RS1 & ~~0--10  & 2.28 & ~~1.9 & ~$-$60.7~ &  ~65.1 &  ~~2.0 & 102.9 \\
RS1 & ~10--20  & 1.07 & ~~4.9 & ~$-$88.1~ &  ~95.8 &  ~~3.2 & 132.9 \\
RS1 & ~20--40  & 0.48 & ~~2.4 & $-$107.3~ &  113.8 &  ~~3.8 & 158.6 \\
RS1 & ~40--80  & 0.15 & ~~5.8 & $-$114.7~ &  128.6 &  ~~8.5 & 169.1 \\
RS1 & ~80--160 & 0.03 & ~~1.1 & $-$113.7~ &  125.3 &  ~10.6 & 170.4 \\
RS3 & ~~0--10  & 2.30 & ~10.1 & $-$51.4~ &  ~75.6 &  ~12.9 & 103.6 \\
RS3 & ~10--20  & 1.11 & ~27.7 & $-$64.4~ &  121.2 &  ~29.1 & 135.4 \\
RS3 & ~20--40  & 0.60 & ~59.9 & $-$51.1~ &  172.1 &  ~61.5 & 159.0 \\
RS3 & ~40--80  & 0.28 & 106.7 & $-$10.5~ &  219.9 &  107.1 & 154.3 \\
RS3 & 80--160  & 0.03 & 187.6 &   ~67.9~ &  314.1 &  192.8 & 143.9 \\
\\
RD1 & ~~0--10  & 0.27 &  ~~9.7& $-$120.0 &  130.8 &  ~~6.1 & 171.7 \\
RD1 & ~10--20  & 0.11 & $-$7.3& $-$126.6 &  124.2 & $-$1.1 & 172.2 \\
RD1 & ~20--40  & 0.05 &  ~15.3& $-$115.9 &  141.4 &  ~12.7 & 185.5 \\
RD1 & ~40--80  & 0.02 &  ~12.0& $-$112.2 &  157.4 &  ~25.2 & 201.5 \\
RD1 & 80--160  & 0.01 &  ~30.9& $-$118.5 &  151.8 &  ~25.9 & 194.5 \\
RD3 & ~~0--10  & 0.26 & ~~4.5 & $-$121.0~ & 128.7 &  ~~5.1 & 173.6 \\
RD3 & ~10--20  & 0.12 & ~35.8 &  $-$90.1~ & 162.3 &  ~41.5 & 181.7 \\
RD3 & ~20--40  & 0.06 & ~96.2 &  $-$44.7~ & 239.4 &  108.9 & 211.5 \\
RD3 & ~40--80  & 0.04 & 188.5 &    ~54.9~ & 331.7 &  202.4 & 205.0 \\
RD3 & 80--160  & 0.02 & 341.4 &    214.9~ & 471.4 &  347.4 & 184.9 \\
\\
RT1 & ~~0--10  & 2.43 & ~~5.2 & $-$170.7~ & 176.6 &  ~~3.6 & 216.7 \\
RT1 & ~~0--20  & 4.72 & ~~4.4 & $-$161.2~ & 172.2 &  ~~5.6 & 214.9 \\
RT1 & ~~0--40  & 6.06 & ~~7.3 & $-$141.4~ & 157.1 &  ~~8.2 & 206.9 \\
RT1 & ~~0--80  & 5.29 & ~~9.2 & $-$124.6~ & 145.1 &  ~11.6 & 195.1 \\
RT1 & 80--160  & 3.54 & ~20.5 & $-$105.0~ & 150.7 &  ~25.5 & 186.4 \\
RT3 & ~~0--10  & 2.67 & ~86.1 & $-$100.8~ & 255.7 &  ~77.2 & 225.6 \\
RT3 & ~10--20  & 5.19 & 129.4 &  $-$49.1~ & 286.3 &  118.7 & 220.6 \\
RT3 & ~20--40  & 7.96 & 167.2 &    ~~6.0~ & 323.0 &  165.4 & 214.3 \\
RT3 & ~40--80  &10.46 & 204.9 &    ~80.3~ & 338.2 &  213.4 & 184.6 \\
RT3 & 80--160 & 6.64 & 301.3 &    194.4~ & 418.9 &  308.5 & 156.1 \\
\enddata
\tablecomments{
$f_d$: fraction of stars;
$v_{med}$: median velocity;
$v_{q1}, v_{q3}$: inter-quartile range;
$v_{avg}$: average velocity;
$\sigma_r$: standard deviation in radial velocity;
HV1, HV3: 1~\msun, 3~\msun\ HVS models;
RS1, RS3: 1~\msun, 3~\msun\ supernova-induced runaway models;
RD1, RD3: 1~\msun, 3~\msun\ dynamically generated runaway models;
RT1, RT3: 1~\msun, 3~\msun\ toy runaway models}
\label{tab: vr-stats}
\end{deluxetable}
\clearpage

\begin{deluxetable}{lccccccc}
\tablecolumns{8}
\tablecaption{Tangential Velocity Statistics for Runaway and Hypervelocity Stars}
\tabletypesize{\scriptsize}
\tablehead{
{Model} &
{~~d (kpc)~~} &
{~~~~$f_d$ (\%)~~~~} &
{$v_{med}$ (km~s$^{-1}$) } &
{$v_{q1}$ (km~s$^{-1}$) } &
{$v_{q3}$ (km~s$^{-1}$) } &
{$v_{avg}$ (km~s$^{-1}$) } &
{$\sigma_t$ (km~s$^{-1}$) }
}
\startdata
HV1 & ~~0--10 & 2.96 & 243.8 & 218.8 & 285.0 & 257.1 & ~74.3 \\
HV1 & ~10--20 & 2.54 & 232.8 & 205.7 & 248.4 & 230.3 & ~50.3 \\
HV1 & ~20--40& 1.73 & 219.4 & 190.7 & 234.7 & 211.5 & ~36.4 \\
HV1 & ~40--80 & 1.12 & 214.2 & 185.1 & 231.4 & 205.8 & ~31.5 \\
HV1 & 80--160 & 0.86 & 214.7 & 186.6 & 230.9 & 205.5 & ~30.0 \\
HV3 & ~~0--10 & 2.80 & 251.6 & 221.5 & 325.1 & 299.6 & 156.7 \\
HV3 & ~10--20 & 3.10 & 242.2 & 216.7 & 297.2 & 274.1 & 112.5 \\
HV3 & ~20--40 & 3.74 & 232.6 & 206.4 & 252.4 & 235.9 & ~56.9 \\
HV3 & ~40--80 & 3.77 & 226.4 & 198.8 & 238.3 & 219.1 & ~35.5 \\
HV3 & 80--160 & 2.77 & 222.6 & 195.5 & 235.5 & 212.9 & ~30.0 \\
\\
RS1 & ~~0--10 & 2.28 & 129.5 & ~79.9 & 195.0 & 145.3 & ~87.8 \\
RS1 & ~10--20 & 1.07 & 199.3 & 134.0 & 293.6 & 217.8 & 107.9 \\
RS1 & ~20--40 & 0.48 & 204.7 & 146.2 & 296.2 & 220.8 & ~93.7 \\
RS1 & ~40--80 & 0.15 & 181.7 & 146.4 & 263.9 & 201.7 & ~70.1 \\
RS1 & 80--160 & 0.03 & 187.3 & 155.4 & 246.3 & 197.6 & ~56.3 \\
RS3 & ~~0--10 & 2.30 & 129.2 &  80.1 & 193.6 & 145.2 & ~87.3 \\
RS3 & ~10--20 & 1.11 & 197.7 & 133.4 & 291.2 & 215.6 & 106.4 \\
RS3 & ~20--40 & 0.60 & 189.9 & 140.5 & 289.2 & 213.8 & ~93.7 \\
RS3 & ~40--80 & 0.28 & 185.0 & 148.9 & 269.5 & 205.9 & ~72.7 \\
RS3 & 80--160 & 0.03 & 186.3 & 150.3 & 265.5 & 204.3 & ~64.0 \\
\\
RD1 & ~~0--10 & 0.27 & 250.2 & 152.6 & 360.2 & 260.8 & 134.6 \\
RD1 & ~10--20 & 0.11 & 229.3 & 142.7 & 324.3 & 236.1 & 115.2 \\
RD1 & ~20--40 & 0.05 & 204.5 & 148.9 & 281.7 & 216.8 & ~85.5 \\
RD1 & ~40--80 & 0.02 & 211.7 & 169.4 & 253.5 & 212.4 & ~58.9 \\
RD1 & 80--160 & 0.01 & 210.8 & 174.5 & 235.0 & 206.6 & ~40.6 \\
RD3 & ~~0--10 & 0.26 & 255.2 & 153.3 & 365.7 & 262.1 & 134.0 \\
RD3 & ~10--20 & 0.12 & 232.6 & 142.5 & 327.4 & 237.8 & 116.2 \\
RD3 & ~20--40 & 0.06 & 209.9 & 148.5 & 286.8 & 218.7 & ~90.6 \\
RD3 & ~40--80 & 0.04 & 198.8 & 161.6 & 252.4 & 206.5 & ~61.4 \\
RD3 & 80--160 & 0.02 & 203.2 & 171.8 & 239.9 & 204.8 & ~45.4 \\
\\
RT1 & ~~0--10 & 2.43 & 316.5 & 244.5 & 393.1 & 320.0 & 111.7 \\
RT1 & ~10--20 & 4.72 & 243.7 & 180.8 & 312.6 & 252.8 & 104.1 \\
RT1 & ~20--40 & 6.06 & 218.9 & 177.0 & 265.2 & 221.2 & ~66.9 \\
RT1 & ~40--80 & 5.29 & 211.8 & 180.8 & 245.7 & 211.6 & ~47.4 \\
RT1 & 80--160 & 3.54 & 208.3 & 180.2 & 233.8 & 205.4 & ~38.0 \\
RT3 & ~~0--10 &~ 2.67 & ~86.1 & $-$100.8~ & 255.7 &  77.2 & 225.6 \\
RT3 & ~10--20 &~ 5.19 & 129.4 & ~$-$49.1~ & 286.3 & 118.7 & 220.6 \\
RT3 & ~20--40 &~ 7.96 & 167.2 &    ~~6.0~ & 323.0 & 165.4 & 214.3 \\
RT3 & ~40--80 & 10.46 & 204.9 &    ~80.3~ & 338.2 & 213.4 & 184.6 \\
RT3 & 80--160 &~ 6.64 & 301.3 &    194.4~ & 418.9 & 308.5 & 156.1 \\
\enddata
\tablecomments{
$f_d$: fraction of stars;
$v_{med}$: median velocity;
$v_{q1}, v_{q3}$: inter-quartile range;
$v_{avg}$: average velocity;
$\sigma_t$: standard deviation in tangential velocity;
HV1, HV3: 1~\msun, 3~\msun\ HVS models;
RS1, RS3: 1~\msun, 3~\msun\ supernova-induced runaway models;
RD1, RD3: 1~\msun, 3~\msun\ dynamically generated runaway models;
RT1, RT3: 1~\msun, 3~\msun\ toy runaway models}
\label{tab: vt-stats}
\end{deluxetable}
\clearpage

\begin{deluxetable}{lccccccc}
\tablecolumns{8}
\tablecaption{Proper Motion Statistics for Runaway and Hypervelocity Stars}
\tabletypesize{\scriptsize}
\tablehead{
{Model} &
{~~~~d (kpc)~~~~} &
{~~~~$f_d$ (\%)~~~~} &
{~~~log~$\mu_{med}$~~~} &
{~~~log~$\mu_{q1}$~~~} &
{~~~log~$\mu_{q3}$~~~} &
{~~~(${\rm log}~\mu)_{avg}$~~~} &
{~~~$\sigma({\rm log}~\mu)$~~~}
}
\startdata
HV1 & ~~0--10 & 2.96 & ~~0.89 & ~~0.77 & ~~1.04 & ~~0.92 & 0.22 \\
HV1 & ~10--20 & 2.54 & ~~0.54 & ~~0.44 & ~~0.63 & ~~0.54 & 0.14 \\
HV1 & ~20--40 & 1.73 & ~~0.20 & ~~0.12 & ~~0.29 & ~~0.20 & 0.12 \\
HV1 & ~40--80 & 1.12 &$-$0.11 &$-$0.19 &$-$0.02 &$-$0.11 & 0.11 \\
HV1 & 80--160 & 0.86 &$-$0.42 &$-$0.50 &$-$0.33 &$-$0.42 & 0.11 \\
HV3 & ~~0--10 & 2.80 & ~~0.94 & ~~0.80 & ~~1.11 & ~~0.98 & 0.25 \\
HV3 & ~10--20 & 3.10 & ~~0.56 & ~~0.46 & ~~0.67 & ~~0.58 & 0.18 \\
HV3 & ~20--40 & 3.74 & ~~0.23 & ~~0.14 & ~~0.32 & ~~0.23 & 0.14 \\
HV3 & ~40--80 & 3.77 &$-$0.09 &$-$0.17 & ~~0.00 &$-$0.09 & 0.12 \\
HV3 & 80--160 & 2.77 &$-$0.39 &$-$0.47 &$-$0.31 &$-$0.39 & 0.11 \\
\\
RS1 & ~~0--10 & 2.28 & ~~0.75 & ~~0.55 & ~~0.95 & ~~0.74 & 0.34 \\
RS1 & ~10--20 & 1.07 & ~~0.49 & ~~0.30 & ~~0.65 & ~~0.46 & 0.26 \\
RS1 & ~20--40 & 0.48 & ~~0.21 & ~~0.04 & ~~0.37 & ~~0.20 & 0.23 \\
RS1 & ~40--80 & 0.15 &$-$0.12 &$-$0.25 & ~~0.01 &$-$0.12 & 0.17 \\
RS1 & 80--160 & 0.03 &$-$0.41 &$-$0.52 &$-$0.30 &$-$0.41 & 0.15 \\
RS3 & ~~0--10 & 2.30 & ~~0.75 & ~~0.55 & ~~0.95 & ~~0.74 & 0.34 \\
RS3 & ~10--20 & 1.11 & ~~0.48 & ~~0.30 & ~~0.64 & ~~0.46 & 0.26 \\
RS3 & ~20--40 & 0.60 & ~~0.17 & ~~0.01 & ~~0.34 & ~~0.17 & 0.23 \\
RS3 & ~40--80 & 0.28 &$-$0.12 &$-$0.24 & ~~0.02 &$-$0.11 & 0.18 \\
RS3 & 80--160 & 0.03 &$-$0.36 &$-$0.46 &$-$0.22 &$-$0.35 & 0.15 \\
\\
RD1 & ~~0--10 & 0.27 & ~~0.99 & ~~0.73 & ~~1.20 & ~~0.97 & 0.41 \\
RD1 & ~10--20 & 0.11 & ~~0.55 & ~~0.33 & ~~0.71 & ~~0.50 & 0.29 \\
RD1 & ~20--40 & 0.05 & ~~0.20 & ~~0.05 & ~~0.34 & ~~0.19 & 0.22 \\
RD1 & ~40--80 & 0.02 &$-$0.10 &$-$0.21 & ~~0.02 &$-$0.10 & 0.16 \\
RD1 & 80--160 & 0.01 &$-$0.38 &$-$0.51 &$-$0.30 &$-$0.41 & 0.14 \\
RD3 & ~~0--10 & 1.04 & ~~0.98 & ~~0.74 & ~~1.19 & ~~0.96 & 0.39 \\
RD3 & ~10--20 & 0.45 & ~~0.55 & ~~0.33 & ~~0.71 & ~~0.50 & 0.29 \\
RD3 & ~20--40 & 0.24 & ~~0.19 & ~~0.03 & ~~0.34 & ~~0.18 & 0.23 \\
RD3 & ~40--80 & 0.17 &$-$0.13 &$-$0.24 &$-$0.01 &$-$0.12 & 0.16 \\
RD3 & 80--160 & 0.09 &$-$0.40 &$-$0.49 &$-$0.31 &$-$0.40 & 0.13 \\
\\
RT1 & ~~0--10 & 2.43 & ~~0.99 & ~~0.83 & ~~1.15 & ~~0.99 & 0.28 \\
RT1 & ~10--20 & 4.72 & ~~0.53 & ~~0.39 & ~~0.67 & ~~0.52 & 0.23 \\
RT1 & ~20--40 & 6.06 & ~~0.20 & ~~0.09 & ~~0.31 & ~~0.19 & 0.17 \\
RT1 & ~40--80 & 5.29 &$-$0.10 &$-$0.20 &$-$0.01 &$-$0.11 & 0.14 \\
RT1 & 80--160 & 3.54 &$-$0.41 &$-$0.50 &$-$0.32 &$-$0.41 & 0.12 \\
RT3 & ~~0--10 & 2.67 & ~~0.99 & ~~0.83 & ~~1.16 & ~~0.99 & 0.29 \\
RT3 & ~10--20 & 5.19 & ~~0.53 & ~~0.38 & ~~0.68 & ~~0.52 & 0.24 \\
RT3 & ~20--40 & 7.96 & ~~0.20 & ~~0.07 & ~~0.32 & ~~0.19 & 0.19 \\
RT3 & ~40--80 &10.46 &$-$0.12 &$-$0.22 &$-$0.01 &$-$0.12 & 0.15 \\
RT3 & 80--160 & 6.64 &$-$0.39 &$-$0.48 &$-$0.30 &$-$0.40 & 0.13 \\
\enddata
\tablecomments{
$f_d$: fraction of stars;
$\mu_{med}$: median proper motion;
$\mu_{q1}, \mu_{q3}$: inter-quartile range;
$({\rm log}~\mu)_{avg}$: average log proper motion;
$\sigma({\rm log}~\mu)$: standard deviation in log proper motion;
all proper motions in units of \masyr;
HV1, HV3: 1~\msun, 3~\msun\ HVS models;
RS1, RS3: 1~\msun, 3~\msun\ supernova-induced runaway models;
RD1, RD3: 1~\msun, 3~\msun\ dynamically generated runaway models;
RT1, RT3: 1~\msun, 3~\msun\ toy runaway models}
\label{tab: mu-stats}
\end{deluxetable}
\clearpage

\clearpage

\clearpage

\clearpage

\clearpage

\clearpage

\clearpage

\clearpage

\clearpage

\clearpage

\clearpage

\clearpage

\clearpage

\clearpage

\clearpage

\clearpage

\clearpage

\clearpage

\clearpage

\clearpage

\clearpage

\clearpage

\clearpage

\clearpage

\clearpage

\clearpage

\clearpage

\clearpage

\clearpage

\clearpage

\end{document}